\documentclass[a4paper,10pt]{article}
\usepackage[utf8]{inputenc}
\usepackage{authblk} 
\usepackage{amsmath} 
\usepackage{amssymb} 
\usepackage{graphicx}
\usepackage{hyperref} 
\usepackage{cite} 
\usepackage{color}
\usepackage{adjustbox}
\usepackage{subfig}
\usepackage{floatrow}
\usepackage{array}
\usepackage{multirow}
\usepackage{arydshln}
\usepackage{cancel}
\usepackage{soul}
\usepackage{wrapfig}
\usepackage{feynmf}
\usepackage[table]{xcolor}
\usepackage{color}
\usepackage{colortbl}
\usepackage{multirow}
\usepackage{graphicx}
\usepackage{caption}
\usepackage{adjustbox}
\usepackage{dashrule}
\usepackage{arydshln}
\usepackage{url}
\usepackage[super]{nth}
\usepackage{enumitem}
\setlist[itemize]{noitemsep, topsep=0pt}
\usepackage{orcidlink} 

\usepackage[top=2cm, bottom=2cm, left=1.5cm, right=1.5cm]{geometry} 
\usepackage{abstract} 
\title{Alternative framework for the left-right symmetric model including vector-like fermions}
\author[1]{Yassine Bouzeraib \thanks{Email: \href{mailto:yassine.bouzeraib@univ-jijel.dz}{yassine.bouzeraib@univ-jijel.dz}}}
\author[1]{Mohamed Sadek Zidi \thanks{Email: \href{mailto:mohamed.sadek.zidi@univ-jijel.dz}{mohamed.sadek.zidi@univ-jijel.dz}}}
\affil[1]{LPTh, Department of Physics, Faculty of Exact and Computer Sciences, University of Jijel,\\B. P. 98 Ouled Aissa, 18000 Jijel, Algeria}

\begin{document}

\maketitle
\begin{abstract}
We extend the left-right symmetric model with an additional non-abelian $SU(2)$ gauge symmetry. The particle content is augmented by one generation of vector-like fermions transforming under the fundamental representation of this new gauge group. Consequently, new self-dual scalar fields have been introduced for the sake of breaking the symmetry and invoking the mixing of vector-like fermions with the chiral fermions. This model explains the smallness of the neutrinos masses, showing that the 1rst and 2nd generation neutrinos masses are governed by the seesaw relation of the LRSM, while the 3rd generation neutrino mass is controlled by a new seesaw relation which involves the VLN.
We investigated the production of a resonant extra charged gauge boson which decays into vector-like quarks in association with 3rd generation or into heavy Majorana neutrinos. We exploited the {\tt run II} LHC data to set lower limits on the mass of $W^{\prime}$, and consequentially on the $Z^{\prime}$. We find that the most restrictive constraints come from the decay of the $W^{\prime}$ to the 2nd generation heavy Majorana neutrinos. We also examine the single production of the top vector-like quark with $t$ and $b$ quarks, where parton-shower is taken into account.

\vspace{0.25cm}
\end{abstract}


\section{Introduction}
\label{int}

\noindent
 The Standard Model (SM) of particle physics is an extremely successful theory from theoretical and experimental perspectives as it has predicted and shown a significant compatibility with many experimental results. Even though, it is still an incomplete theory where it doesn't give an answer for several problems, such as parity violation in weak interactions, the neutrino masses, the quadratic divergence in the Higgs boson mass and the dark matter, which emphasizes the need of for physics beyond the SM (BSM).\\

 \noindent
 Many extensions have been introduced to solve these problems, and one of the most important extensions that has been considered for a long time is the left-right symmetric model (LRSM) \cite{Pati:1974yy,Mohapatra:1974gc,Senjanovic:1975rk,P,a,D,M}. This model belongs to the category of models with extended gauge symmetry, where it gives a possible solution for neutrino mass problem and explains their smallness naturally through the see-saw mechanism \cite{seesaw,seesaw1}. Moreover, it provides an explanation for the parity violation at the weak scale. As several BSM models, the LRSM predicts the existence of extra gauge bosons, heavy neutrinos (HNs) and extra scalars. Another category of extensions is based on adding new matter content to the SM, and one of the most attractive exotic particles is the hypothetical vector-like fermions (VLFs). Since both left- and right-handed parts of VLFs fields belong to the same representation of the gauge group, the VLFs can be added to the fermion content of the model without causing any axial anomalies. Mixing of the VLFs with the SM fermions has been studied extensively in recent years, see for example~\cite{La,Ba,Aguilar1,Aguilar2}. Notably, the existence of the new vector-like top quark would resolve the quadratic divergence in the Higgs boson mass \cite{Higgs-mass} and explains the observed $B$-meson anomalies~\cite{s}. Alternatively, the vector-like neutrinos could serve as a dark matter candidate~\cite{DM1,DM,BBZ}.\\

\noindent
In this work, we present an alternative model based on the LRSM. We extend the gauge group by an additional $SU(2)$ factor, which contains one generation of VLFs. Due to the left-right symmetry of the model, parity violation is naturally explained: parity is conserved before symmetry breaking, while the observed parity violation at the weak scale emerges as a consequence of spontaneous symmetry breaking (SSB). The dedicated model predicts an asymmetry in the VLFs mixing angles to the chiral fermions. More precisely, the SSB leads to an asymmetry between the mixing of the VLFs doublet with the SM chiral fermions, where they would mix just through the right-sector mixing angles. This feature shared by many BSM theories~\cite{Han:2003wu,Buchkremer:2013bha} and it is studied extensively in several phenomenological models involving vector-like quarks (VLQs)~\cite{Aguilar1,Aguilar2,Zidi:2018avr}. The the structure of the model allows  for the existence of a wide range of hypothetical particles; in particular, extra scalars, extra gauge bosons, heavy neutrinos and VLFs. These new states can be produced in different ways at the LHC and would therefore be the subject of new physics research at collider~\cite{ATLAS:2024auw, Alcaide:2019kdr, ATLAS:2021jol,ATLAS:2018ceg, Solera:2023kwt, Osland:2020onj, ATLAS:2023vxg, ATLAS:2023taw, Kalsi:2018dgi, ATLAS:2019erb, ATLAS:2019lsy, ATLAS:2023cjo, ATLAS:2023tkz, CMS:2022hvh, CMS:2023jqi, CMS:2022fck, CMS:2018wpl, ATLAS:2024zlo, ATLAS:2024mrr, ATLAS:2023sbu, CMS:2022cpe}. One of the main achievements of this model is that it explains naturally the neutrinos masses. Furthermore, the vector-like neutrino would be a good dark matter candidate under certain conditions on Yukawa couplings, where it is compatible with the constraints of direct and indirect detection for a wide range of the parameters. We leave this part to the dedicated article~\cite{BBZ}.\\

\noindent
 A search for a resonance of new gauge bosons decaying through VLQs in association with one of the third generation quarks, or through the decay to HNs with leptons has been investigated recently by the CMS collaboration~\cite{CMS:2022tdo,CMS:2017hwg,CMS:2021dzb, CMS:2017xcw}. In this work, we use this data to derive lower bounds on the lightest extra  charged gauge boson $W^{\prime}$ mass, by analyzing its decay into VLQs and HNs. This consequently lead to setting lower bounds on the mass of the neutral gauge boson $Z^{\prime}$. To simplify the study, we assumed that the scalar sector decouples from the scale of interest, by setting their masses to several TeVs.
The additional VLQ fields might affect their couplings to $Z$ boson. However, such deviation is highly constrained from the atomic parity violation experiments and the measurement of $R_c$ from LEP~\cite{LEP1, LEP2}, especially in the case VLQs mixing with 1rst and 2nd quark generations.  However, such mixing remain less constrained than the mixing with the two first generation because of the hierarchy on the Yukawa coupling. Thus, it is more convenient to assume that the VLQs mix only with the quarks $t$ and $b$.\\

 \noindent
 The paper is organized as follows. In section~\ref{sec2}, we provide a general description of the model. In section~\ref{sec3}, we discuss the scalar fields vacuum expectation values (vevs) and the patterns of symmetry breaking. The section \ref{sec4} is dedicated to discuss the scalars, gauge bosons  and fermions masses and VLF mixing with ordinary fermions. Next, in section \ref{W_pheno}, we make a comprehensive analysis for the decay widths of $W^{\prime}$ and the VLQs, incorporating the experimental bounds on the VLQs-quarks mixing angles. We check out the validity of the Narrow Width approximation (NWA) during the calculation of the cross section of $W^{\prime}$ production and decay through the channels of interest, where lower limits on $W^{\prime}$ mass have been established. In section \ref{LHC_VLQ_singel}, study the $T$ quark single production in association with SM 3rd generation quarks. Several differential distributions are produced, where the calculation is matched the parton shower (PS), for several benchmark values of the model new gauge coupling. We conclude this paper by giving a summary and outlooks in section~\ref{conc}.\\

\noindent
 For the sake of completeness, we provide three appendices (\ref{appA}, \ref{diagNmass} and \ref{appD}). In the former one, we provide the minimization conditions of the model Higgs potential, which lead to the vev seesaw relation. In the secod one, we diagonalize the Dirac-Majorana neutrino mass matrix. In the latter one, we discuss in detail the scalar sector rotations and mass matrices diagonalization.

\section{Construction of the model}
\label{sec2} 

\subsection{Model gauge group}
\noindent   
We build a LRSM based on the extended gauge group $\mathcal{G}_{_\text{VLRSM}}\equiv SU(2)_V\times SU(2)_L\times SU(2)_R\times U(1)_{B-L}$. The subscript $V$ in the extra gauge factor $SU(2)_V$ stands for vector-like-fermions (VLQs and VLLs) embedded in the fundamental representation of this group. The left- and the right-handed parts of the chiral fermion fields are arranged as doublets under $SU(2)_L$ and $SU(2)_R$, respectively. The abelian factor $U(1)_{B-L}$ is generated by the $B-L$ quantum number~\footnote{$B-L$ denotes the baryonic number minus the leptonic number.}. The chiral and VL fermion fields are assigned to the irreducible representations of $\mathcal{G}_{_\text{VLRSM}}$ as follows:
 \begin{align}
  l^{i}_{L}&\equiv\begin{pmatrix}
\nu^{i} \\
e^{i}
\end{pmatrix}_{L}\sim ({\bf 1},{\bf2},{\bf1})_{-1} & q^{i}_{L}&\equiv\begin{pmatrix}
u^{i} \\
d^{i}
\end{pmatrix}_{L}\sim ({\bf1},{\bf2},{\bf1})_{\frac{1}{3}}
\\
l^{i}_{R}&\equiv\begin{pmatrix}
\nu^{i} \\
e^{i}
\end{pmatrix}_{R}\sim ({\bf1},{\bf1},{\bf2})_{-1} & q^{i}_{R}&\equiv\begin{pmatrix}
u^{i} \\
d^{i}
\end{pmatrix}_{R}\sim ({\bf1},{\bf1},{\bf2})_{\frac{1}{3}}\\
L_{L,R}&\equiv\begin{pmatrix}
N \\
E
\end{pmatrix}_{L,R}\sim ({\bf2},{\bf1},{\bf1})_{-1} & Q_{L,R}&\equiv\begin{pmatrix}
T \\
B
\end{pmatrix}_{L,R}\sim ({\bf2},{\bf1},{\bf1})_{\frac{1}{3}}
 \end{align}
where doublets $l^{i}_{L}$ ($l^{i}_{R}$) and $q^{i}_{L}$ ($q^{i}_{R}$) are the left-handed (right-handed) ordinary leptons and quarks fields of a given generation $i$ (with $i=1,2,3$). The doublets $L_L$ ($L_R$) and $Q_L$ ($Q_R$) stand for the left-handed (right-handed) vector-like leptons and quarks fields, which are assumed to come in one generation.
 The electric charge generator $Q$ is defined as a sum of the diagonal generators of $\mathcal{G}_{_\text{VLRSM}}$, thus we have:
 \begin{align}
  Q=\tilde{T}^{3}_{V}+\tilde{T}^{3}_{L}+\tilde{T}^{3}_{R}+\frac{\left(B-L\right)}{2}
  \label{chargeBmL}
 \end{align}
 where the $\tilde{T}^{3}_{V/L/R}$ are the third generators the sub-groups $SU(2)_{V/L/R}$ and $(B-L)/2$ is the generator of $U(1)_{_{B-L}}$.\\

\noindent
We denote the gauge bosons and couplings associated to each gauge factor of the model by:
 \begin{align}
  &SU(2)_V: W^j_{V}, g_{_V}&
  &SU(2)_L: W^j_{L}, g_{_L}&
  &SU(2)_R: W^j_{R}, g_{_R}&
  &U(1)_{B-L}: B, g^{\prime}
 \end{align}
whit $j=1,2,3$. \\

\noindent
 We recall that one of the model important characteristic is the parity conservation, which arises from its left-right symmetry before the SSB. This translates to the invariance of the Lagrangian under the exchange of the left and right components of all fermion ($\Psi$) and gauge boson ($W$) fields, i.e. 
\begin{align}
  \Psi_L \longleftrightarrow \psi_R && \vec W_{L\mu}\longleftrightarrow\vec W_{R\mu}
 \end{align}
which leads to $g_{_L}=g_{_R}=g$, where $g$ is the usual SM gauge coupling. It should be noted that the coupling $g_{_V}$ is not constrained by this symmetry, since the gauge group $SU(2)_V$ contains both left and right parts of the VLFs fields. \\

\noindent
We note that all the fields introduced here belong to the weak basis, meaning that all gauge and chiral fermionic fields are massless. The generation of the mass will be addressed in the next sections.
\subsection{Model scalar fields}
\label{sf}
 We need to introduce two scalar self-dual bi-doublets, $\Phi_{_{VL}}$ and $\Phi_{_{VR}}$ to allow the VLFs (belonging to $SU(2)_{V}$) to mix with the ordinary fermions (belonging to $SU(2)_{L}\times SU(2)_{R}$)~\footnote{Each of these fields satisfy the self-duality condition: $\tilde{\Phi}=\sigma_2\,\,\Phi^{*}\sigma_2=\Phi$, where $\sigma_2$ is the 2nd Pauli matrix.}. The remaining scalar fields are similar to those required to break the symmetry of the usual LRSM. Thus, we introduce a bi-doublet $\Phi$ which enables to mix the right- and left-handed chiral fermions via the Yukawa interaction, and two scalar triplets $\Delta_L$ and $\Delta_R$ embedded in the adjoint representation of $SU(2)_L$ and $SU(2)_R$, respectively.  The latter two triplets are necessary for introducing the neutrino Majorana states, which are essential to explain the smallness of the neutrinos masses through {\it see-saw} mechanism. Following the notation of ref.~\cite{vladimir}, the scalar fields matrix representation and their assignment under $\mathcal{G}_{_\text{VLRSM}}$ are:
   \begin{align}
  \Phi_{_{VL}}=\begin{pmatrix}
  \phi^0_{_{1}} &  -\phi^{+}_{_{1}}\\
   \phi^{-}_{_{1}} & \phi^{0*}_{_{1}}\\
   \end{pmatrix}\sim({\bf2},{\bf2},{\bf1})_{0}
   &&
  \Phi_{_{VR}}=\begin{pmatrix}
  \phi^0_{_{2}} &  -\phi^{+}_{_{2}}\\
   \phi^{-}_{_{2}} & \phi^{0*}_{_{2}}\\
   \end{pmatrix}\sim({\bf2},{\bf1},{\bf2})_{0}
   &&
  \Phi=\begin{pmatrix}
  \phi^0_{_{3}} &  -\phi^{+}_{_{4}}\\
   \phi^{-}_{_{3}} & \phi^{0*}_{_{4}}\\
   \end{pmatrix}\sim({\bf1},{\bf2},{\bf2})_{0}
    \end{align}
 \begin{align}
\Delta_{_{L}}=\begin{pmatrix}
\delta^{+}_{_{L}}/\sqrt{2} & \delta_{_{L}}^{++} \\
 \delta_{_{L}}^{0} & -\delta^{+}_{_{L}}/\sqrt{2}
\end{pmatrix}\sim({\bf1},{\bf3},{\bf1})_{2}
&&
\Delta_{_{R}}=\begin{pmatrix}
\delta^{+}_{_{R}}/\sqrt{2} & \delta_{_{R}}^{++} \\
 \delta_{_{R}}^{0} & -\delta^{+}_{_{R}}/\sqrt{2}
\end{pmatrix}\sim({\bf1},{\bf1},{\bf3})_{2}
\end{align}

\noindent
The symmetry of the model is broken into $U(1)_{\text{em}}$ in three stages, by the following scalar field vevs~\footnote{These vevs can simply obtained by giving the electric neutral components of the scalar fields non zero vevs.}~\cite{Bhupal:Dev,
Hsieh}:
\begin{align}
 <\Phi_{_{VL}}>&=\frac{1}{\sqrt{2}}\begin{pmatrix}
u_{_L}e^{i\theta_1} & 0 \\
0 & u_{_L}e^{i\theta_2}
\end{pmatrix}&
<\Phi_{_{VR}}>&=\frac{1}{\sqrt{2}}\begin{pmatrix}
u_{_R}e^{i\theta_2} & 0 \\
0 & u_{_R}e^{i\theta_2}
\end{pmatrix}
&
 <\Phi>&=\frac{1}{\sqrt{2}}\begin{pmatrix}
k_1e^{i\theta_3} & 0 \\
0 & k_2e^{i\theta_4}
\end{pmatrix}\nonumber\\
<\Delta_{L}>&=\frac{1}{\sqrt{2}}\begin{pmatrix}
0 & 0 \\
v_{_L}e^{i\theta_L} & 0
\end{pmatrix}&
<\Delta_{R}>&=\frac{1}{\sqrt{2}}\begin{pmatrix}
0 & 0 \\
v_{_R}e^{i\theta_R} & 0
\end{pmatrix}
\label{vevs}
\end{align}

\noindent
Here, we have:
 $k_1=v_{_{EW}}\,c_\beta$,  $k_2=v_{_{EW}}\,s_\beta$ (i.e. $k_1^2+k_2^2=v_{_{EW}}^2$),
where the angle $\beta$ is treated as a free parameter and $v_{_{EW}}$ being the SM vev (with $c_\beta=\cos(\beta)$ and $s_\beta=\sin(\beta)$). To ensure that the SM top quark is heavier than bottom quark, we assume that $k_2\ll k_1$ or equivalently $s_{\beta}\ll 1$. The vevs $u_{_{R}}$ and $v_{_R}$ are on the order of TeV, while $u_{_{L}}$ and $v_{L}$ may vanish as we will show later. We emphasize that the non-vanishing vevs are assumed to respect the hierarchy $v_{_{EW}}\ll v_{_R} < u_{_R}$.  We note that the phases $\theta_1$, $\theta_2$, $\theta_3$ and $\theta_R$, cf.~eq.~(\ref{vevs}), can be absorbed by imposing some underlying unitary transformations involving the third generators of the $SU(2)$ gauge factors~\cite{D,Bhupal:Dev,Z}.\\

\noindent
It is crucial to notice that the breakdown of parity is directly linked to the violation of $B-L$ quantum number. Slightly above the electroweak scale, all the generators of the $\mathcal{G}_{_\text{VLRSM}}$ are broken except $\tilde{T}_L^3$ (i.e. $\Delta \tilde{T}_L^3\approx 0$). Thus, using the electric charge formula eq.~(\ref{chargeBmL}) and requiring $\Delta Q=0$, we can write: $\Delta\tilde{T}_V^3+\Delta\tilde{T}_R^3=-\Delta(B-L)/2$. This means that the breakdown of $B-L$, parity (and the extra symmetry $V$) are related ~\cite{Mohapatra:1980yp} for more detail. 
 
 \subsection{Model Lagrangian}
The Model Lagrangian takes the following general form:
\begin{align}
 {\cal{L}}= {\cal{L}}_{F}+{\cal{L}}_{G}+{\cal{L}}_{S}+{\cal{L}}_{Y}
 \label{l1}
\end{align}
where ${\cal{L}}_{F}$, ${\cal{L}}_{G}$, ${\cal{L}}_{S}$ and ${\cal{L}}_{Y}$ are the fermion, the gauge bosons, the scalar and the Yukawa Lagrangians, respectively.\\

\noindent
The fermion Lagrangian is given by:
  \begin{align}
 {\cal{L}}_{F}&=\sum_{\psi=l,q,L,Q}\,\left[\bar \psi_{L} (i\not{\!\!D}_{\mu}) \psi_{L}+\bar \psi_{R} (i\not{\!\!D}_{\mu}) \psi_{R}\right].
 \label{LFerm}
\end{align}
The covariant derivatives acting on the chiral fermions and the VLFs are, respectively, expressed as:
\begin{align}
D_{\mu}=\partial_{\mu}-ig_{\left(L,R\right)}\,\frac{\vec{\sigma}}{2}\cdot\vec{W}_{\left(L,R\right)\mu}-ig'\frac{\left(B-L\right)}{2}B_{\mu}
&&
D_{\mu}=\partial_{\mu}-ig_{V}\,\frac{\vec{\sigma}}{2}\cdot\vec{W}_{V\mu}-ig'\frac{\left(B-L\right)}{2}B_{\mu}
\end{align}
\noindent
where the $\vec{\sigma}=(\sigma_1, \sigma_2, \sigma_3)$ is a vector where the components are the Pauli matrices.\\

\noindent
The gauge field Lagrangian ${\cal{L}}_G$ is
\begin{align}
 {\cal{L}}_G=-\frac{1}{4} W_{V_i\mu\nu}W_{V_i}^{\mu\nu}-\frac{1}{4}W_{L_i\mu\nu}W_{L_i}^{\mu\nu}-\frac{1}{4}W_{R_i\mu\nu}W_{R_i}^{\mu\nu}-\frac{1}{4}B_{\mu\nu}B^{\mu\nu}
\end{align}
where the gauge field strength tensors are defined by:
\begin{align}
W_{\bullet_i\mu\nu}&=\partial_{\mu}W_{\bullet_i\nu}-\partial_{\nu}W_{\bullet_i\mu}+g_{\bullet_i}\varepsilon_{ijk}\, W_{\bullet_j\mu}W_{\bullet_k\nu},
&
B_{\mu\nu}&=\partial_{\mu}B_{\nu}-\partial_{\nu}B_{\mu}.
\end{align}
with "$\bullet$" refs to $V$, $L$ or $R$.

\noindent
The Yukawa Lagrangian ${\cal{L}}_Y$ is constructed from the most general mixing of the different fermions of the model via the scalar multiplets. The general form of the Yukawa Lagrangian, which enables to generate the masses of the fermions, is given by:
\begin{align}
 {\cal{L}}_Y={\cal{L}}_{\Phi_{_{VL}}}+{\cal{L}}_{\Phi_{_{VR}}}+{\cal{L}}_{\Phi}+ {\cal{L}}_{\Delta_{_{L,R}}}+{\cal{L}}_M
 \label{Yuk}
\end{align}

\noindent
The Yukawa term, that describes the interaction of the right-handed VLFs and the left chiral fermions via the field $\Phi_{VL}$, is given by:
 \begin{align}
  {\cal{L}}_{\Phi_{VL}}=-\overline{Q}_R \lambda^{\dagger}_q\,\Phi_{_{VL}}\,q_L-\overline{L}_R \lambda^{\dagger}_l\,\Phi_{_{VL}}\,l_L+\text{h.c}
 \end{align}
 while the left-handed part of the VLFs and the right chiral fermions via the field $\Phi_{_{VR}}$, is:
  \begin{align}
  {\cal{L}}_{\Phi_{_{VR}}}=-\overline{Q}_L \lambda^{\prime}_q\,\Phi_{_{VR}}\,q_R-\overline{L}_L \lambda^{\prime}_l\,\Phi_{_{VR}}\,l_R+\text{h.c}
 \end{align}
 where $\lambda_l$, $\lambda^{\prime}_l$, $\lambda_q$ and $\lambda^{\prime}_q$ are $1\times3$ matrices.

 \noindent
 The mixing between the left- and right-handed chiral fermions is done through the field $\Phi$,
 \begin{align}
  {\cal{L}}_{\Phi}=-\bar l_{L}\,\left(y_l\,\Phi+\tilde{y}_l\,\tilde{\Phi}\right)l_{R}-\bar q_{L}\,\left(y_q\,\Phi+\tilde{y}_q\,\tilde{\Phi}\right)q_{R}+\text{h.c}
 \end{align}
Until now, all the fermionic fields are of Dirac states. As a consequence, neutrinos masses after breaking the symmetry will be at the same scale of the corresponding charged leptons, unless if we finetuned the Yukawa couplings to be unnaturally small. Otherwise, including three heavy Majorana states will fix the problem. The Yukawa Lagrangian of this part is written as:
 \begin{align}
  {\cal{L}}_{\Delta_{_{L,R}}}=-\overline{\left(l_{L}\right)^c}\Sigma_{_L}\left(y_M\right) l_{L}-\overline{\left(l_{R}\right)^c}\Sigma_{_R}\left(y_M\right)l_{R}+h.c
 \end{align}
 where $\Sigma_{_{L,R}}=i\sigma_2\Delta_{_{L,R}}$ and the relations $\left(l_{L,R}\right)^c=i\sigma_2l_{L,R}$, $\overline{\left(l_{L,R}\right)^c}=(\bar {l}^c)_{R,L}$ are satisfied. $y_{l,q}$, $\tilde{y}_{l,q}$ and $y_M$ are $3\times 3$ Yukawa matrices in flavor space while $\lambda_{l,q}$, $\lambda^{\prime}_{l,q}$ are $3\times 1$ and $1\times 3$ Yukawa matrices, respectively~\cite{P,a,D,M}.\\

\noindent
The VLFs masses are not generated through Yukawa term. Instead of that, we are able to add a bare Dirac mass term without spoiling the gauge invariance. The mass term can be wriiten as:
\begin{align}
 {\cal{L}}_M=-\overline{Q}_L\,M_Q\,Q_R-\overline{L}_L\,M_L\,L_R+\text{h.c}
 \label{massVLFs}
\end{align}
where $M_Q$ and $M_L$ are the VLQs and VLLs bare masses, respectively.

\section{Patterns of symmetry breaking}
\label{sec3}
\noindent
The scalar Lagrangian has the following compact form: 
\begin{align}
{\cal{L}}_S&=\text{Tr}[\left(D_{\mu}\Phi_{_{VL}}\right)^{\dagger}\left(D_{\mu}\Phi_{_{VL}}\right)]+
  \text{Tr}[\left(D_{\mu}\Phi_{_{VR}}\right)^{\dagger}\left(D_{\mu}\Phi_{_{VR}}\right)]+\text{Tr}[\left(D_{\mu}\Phi\right)^{\dagger}\left(D_{\mu}\Phi\right)]\nonumber\\&+ \text{Tr}[\left(D_{\mu}\Delta_{_{L}}\right)^{\dagger}\left(D_{\mu}\Delta_{_{L}}\right)]+ \text{Tr}[\left(D_{\mu}\Delta_{_{R}}\right)^{\dagger}\left(D_{\mu}\Delta_{_{R}}\right)]-V
  \label{Lscalar}
\end{align}
where the explicit formulae of the covariant derivatives acting on each scalar multiplet are given by:
 \begin{align}
&D_{\mu}\Phi_{_{V\{L,R\}}}=\partial_{\mu}\Phi_{_{V\{L,R\}}}-ig_{_{V}}\,\mathbb{W}_{V\mu}\Phi_{_{V\{L,R\}}}+ig_{_{\{L,R\}}}\,\Phi_{_{V\{L,R\}}}\mathbb{W}_{\{L,R\}\mu}\nonumber\\
&D_{\mu}\Phi=\partial_{\mu}\Phi-ig_{_{L}}\,\mathbb{W}_{L\mu}\Phi+ig_{_{R}}\,\Phi\mathbb{W}_{R\mu}\nonumber\\ 
&D_{\mu}\Delta_{_{L,R}}=\partial_\mu\Delta_{_{L,R}}-i\,g_{_{L,R}}\,\left[\mathbb{W}_{L,R\mu},\Delta_{_{L,R}}\right]-i\,g^{\prime}B_\mu\,\Delta_{_{L,R}}
\end{align}
 with $\mathbb{W}\equiv \vec{\sigma}\cdot\vec{W}/2$. The Higgs potential $V$ consists of the ordinary LRSM potential (denoted $V_{_\text{LRSM}}$) and the potential associated with the self dual bi-doublet fields and their interaction  with the standard LRSM scalars (denoted $V_{_\text{VLRSM}}$). Thus, we write
 \begin{align}
  V&=V_{_\text{LRSM}}+V_{_\text{VLRSM}}
  \label{potentialTot}
  \end{align}
where
\begin{align}
 V_{_\text{LRSM}}=&-\mu_1^2\left(\text{Tr}[\Phi\Phi^{\dagger}]\right)-\mu_2^2\left(\text{Tr}[\tilde{\Phi}\Phi^{\dagger}]+\text{Tr}[\tilde{\Phi}^{\dagger}\Phi]\right)-
 \mu_3^2\left(\text{Tr}[\Delta_L\Delta^{\dagger}_L]+\text{Tr}[\Delta_R\Delta^{\dagger}_R]\right)\nonumber\\
 &+\lambda_1 \text{Tr}[\Phi\Phi^{\dagger}]^2+\lambda_2\left(\text{Tr}[\tilde{\Phi}\Phi^{\dagger}]^2+\text{Tr}[\tilde{\Phi}^{\dagger}\Phi]^2\right)+
 \lambda_3 \text{Tr}[\tilde{\Phi}\Phi^{\dagger}]\text{Tr}[\tilde{\Phi}^{\dagger}\Phi]+\lambda_4 \text{Tr}[\Phi\Phi^{\dagger}](\text{Tr}[\tilde{\Phi}\Phi^{\dagger}]+\text{Tr}[\tilde{\Phi}^{\dagger}\Phi]) \nonumber\\
 &+\rho_1\left((\text{Tr}[\Delta_L\Delta^{\dagger}_L])^2+(\text{Tr}[\Delta_R\Delta^{\dagger}_R])^2\right)+\rho_2\left(\text{Tr}[\Delta_L\Delta_L]\text{Tr}[\Delta^{\dagger}_L\Delta^{\dagger}_L]+
 \text{Tr}[\Delta_R\Delta_R]\text{Tr}[\Delta^{\dagger}_R\Delta^{\dagger}_R]\right)\nonumber\\
 &+\rho_3\left(\text{Tr}[\Delta_L\Delta^{\dagger}_L]\text{Tr}[\Delta_R\Delta^{\dagger}_R]\right)+
 \rho_4\left(\text{Tr}[\Delta_L\Delta_L]\text{Tr}[\Delta^{\dagger}_R\Delta^{\dagger}_R]+\text{Tr}[\Delta^{\dagger}_L\Delta^{\dagger}_L]\text{Tr}[\Delta_R\Delta_R]\right)\nonumber\\
 &+\alpha_1\left(\text{Tr}[\Phi\Phi^{\dagger}](\text{Tr}[\Delta_L\Delta^{\dagger}_L]+\text{Tr}[\Delta_R\Delta^{\dagger}_R])\right)\nonumber\\
 &+\alpha_2\left(\text{Tr}[\Phi\tilde{\Phi}^{\dagger}]\text{Tr}[\Delta_R\Delta^{\dagger}_R]+\text{Tr}[\Phi^{\dagger}\tilde{\Phi}]\text{Tr}[\Delta_L\Delta^{\dagger}_L]+\text{hc.}\right)+\alpha_3\left(\text{Tr}[\Phi\Phi^{\dagger}\Delta_L\Delta_L^{\dagger}]+\text{Tr}[\Phi^{\dagger}\Phi\Delta_R\Delta_R^{\dagger}]\right)\nonumber\\
 &+\beta_1\left(\text{Tr}[\Phi\Delta_R\Phi^{\dagger}\Delta^{\dagger}_L]+\text{Tr}[\Phi^{\dagger}\Delta_L\Phi\Delta^{\dagger}_R]\right)+\beta_2\left(\text{Tr}[\tilde{\Phi}\Delta_R\Phi^{\dagger}\Delta^{\dagger}_L]+\text{Tr}[\tilde{\Phi}^{\dagger}\Delta_L\Phi\Delta^{\dagger}_R]\right)\nonumber\\
 &+\beta_3\left(\text{Tr}[\Phi\Delta_R\tilde{\Phi}^{\dagger}\Delta^{\dagger}_L]+\text{Tr}[\Phi^{\dagger}\Delta_L\tilde{\Phi}\Delta^{\dagger}_R]\right)
 \label{Potential1}
\end{align}
and
\begin{align}
 V_{_\text{VLRSM}}&=-\mu_4^2\left( \text{Tr}[\Phi_{_{VL}}\Phi_{_{VL}}^{\dagger}]+ \text{Tr}[\Phi_{_{VR}}\Phi_{_{VR}}^{\dagger}]\right)+\lambda_5\left( (\text{Tr}[\Phi_{_{VL}}\Phi_{_{VL}}^{\dagger}])^2+ (\text{Tr}[\Phi_{_{VR}}\Phi_{_{VR}}^{\dagger}])^2\right)\nonumber\\
 &+\alpha_4\left( \text{Tr}[\Phi_{_{VL}}\Phi_{_{VL}}^{\dagger}]\text{Tr}[\Phi_{_{VR}}\Phi_{_{VR}}^{\dagger}]\right)+\alpha_5\left(\text{Tr}[\Phi\Phi^{\dagger}](\text{Tr}[\Phi_{_{VL}}\Phi_{_{VL}}^{\dagger}]+\text{Tr}[\Phi_{_{VR}}\Phi_{_{VR}}^{\dagger}]\right)\nonumber\\
 &+\alpha_6\left(\text{Tr}[\Phi\tilde{\Phi}^{\dagger}](\text{Tr}[\Phi_{_{VL}}\Phi_{_{VL}}^{\dagger}]+\text{Tr}[\Phi_{_{VR}}\Phi_{_{VR}}^{\dagger}])+\text{hc.}\right)\nonumber\\
&+\alpha_7\left( \text{Tr}[\Phi_{_{VL}}\Phi_{_{VL}}^{\dagger}]+\text{Tr}[\Phi_{_{VR}}\Phi_{_{VR}}^{\dagger}]\right)\left(\text{Tr}[\Delta_L\Delta^{\dagger}_L)]+\text{Tr}[\Delta_R\Delta^{\dagger}_R)]\right)\nonumber\\
 &+\alpha_8\left( \text{Tr}[\Phi_{_{VL}}^{\dagger}\Phi_{_{VL}}\Delta_L\Delta^{\dagger}_L]+\text{Tr}[\Phi_{_{VR}}^{\dagger}\Phi_{_{VR}}\Delta_R\Delta^{\dagger}_R]\right)+\alpha_9\left( \text{Tr}[\Phi_{_{VL}}\Phi_{_{VL}}^{\dagger}\Phi_{_{VR}}\Phi_{_{VR}}^{\dagger}]\right)\nonumber\\
 &+\alpha_{10}\left( \text{Tr}[\Phi_{_{VL}}^{\dagger}\Phi_{_{VL}}\Phi\Phi^{\dagger}]+\text{Tr}[\Phi_{_{VR}}^{\dagger}\Phi_{_{VR}}\Phi^{\dagger}\Phi]\right)
 \label{Potential2}
\end{align}
This potential is the most general renormalizable and gauge invariant under the group $\mathcal{G}_{_\text{VLRSM}}$ that one can construct.  We note that it can be invariant under the discrete symmetries of the parity $\mathcal{P}$ or the parity $\mathcal{C}$~\cite{w,Bhupal:Dev}:
\begin{align}
 P:&&  \Delta_{_{L}}\longleftrightarrow \Delta_{_{R}}&& \Phi\longleftrightarrow\Phi^{\dagger}&&\Phi_{_{VL}}\longleftrightarrow\Phi_{_{VR}}\label{Par}\\
  C:&&  \Delta_{_{L}}\longleftrightarrow \Delta^{*}_{_{R}}&& \Phi\longleftrightarrow\Phi^{T}&&\Phi_{_{VL}}\longleftrightarrow\Phi^{*}_{_{VR}}
 \end{align}

\noindent
We assume that the vevs defined in eq.~(\ref{vevs}) minimize the potential (\ref{potentialTot}). Thus, one can replace the scalar multiplets with their vevs (after absorbing the phases $\theta_1$,$\theta_2$, $\theta_3$ and $\theta_R$) and imposing the vicinity of all potential first derivatives over the remaining parameters. This leads consequently to the following eight minimization conditions:
\begin{align}
\frac{\partial V}{\partial u_{_R}}=\frac{\partial V}{\partial u_{_L}}=\frac{\partial V}{\partial k_1}=\frac{\partial V}{\partial k_2}=\frac{\partial V}{\partial v_{_R}}=\frac{\partial V}{\partial v_{_L}}=\frac{\partial V}{\partial \theta_4}=\frac{\partial V}{\partial \theta_L}=0
\label{minc1}
\end{align}
The explicit formulae of these derivatives are provided in eqs~(\ref{minimcond1}-\ref{minimcond8}) (cf. appendix~\ref{appA}).
\noindent
From the first condition, we can extract the analytical formula of $\mu_4^2$ (see eq.~(\ref{mu4param})). Inserting the latter one in the second condition (cf. eq.~(\ref{minimcond2})), we obtain the following relation between the self-dual bi-doublet and triplet scalar fields vevs:
\begin{align}
\left(u_{_R}^2-u_{_L}^2\right)=\omega\, \left(v_{R}^2-v_{L}^2\right), &&\text{with}&&\omega=\frac{\alpha_8}{2\left(2 \alpha_4 + \alpha_9-4 \lambda_5\right)}. 
\label{seesaawv2}
\end{align}

\noindent
The 3rd, 4th and 5th conditions can be used to express the mass parameters $\mu_1^2$, $\mu_2^2$ and $\mu_3^2$ as a function of the quartic parameters and the vevs, see appendix~\ref{appA}. Inserting $\mu_3^2$ with its expression (cf. eq.~(\ref{mu3param})) in the 6th condition (cf. eq.~(\ref{minimcond6})), we get:
\begin{align}
(v_{_R}^2-v_{_L}^2)\, \left[v_{_L}v_{_R} (2\rho_1-\rho_3)-\left\{\beta_1 k_1 k_2 \cos(\theta_4-\theta_L)+\beta_2 k_1^2 \cos(\theta_L)+\beta_3 k_2^2 \cos(2\theta_4-\theta_L)\right\}\right]+(u_{_R}^2-u_{_L}^2)v_{_L}v_{_R}\alpha_8
\label{seesaawv1}
\end{align}
Inserting eq.~(\ref{seesaawv2}) in eq.~(\ref{seesaawv1}), we obtain the modified {\it seesaw} vevs relation:
\begin{align}
\beta_2k_1^2 \cos\left(\theta_L\right)+\beta_1k_1k_2 \cos\left(\theta_4-\theta_L\right)+\beta_3k_2^2 \cos\left(2\theta_4-\theta_L\right)-\Bigl(2\rho_1-\rho_3+\frac{\,\alpha^2_8}{2\left(2 \alpha_4 + \alpha_9-4 \lambda_5\right)}\Bigr)v_{_L}v_{_R}=0
\label{seesaw}
\end{align}
We note that the assumptions  $u_R\neq u_L$ and $v_R\neq v_L$ are mandatory to get the seesaw relation (\ref{seesaw}) and also, they are required to insure parity violation at the weak scale. \\

\noindent
The assumption that CP invariance in the potential to be preserved, restrict the values of the vevs to be real in aim to get a hermitian quark mass matrices~\cite{Z}. The $\theta_4=0$ feature can be obtained naturally from the 7th minimization condition in eq.~(\ref{minc1}) ($\partial V/\partial \theta_4=0$), where we can show that $\theta_4$ must vanish, see appendix~\ref{appA}. Moreover, the remaining phase  $\theta_L$ should vanish in order to conserve CP in the potential, as mentioned. Therefore, the vevs seesaw relation (\ref{seesaw}) simply becomes:
\begin{align}
 v_{_R}v_{_L}=\gamma\left(k_1^2+k_2^2\right)&& \text{with}
 && \gamma&=\frac{\beta_2k_1^2+\beta_1k_1k_2+\beta_3k_2^2}{\Bigl(2\rho_1-\rho_3
 +\frac{\,\alpha^2_8}{2\left(2 \alpha_4 + \alpha_9-4 \lambda_5\right)}\Bigr)\left(k_1^2+k_2^2\right)}
 \label{seevev}
\end{align}
If the parameter $\gamma$ is of order unity (i.e. $\gamma\sim 1$), the vev $v_R$ can become extremely large (of order $~10^{8}$ GeV), since the neutrinos masses (which are proportional to $v_{_L}$) are experimentally constrained to be bellow a few eV~\cite{Battye:2013xqa, DayaBay:2014fct}. This would in turn give very large masses to the heavy neutrinos and the extra gauge and scalar bosons, unless one fine-tunes the couplings and $v_R$ to keep these masses at the TeV scale. To avoid such fine-tuning, we can eliminate the parameters $\beta_i$ assuming this is justified by higher symmetries than the symmetries of this model~\cite{a, P}. The resulting seesaw relation is then:
\begin{align}
 \Bigl(2\rho_1-\rho_3
 +\frac{\,\alpha^2_8}{2\left(2 \alpha_4 + \alpha_9-4 \lambda_5\right)}\Bigr)v_Lv_R=0
 \label{seesa}
\end{align}
 There are three possible ways to solve this equation. The 1rst possibility is to put $v_R=0$, but this leads to extra gauge masses of the order of the SM gauge boson masses ($m_{W^\prime}\sim m_W$), which must be excluded. The 2nd possibility is to take $\left( 2\rho_1-\rho_3+\,\alpha^2_8\right)/\left[2\left(2 \alpha_4 + \alpha_9-4 \lambda_5\right)\right]=0$, however this leads to massless scalar and pseudo scalar bosons~\footnote{ The scalar and pseudo scalar bosons are coming from the components of $\Delta_L^{0R}$ and $\Delta_L^{0I}$, respectively.}~\cite{O}. Thus, the only consistent choice to solve eq.~(\ref{seesa}) is to set $v_{_L}=0$ and $v_{_R}\neq 0$, where in this case, $v_{_R}$ would be at the TeV scale.\\

\noindent
In eq.~(\ref{seesaawv2}), we assume that coefficient $\omega$ is of order $\mathcal{O}(1)$, which is justified by the fact that the quartic parameters should be chosen at the order of unity. Thus, the difference $u_{_R}^2-u_{_L}^2$ should behave as $v_{R}^2-v_{L}^2$. This implies that $u_{_L}\ll u_{_R}$ and $u_{_R}$ at the TeV scale. \footnote{If we fine-tuned $\alpha_8$ to be zero or very near. This would make $u_{_L}$ and $u_{_R}$ at the same scale. This is actually an excluded solution because it leads to a large deviation in $W$ and $Z$ masses.}.\\

\noindent
In the rest of this paper, we set $u_{_L}=v_{_L}=0$ and assume that $u_{_R}$ and $v_{_R}$ at the TeV scale.
As a result, we can distinguish two symmetry breaking patterns:
\begin{align}
\textbf{Pattern 1}\,\, (u_{_{R}}>v_{_{R}}):\qquad SU(2)_V\times SU(2)_L&\times SU(2)_R\times U(1)_{B-L}\xrightarrow[<\Phi_{_{VL}}>=0]{<\Phi_{_{VR}}>=\mathfrak{u}_{_{R}}} SU(2)_L\times SU(2)_R\times U(1)_{B-L}\nonumber\\& \xrightarrow[<\Delta_L>=0]{<\Delta_R>=v_R}SU(2)_L\times U(1)_{Y}\xrightarrow{<\Phi>\neq 0}U(1)_{EM}\\
\label{patern1}
\textbf{Pattern 2}\,\, (v_{_{R}}>u_{_{R}}):\qquad
SU(2)_V\times SU(2)_L&\times SU(2)_R\times U(1)_{B-L}\xrightarrow[<\Delta_L>=0]{<\Delta_R>=v_R}SU(2)_V\times SU(2)_L\times U(1)_{Y}\nonumber\\
&\xrightarrow[<\Phi_{_{VL}}>=0]{<\Phi_{_{VR}}>=\mathfrak{u}_{_R}}SU(2)_L\times U(1)_{Y}\xrightarrow{<\Phi>\neq 0}U(1)_{EM}
 \end{align}
 where in both cases the scalar $\Phi$ will bring us from the SM symmetry to the $U(1)_{_\text{EM}}$ symmetry.

\section{Scalars, gauge bosons and fermions masses}
\label{sec4}
\subsection{Scalar bosons masses}
\label{sec4.1}
The scalar fields required to break the original symmetry down to $U(1)_{_\text{EM}}$ and allow the mixing of the VLFs with the chiral ones, cf. subsection~\ref{sf}, have $28$ degrees of freedom. Three neutral and three pairs of singly charged states will be absorbed to form the longitudinal components of the massive gauge bosons. The remaining fields will constitute the physical scalars. Let us reparameterize the complex neutral components of the scalars and shift them by theirs vevs, we have:
\begin{align}
\Phi_{_\text{VL}}&\rightarrow \frac{\text{diag}[u_{_L}\, ,\, u_{_L}]}{\sqrt{2}}+\Phi_{_\text{VL}},&
\Phi_{_\text{VR}}&\rightarrow \frac{\text{diag}[u_{_R}\, ,\, u_{_R}]}{\sqrt{2}}+\Phi_{_\text{VR}},&
\Phi&\rightarrow \frac{\text{diag}[k_1\, ,\, k_2]}{\sqrt{2}} +\Phi.
\end{align}
So that the neutral components shift as:
\begin{align}
\phi_{1}^0=\frac{1}{\sqrt{2}}\left(u_{_{L}}+\phi_{1}^{0\, \text{re}}+i\,\phi_{1}^{0\, \text{im}}\right)&&
\phi_{2}^0=\frac{1}{\sqrt{2}}\left(u_{_{R}}+\phi_{2}^{0\, \text{re}}+i\,\phi_{2}^{0\, \text{im}}\right)\nonumber
\\
\phi_{3}^0=\frac{1}{\sqrt{2}}\left(k_1+\phi_{3}^{0\, \text{re}}+i\,\phi_{3}^{0\, \text{im}}\right)&&
\phi_{4}^0=\frac{1}{\sqrt{2}}\left(k_2+\phi_{4}^{0\, \text{re}}+i\,\phi_{4}^{0\, \text{im}}\right)\nonumber
\\
\delta_{L}^0=\frac{1}{\sqrt{2}}\left(v_{_L}+\delta_{L}^{0\, \text{re}}+i\,\delta_{L}^{0\, \text{im}}\right)&&
\delta_{R}^0=\frac{1}{\sqrt{2}}\left(v_{_R}+\delta_{R}^{0\, \text{re}}+i\,\delta_{R}^{0\, \text{im}}\right)\nonumber \end{align}
The scalar mass matrix corresponds to the second derivative of the potential with respect to all scalars degrees of freedom, we write:
\begin{align}
 \frac{\partial{^2 V}}{\partial{\phi_i}\partial{\phi_j}}\biggl\vert_{\phi_i,\phi_j=0}=m^2_{ij}
\end{align}
This mass matrix can be decomposed into three independent blocks: neutral, singly-charged and doubly-charged scalar mass matrices, which are, respectively, associated to the scalar fields in the weak basis:
\begin{align}
\text{neutral:}&&  \{\phi_{i}^{0}\}&\equiv \{\phi_{3}^{0\, \text{re}},\phi_{4}^{0\, \text{re}},\delta_{R}^{0\, \text{re}},\phi_{2}^{0\, \text{re}},\phi_{1}^{0\, \text{re}},\delta_{L}^{0\, \text{re}},\phi_{3}^{0\, \text{im}},\phi_{4}^{0\, \text{im}},\delta_{R}^{0\, \text{im}},\phi_{2}^{0\, \text{im}},\phi_{1}^{0\, \text{im}},\delta_{L}^{0\, \text{im}}\}
  \label{neutralBase}\\
 \text{singly-charged:}&&\{\phi_{i}^{\pm}\}&\equiv\{\phi_{3}^{\pm},\phi_{4}^{\pm},\delta_{R}^{\pm},\phi_{1}^{\pm},\delta_{L}^{\pm},\phi_{2}^{\pm}\}\label{SingChargBase}\\
 \text{doubly-charged:}&&\{\phi_{i}^{\pm\pm}\}&\equiv\{\delta_{L}^{\pm\pm},\delta_{R}^{\pm\pm}\}\label{DoubChargBase}
\end{align}
Due to the large number of degrees of freedom, the mass matrices are complicated and their exact eigenvalues are hard to extract (or impossible to get). Thus, it is more convenient to take into account the fact that $v_{_{EW}}/v_R\ll1$, $v_{_{EW}}/u_R\ll1$ and $k_2\ll k_1$~\cite{P,Dev,Hm} and extract the leading terms of the scalars masses. The neutral sector contains six scalars (including the Higgs boson) and three pseudo-scalars, with masses:
\begin{align}
m^2_{h}&\simeq 2v_{_{EW}}^2\Bigl(\lambda_1+\left(\frac{2k_1k_2}{v_{_{EW}}^2}\right)^2\left(2\lambda_2+\lambda_3\right)+\frac{4k_1k_2}{v_{_{EW}}^2}\lambda_4\Bigr)&
m^2_{H_1^{0}}&\simeq \frac{1}{2}\alpha_3 v_R^2\frac{v_{_{EW}}^2}{k_{-}^2}\nonumber\\
m^2_{H_2^{0}}&\simeq 4 u_{_{R}}^2\lambda_5+v_R^2\rho_1+\sqrt{\left(4u_{_{R}}^2\lambda_5-v_R^2\rho_1\right)^2+v_R^2u_{_{R}}^2\left(2\alpha_7+\alpha_8\right)^2} &  m^2_{H^0_3}&=\left(2\alpha_4+\alpha_9-4\lambda_5\right)u^2_{_{R}}-\frac{1}{2}\alpha_8 v_R^2 \nonumber\\
m^2_{H_4^{0}}&\simeq 4 u_{_{R}}^2\lambda_5+v_R^2\rho_1-\sqrt{\left(4 u_{_{R}}^2\lambda_5-v_R^2\rho_1\right)^2+v_R^2 u_{_{R}}^2\left(2\alpha_7+\alpha_8\right)^2} &  m^2_{H^0_5}&=\frac{1}{2}\left(\left(\rho_3-2\rho_1\right)v_R^2-\alpha_8 u^2_{_{R}}\right)
\end{align}
three pseudo-scalars and three charged scalars with masses,
 \begin{align}
 m^2_{A^{0}_1}&= \frac{1}{2}v_{_{EW}}^2\left(\frac{\alpha_3\,v_R^2}{k_{-}^2}-8\lambda_2+4\lambda_3\right) &&  m^2_{H^{\pm}_1}=\left(\frac{1}{4}\,k_{-}^2+\frac{v_{_{EW}}^2}{2\,k_{-}^2}\,v_R^2\right)\alpha_3\nonumber\\
 m^2_{A^0_2}&=\left(2\alpha_4+\alpha_9-4\lambda_5\right) u^2_{_{R}}-\frac{1}{2}\alpha_8 v_R^2 &&
 m^2_{H^{\pm}_2}=u^2_{_{R}}\left(2\alpha_4+\alpha_9-4\lambda_5\right)-\frac{1}{2}\alpha_8 v_R^2\nonumber\\
 m^2_{A^0_3}&=\frac{1}{2}\left(\left(\rho_3-2\rho_1\right)v_R^2-\alpha_8 u^2_{_{R}}\right) &&  m^2_{H^{\pm}_3}=\frac{1}{4}k_{-}^2\alpha_3+\frac{1}{2}\left(\rho_3-2\rho_1\right)v_R^2-\frac{1}{2}\alpha_8 u^2_{_{R}}
\end{align}
 \begin{align}
 m^2_{H^{\pm}_1}&=\left(\frac{1}{4}\,k_{-}^2+\frac{v_{_{EW}}^2}{2\,k_{-}^2}\,v_R^2\right)\alpha_3 
\nonumber\\
 m^2_{H^{\pm}_2}&=u^2_{_{R}}\left(2\alpha_4+\alpha_9-4\lambda_5\right)-\frac{1}{2}\alpha_8 v_R^2
 \nonumber\\ 
 m^2_{H^{\pm}_3}&=\frac{1}{4}k_{-}^2\alpha_3+\frac{1}{2}\left(\rho_3-2\rho_1\right)v_R^2-\frac{1}{2}\alpha_8 u^2_{_{R}}.
\end{align}
and two doubly charged scalars,
\begin{align}
 m^2_{H_1^{\pm\pm}}&=\frac{1}{2}k_{-}^2\alpha_3+\frac{1}{2}\left(\rho_3-2\rho_1\right)\,v_R^2- \frac{1}{2}\,u_{_{R}}^2\alpha_8 &
 m^2_{H_2^{\pm\pm}}&=\frac{1}{2} k_{-}^2\alpha_3+2 \rho_2\,v_R^2
\end{align}
A detailed description of the scalar mass matrices diagonalization is provided in appendix~\ref{appD}.

\subsection{Gauge boson masses}
\label{sec4.2}
From the kinetic term of the scalar fields defined in eq.~(\ref{Lscalar}), one can extract the gauge boson masses by replacing the scalar fields with their vevs. The charged gauge bosons matrix in the basis $(W^{+}_{V}, W^{+}_{R}, W_{L})$ is given by:
\begin{align}
 {\cal{M}}^{2}_{V^{\pm}}=\frac{1}{4}\begin{pmatrix}
 2g^2_{_{V}}u^2_{_{R}} && -2gg_{_{V}}u^2_{_{R}} && 0  \\ \\
-2gg_{_{V}}u^2_{_{R}} && g^2\bigl(v_{_{EW}}^2+2(v^2_R+u^2_{_{R}})\bigr) && -2g^2k_1k_2\\ \\
 0 &&   -2g^2k_1k_2 && g^2v_{_{EW}}^2
\end{pmatrix}\nonumber\\
\end{align}
and the neutral gauge bosons mass matrix in the basis $(W^{3}_{V}, W^{3}_{R}, W^{3}_{L}, B)$ is:
\begin{align}
 {\cal{M}}^{2}_{V^{0}}=\frac{1}{2}\begin{pmatrix}
 g^2_{_{V}}u^2_{_{R}} && -gg_{_{V}}u^2_{_{R}} && 0 && 0 \\ \\
-gg_{_{V}}u^2_{_{R}} && \frac{1}{2} g^2\bigl(v_{_{EW}}^2+2u^2_{_{R}}+4v^2_R\bigr)  && -\frac{1}{2}g^2v_{_{EW}}^2 && -2gg^{\prime}v^2_R  \\ \\
0 && -\frac{1}{2}g^2v_{_{EW}}^2 &&  \frac{1}{2} g^2v_{_{EW}}^2  && 0\\ \\
0 && -2gg^{\prime}v^2_R && 0 && 2g^{\prime2}v^2_R
\end{pmatrix}
\end{align}
The mass matrices ${\cal{M}}^{2}_{V^{\pm}}$ and ${\cal{M}}^{2}_{V^{0}}$ are symmetric and can therefore be diagonalized by unitary transformations, rotating the gauge fields from the weak basis to the mass basis. Using the hierarchies $v_{_{EW}}\ll u_{_R}$ (or $\varepsilon_1=v_{_{EW}}^2/u_R^2\ll1$) and $v_{_{EW}}\ll v_{_R}$ (or $\varepsilon_2=v_{_{EW}}^2/v_R^2\ll1$), and assuming $k_2\ll k_1$ (i.e. $s^2_\beta\sim0$)~\cite{Zhang:2007fn},  we find that the masses of the charged and neutral gauge bosons (the eigenvalues of these matrices) are given, to first order in $\epsilon_{1}$ and $\epsilon_{2}$, by:
\begin{align}
m^2_W\simeq&\frac{1}{4}g^2\,v^2_{_{EW}}\nonumber\\
m^2_{W^{\prime}}\simeq&\frac{g^2\,g_{_V}^2}{4\left(g^2+g_{_V}^2\right)}\left(v^2_{_{EW}}+2\,v^2_{R}\right) \nonumber\\
m^2_{W^{\prime\prime}}\simeq&\frac{1}{4\left(g^2+g_{_V}^2\right)}\Bigl(2(g^2+g_{_V}^2)^2\,u^2_{_{R}}+g^4\left(v^2_{_{EW}}+2\,v^2_{R}\right)\Bigr)\nonumber\\
m^2_Z\simeq&\frac{1}{4}g^2\,\frac{v^2_{_{EW}}}{c^2_{\theta_w}}
\nonumber\\
m^2_{Z^{\prime}}\simeq& \frac{1}{\left(g^2+g_{_V}^2\right)}\Bigl(\frac{g^2\bigl(c^2_{\theta_{w}}\,g^2_{_V}-s^2_{\theta_{w}}(g^2+g_{_V}^2)\bigr)}{4 c^2_{\theta_{w}}} v^2_{_{EW}} +\frac{g^2 g^4_{_V}c^2_{\theta_{w}}}{\bigl(c^2_{\theta_{w}}\,g^2_{_V}-s^2_{\theta_{w}}\left(g^2+g_{_V}^2\right)\bigr)} v^2_{R}\Bigr)
\nonumber\\
m^2_{Z^{\prime\prime}}\simeq& \frac{1}{4\left(g^2+g_{_V}^2\right)}\Bigl(2\left(g^2+g_{_V}^2\right)^2\,u^2_{_{R}}+g^4\left(v^2_{_{EW}}+4\,v^2_{R}\right)\Bigr)
\label{masseq}
\end{align}
where $\theta_w$ is the Weinberg mixing angle. To obtain eq.~(\ref{masseq}), we adopt the 1rst symmetry breaking pattern ($u_{_R}>v_{_{R}}$), which implies that the 2nd-generation extra-gauge bosons ($W^{\prime\prime}$ and $Z^{\prime\prime}$) are heavier than their 1st-generation counterparts ($W^{\prime}$ and $Z^{\prime}$). To simplify the expressions in eq.~(\ref{masseq}), we have expressed the couplings $g_{_L}$, $g_{_R}$ and $g^{\prime}$ in terms the electromagnetic coupling $e$, the Weinberg mixing angle and the coupling $g_{_V}$, using the following relations:
\begin{align}
g_L=g_R=g=\frac{e}{s_{\theta_{w}}} && g^{\prime}=\frac{g\,g_{_V}\,s_{\theta_{w}}}{\sqrt{c^2_{\theta_{w}}\,g_{_V}^2-s^2_{\theta_{w}}\left(g^2+g_{_V}^2\right)}}
\label{eq1}
\end{align}
We note that the gauge coupling $g_{_V}$ is considered as a free parameter, which can be constrained from above by the perturbative unitarity condition $g_{_V}<\sqrt{4\pi}$, and from bellow by requiring that $g^{\prime}$, defined in eq.~(\ref{eq1}), must be real. Thus, we get:
\begin{align}
 g_{_V}>\frac{2\,\sqrt{\pi}\,s_{\theta_w}\,g}{\sqrt{4\,\pi\,\left(c^2_{\theta_w}-s^2_{\theta_w}\right)-s^2_{\theta_w}\,g^2}}
\end{align}
 This last relation ensure that $g^{\prime}$ is simultaneously real and constrained from above by the perturbative unitarity condition.\\

\noindent
The charged and neutral physical gauge bosons fields are related to the nonphysical ones, at the zero order in $\epsilon_{1}$ and $\epsilon_{2}$, by the transformations:
\begin{align}
\begin{array}{lr}
\begin{pmatrix}
W^{\prime\prime\pm}_{(0)\mu} \\
W^{\prime\pm}_{(0)\mu} \\
W^{\pm}_{(0)\mu}
\end{pmatrix}=
\begin{pmatrix}
 c_1 & -s_1 & 0\\
 s_1 & c_1 & 0 \\
 0 & 0 & 1
\end{pmatrix}
\begin{pmatrix}
W^{\pm}_{V\mu}  \\
W^{\pm}_{R\mu} \\
W^{\pm}_{\mu}
\end{pmatrix}
\end{array}
&&
\begin{array}{lr}
\begin{pmatrix}
Z^{\prime\prime}_{(0)\mu} \\
Z^{\prime}_{(0)\mu} \\
Z_{(0)\mu} \\
A_{(0)\mu}
\end{pmatrix}=
\begin{pmatrix}
 c_1 & -s_1 & 0 & 0 \\
 s_1\,c_2 & c_1\,c_2 & 0 & -s_2 \\
 -s_1\,s_2\,s_{\theta_w} & -c_1\,s_2\,s_{\theta_w} & c_{\theta_w} & -c_2\,s_{\theta_w} \\
 s_1\,s_2\,c_{\theta_w} & c_1\,s_2\,c_{\theta_w} & s_{\theta_w} & c_2\,c_{\theta_w}
\end{pmatrix}
\begin{pmatrix}
 W^{3}_{V\mu} \\
 W^{3}_{R\mu} \\
 W^{3}_{\mu}  \\
 B_{\mu}
\end{pmatrix}
\end{array}
\label{rotations}
\end{align}
where the cosine (sine) of the rotation angles involved in eq.~(\ref{rotations}) are defined as the following~\footnote{We note that these transformations are constructed from the eigenvectors of the CC and NC mass matrices.
}:
\begin{align}
 c_1 = \frac{g_{_V}}{\sqrt{g^2+g^2_{_V}}} &&
 c_2 = \frac{g\,g_{_V}}{\sqrt{g^{\prime 2}\left(g^2+g_{_V}^2\right)+g^2\,g_{_V}^2}} &&
 s_{1,2} =\sqrt{1-c^2_{1,2}}
\end{align}

\noindent
Moreover, the CC and NC covariant derivatives associated to the $W$ and $Z$ boson fields, at order $\epsilon_{1}^0$ and $\epsilon_{2}^0$, correspond exactly to the SM ones, they are given by:
\begin{align}
 D^{\text{CC}}_{(0)\mu}=\partial_{\mu}-i\frac{e}{s_w}\tilde{T}^{\pm}\,W^{\pm}_{\mu}&&
D^{\text{NC}}_{(0)\mu}=\partial_{\mu}-i\frac{e}{s_w\,c_w}\left(c_w^2\,\tilde{T}^{3}-s_w^2\frac{Y}{2}\right)\,Z_{\mu}
\label{derives}
\end{align}
where $\tilde{T}^{\pm}=\tilde{T}^{1}\pm i\tilde{T}^{2}$, $\tilde{T}^{3}$ and $Y$ are the generators of $SU(2)_L$ and $U(1)_Y$, respectively\footnote{We have shown that the higher order contributions in the physical fields expressions to the covariant derivatives in eq.~(\ref{derives}) are extremely suppressed. Thus, the SM NC and CC processes are not effected.}

\subsection{Fermions masses and mixing}
\label{sec5}
After substituting the scalar fields with their vevs in the Yukawa Lagrangian (cf.~eq~(\ref{Yuk})) and adding the VLFs bare mass term (cf.~eq~(\ref{massVLFs})), the mass term of all the fermions involved in this model can be expressed as:
 \begin{align}
  {\cal{L}}_{\text{mass}}^{F}=-\overline{\mathcal{U}}_{L}\mathcal{M}_{\mathcal{U}}\mathcal{U}_{R}-\overline{\mathcal{D}}_{L}\mathcal{M}_{\mathcal{D}}\mathcal{D}_{R}-\overline{\mathcal{E}}_{L}\mathcal{M}_{\mathcal{E}}\mathcal{E}_{R}-\frac{1}{2}\overline{\mathcal{N}}_{L}\mathcal{M}_{\mathcal{N}}\mathcal{N}_{R}+\text{h.c}
 \label{Lmass}
 \end{align}
In eq.~(\ref{Lmass}), we introduced the combined basis of the chiral fermions and extra VLFs, which are defined by:
\begin{align}
 \mathcal{U}^k_{L,R}=\begin{pmatrix}
u^{i\prime}_{L,R} \\
T^{\prime}_{L,R}
\end{pmatrix}&&
 \mathcal{D}^k_{L,R}=\begin{pmatrix}
d^{i\prime}_{L,R} \\
B^{\prime}_{L,R}
\end{pmatrix}\\
 \mathcal{E}^k_{L,R}=\begin{pmatrix}
e^{i\prime}_{L,R} \\
E^{\prime}_{L,R}
\end{pmatrix}&&
 \mathcal{N}^k_{L}=\begin{pmatrix}
\nu^{i\prime}_{L} \\
(\nu^c)^{i\prime}_{L} \\
N^{\prime}_{L}
\end{pmatrix}&&
\mathcal{N}^k_{R}=\begin{pmatrix}
(\nu^c)^{i\prime}_{R} \\
\nu^{i\prime}_{R} \\
N^{\prime}_{R}
\end{pmatrix}
\end{align}
where the weak basis mass matrices $\mathcal{M}_{\mathcal{U}}(4\times4)$, $\mathcal{M}_{\mathcal{D}}(4\times4)$, $\mathcal{M}_{\mathcal{E}}(4\times4)$ and $\mathcal{M}_{\mathcal{N}}(7\times7)$ are expressed by:
 
 \begin{align}
 & \mathcal{M}_{\mathcal{U}}=
  \left(
  \begin{array}{c!{\vrule width 0.4pt}c}
\frac{v_{_{\text{EW}}}}{\sqrt{2}}\left(y_q c_{\beta}+\tilde{y}_q s_{\beta}\right) & 0\\
\hline
\frac{u_{_{R}}}{\sqrt{2}}\lambda^{\prime}_q & M_Q
\end{array}
\right)
&&
 \mathcal{M}_{\mathcal{U}}=
  \left(
  \begin{array}{c!{\vrule width 0.4pt}c}
\frac{v_{_{\text{EW}}}}{\sqrt{2}}\left(y_q s_{\beta}+\tilde{y}_q c_{\beta}\right) & 0\\
\hline
\frac{u_{_{R}}}{\sqrt{2}}\lambda^{\prime}_q & M_Q
\end{array}
\right)
\nonumber\\
&\mathcal{M}_{\mathcal{E}}=
  \left(
  \begin{array}{c!{\vrule width 0.4pt}c}
\frac{v_{_{\text{EW}}}}{\sqrt{2}}\left(y_l s_{\beta}+\tilde{y}_l c_{\beta}\right) & 0\\
\hline
\frac{u_{_{R}}}{\sqrt{2}}\lambda^{\prime}_l & M_L
\end{array}
\right)
&& \mathcal{M}_{\mathcal{N}}=\left(\begin{array}{ccc}
 0 & \frac{v_{_{\text{EW}}}}{\sqrt{2}}\left(y_l c_{\beta}+\tilde{y}_l s_{\beta}\right) & 0\\
\frac{v_{_{\text{EW}}}}{\sqrt{2}}\left(y_l c_{\beta}+\tilde{y}_l s_{\beta}\right)^T & \sqrt{2}\,y_M\,v_R & 0\\
0 & \sqrt{2}\,\lambda^{\prime}_l\, u_{_{R}}&  2\,M_L
\end{array}\right)
\end{align}
where we used $\left(\bar{\nu}_{R/L}\right)^c=(\bar{\nu}^c)_{L/R}$ and  $\overline{\nu}_{L}\nu_{R}=(\overline{\nu^c})_{L}(\nu^c)_{R}$ to get the matrix $\mathcal{M}_{\mathcal{N}}$.\\

\noindent
The mass matrices of the up, down quarks as well as the charged leptons can be diagonalized by bi-unitary transformations as follows:
\begin{align}
\mathcal{M}^{\mathcal{U},\mathcal{D},\mathcal{E}}_{diag}=U^{^{\mathcal{U},\mathcal{D},\mathcal{E}}}_L\mathcal{M}^{\mathcal{U},\mathcal{D},\mathcal{E}}\left(U^{^{\mathcal{U},\mathcal{D},\mathcal{E}}}_R\right)^{\dagger}
\end{align}
In this work, we assume that the VLFs mix solely with 3rd generation of SM fermions. Therefore, these bi-unitary transformations can be expressed as:
\begin{align}
U^{\mathcal{U,{\mathcal{D}},{\mathcal{E}}}}_{L}=\begin{pmatrix}
V^{u,d,l}_{L} && 0 \\
0 && 1
\end{pmatrix}
\begin{pmatrix}
{1\!\!1}_{2\times2} && 0 \\
0 && W_{L}^{\mathcal{U,{\mathcal{D}},{\mathcal{E}}}}
\end{pmatrix}
&&
U^{\mathcal{U,{\mathcal{D}},{\mathcal{E}}}}_{R}=
\begin{pmatrix}
V^{u,d,l}_{R} && 0 \\
0 && 1
\end{pmatrix}
\begin{pmatrix}
{1\!\!1}_{2\times2} && 0 \\
0 && W_{R}^{\mathcal{U,{\mathcal{D}},{\mathcal{E}}}}
\end{pmatrix}
 \end{align}
where $V^{u,d,l}_{L,R}$ are $3\times3$ unitary matrices that diagonalize the upper diagonal blocks of the up quark, down quark and charged leptons mass matrices, respectively, i.e.
\begin{align}
\left(V_{L}^{u,d,l}\right)^{\dagger}\, \left[\frac{v_{_{\text{EW}}}}{\sqrt{2}}\left(y_{q,l} c_{\beta}+\tilde{y}_{q,l} s_{\beta}\right)\right]\, V_{R}^{u,d,l}&=\text{diag}\left[m_{u,d,e}, m_{c,s,\mu}, m_{t,b,\tau}\right]. 
\end{align}
We define the left-handed and right-handed Cabibbo-Kobayashi-Maskawa (CKM) matrices, $V^{CKM}_L$ and $V^{CKM}_R$ by:
 \begin{align}
  V^{CKM}_L=\left(V^{u}_{L}\right)^{\dagger}V^{d}_{L}
  &&
  V^{CKM}_R=\left(V^{u}_{R}\right)^{\dagger}V^{d}_{R}
 \end{align}
where both of them are similar to the known CKM mixing matrix~\cite{Z}. The unitary matrices $W_{L,R}^{\mathcal{U,{\mathcal{D}},{\mathcal{E}}}}$ allow to mix the VLFs with the ordinary fermions of the 3rd generation. They can be expressed as:
\begin{align}
W_{L,R}^{\mathcal{U,{\mathcal{D}},{\mathcal{E}}}}&=
\begin{pmatrix}
c_{_{\theta}L,R}^{u,d,l} && - s_{_{\theta}L,R}^{u,d,l} \\
s_{_{\theta}L,R}^{u,d,l} && c_{_{\theta}L,R}^{u,d,l}    
\end{pmatrix}
\end{align}
where $c_{_{\theta}L,R}^{u,d,l}$ and $s_{_{\theta}L,R}^{u,d,l}$ will be fixed later on.\\

\noindent
Let's focus on the quark sector. The rotation that brings these quarks from the weak to the mass basis can be expressed as:
 \begin{align}
\begin{pmatrix}
t_{L,R}\\
T_{L,R}
\end{pmatrix}=
\begin{pmatrix}
c_{_{\theta}L,R}^u && - s_{_{\theta}L,R}^u \\
s_{_{\theta}L,R}^u && c_{_{\theta}L,R}^u
\end{pmatrix}
\begin{pmatrix}
t^{\prime}_{L,R}\\
T^{\prime}_{L,R}
\end{pmatrix}
&&
\begin{pmatrix}
b_{L,R}\\
B_{L,R}
\end{pmatrix}=
\begin{pmatrix}
c_{_{\theta}L,R}^d && - s_{_{\theta}L,R}^d \\
s_{_{\theta}L,R}^d  && c_{_{\theta}L,R}^d
\end{pmatrix}
\begin{pmatrix}
b^{\prime}_{L,R}\\
B^{\prime}_{L,R}
\end{pmatrix}
\end{align}
where the fields with superscript ($\prime$) are defined in the weak basis. Following the same strategy of \cite{Aguilar1,Aguilar2}, one can express the mixing angles in terms of the vevs, the Yukawa couplings and $M_{Q}$. We get: 
\begin{align}
 \tan\left(2\theta_{R}^{u}\right)&=\frac{\sqrt{2}\left(\lambda^{\prime}_{q}\right)_{13} u_{_{R}} M_{Q}}{M_{Q}^2-\left(\lambda^{\prime}_{q}\right)^2_{13} u_{_{R}}^2/2-\left(y^{q}_{33} c_{\beta}+\tilde{y}^{q}_{33} s_{\beta}\right)^2 v_{_{\text{EW}}}^2/2},
 &
 \tan\left(2\theta_{R}^{d}\right)&=\frac{\sqrt{2}\left(\lambda^{\prime}_{q}\right)_{13} u_{_{R}} M_{Q}}{M_{Q}^2-\left(\lambda^{\prime}_{q}\right)^2_{13} u_{_{R}}^2/2-\left(y^{q}_{33} s_{\beta}+\tilde{y}^{q}_{33} c_{\beta}\right)^2 v_{_{\text{EW}}}^2/2}.
\nonumber\\
 \tan\left(\theta_{L}^{u}\right)&=\frac{m_{u}}{m_{T}} \tan \left(\theta_{R}^{u}\right),&
 \tan\left(\theta_{L}^{d}\right)&=\frac{m_{d}}{m_{B}} \tan \left(\theta_{R}^{d}\right).
 \label{eqQq}
\end{align}
We should mention that the first two components of the $3\times 1$ coupling matrix $\lambda^{\prime}_{q}$ are taken to be zero, as the mixing with the first two quark generations is neglected. Now, we are able to express the physical masses of the quarks $t$, $T$, $b$ and $B$ in terms of Yukawa couplings, the vevs, the bare mas $M_Q$ and the cosine (sine) of the mixing angle $\theta_{L/R}^{u,d}$. Thus, we get: 
\begin{align}
 m_t&= c_{_{\theta}L}^u c_{_{\theta}R}^u \frac{\left(y^{q}_{33} k_1+\tilde{y}^{q}_{33} k_2\right)}{\sqrt{2}}-s_{_{\theta}L}^u c_{_{\theta}R}^u \left(\lambda^{\prime}_{q}\right)_{13}\frac{u_{_{R}}}{\sqrt{2}}+s_{_{\theta}L}^u s_{_{\theta}R}^u M_Q\nonumber\\
  m_T&= s_{_{\theta}L}^u s_{_{\theta}R}^u \frac{\left(y^{q}_{33} k_1+\tilde{y}^{q}_{33} k_2\right)}{\sqrt{2}}+c_{_{\theta}L}^u s_{_{\theta}R}^u \left(\lambda^{\prime}_{q}\right)_{13}\frac{u_{_{R}}}{\sqrt{2}}+c_{_{\theta}L}^u c_{_{\theta}R}^u M_Q\nonumber\\
  \nonumber\\
 m_b&= c_{_{\theta}L}^d c_{_{\theta}R}^d \frac{\left(y^{q}_{33} k_2+\tilde{y}^{q}_{33} k_1\right)}{\sqrt{2}}-s_{_{\theta}L}^d c_{_{\theta}R}^d \left(\lambda^{\prime}_{q}\right)_{13}\frac{u_{_{R}}}{\sqrt{2}}+s_{_{\theta}L}^d s_{_{\theta}R}^d M_Q\nonumber\\
 m_B&= s_{_{\theta}L}^d s_{_{\theta}R}^d \frac{\left(y^{q}_{33} k_2+\tilde{y}^{q}_{33} k_1\right)}{\sqrt{2}}+c_{_{\theta}L}^d s_{_{\theta}R}^d \left(\lambda^{\prime}_{q}\right)_{13}\frac{u_{_{R}}}{\sqrt{2}}+c_{_{\theta}L}^d c_{_{\theta}R}^d M_Q
 \label{quarkMass}
\end{align}

\noindent
If we invert eq.~(\ref{quarkMass}), we can write the elements of the mass matrices $\mathcal{M}_{\mathcal{U}}$ and $\mathcal{M}_{\mathcal{D}}$ as follows:
\begin{align}
 M_Q=\left[m_{T,B}^2 c_{_{\theta}R}^{2 u,d}+m_{t,b}^2 s_{_{\theta}R}^{2 u,d}\right]^{1/2}
 && \frac{\left(y^{q}_{33} k_1+\tilde{y}^{q}_{33} k_2\right)}{\sqrt{2}}=\left[m_T^2 s_{_{\theta}R}^{2 u}+m_t^2 c_{_{\theta}R}^{2 u}-\left(\left(\lambda^{\prime}_{q}\right)_{13}\frac{u_{_{R}}}{\sqrt{2}}\right)^2\right]^{1/2}
 \nonumber\\
 (\lambda^{\prime}_{q})_{13}\frac{u_{_{R}}}{\sqrt{2}}=\frac{m_{T,B}^2-m_{t,b}^2}{M_Q} s_{_{\theta}R}^{u,d}  c_{_{\theta}R}^{u,d}
 &&
  \frac{\left(y^{q}_{33} k_2+\tilde{y}^{q}_{33} k_1\right)}{\sqrt{2}}=\left[m_B^2 s_{_{\theta}R}^{2 d}+m_b^2 c_{_{\theta}R}^{2 d}-\left(\left(\lambda^{\prime}_{q}\right)_{13}\frac{u_{_{R}}}{\sqrt{2}}\right)^2\right]^{1/2}
  \label{yq}
\end{align}
If we neglect $k_2$, we can set $k_1\approx v_{_\text{EW}}$. Therefore, we use  relations (\ref{yq}), we can express $y^{q}_{33}$ and $\tilde{y}^{q}_{33}$ the following
\begin{align}
 y^{q}_{33}&\approx\frac{\sqrt{2}}{v_{_\text{EW}}}\left[m_T^2 s_{_{\theta}R}^{2 u}+m_t^2 c_{_{\theta}R}^{2 u}-\left(\left(\lambda^{\prime}_{q}\right)_{13}\frac{u_{_{R}}}{\sqrt{2}}\right)^2 \right]^{1/2}\equiv y^{t}_{33}.\nonumber\\
 \tilde{y}^{q}_{33}&\approx\frac{\sqrt{2}}{v_{_\text{EW}}}\left[m_B^2 s_{_{\theta}R}^{2 d}+m_b^2 c_{_{\theta}R}^{2 d}-\left(\left(\lambda^{\prime}_{q}\right)_{13}\frac{u_{_{R}}}{\sqrt{2}}\right)^2 \right]^{1/2}\equiv y^{b}_{33}.
\label{y33tb}
\end{align}

\noindent
The two relations in eq.~(\ref{y33tb}) are very import, as they allow us to constrain the masses of the VLQs. The couplings $y^{q}_{33}$ and $\tilde{y}^{q}_{33}$ are identified to the SM Yukawa couplings of the ordinary top and bottom quarks ($y^{t}_{33}$ and $y^{b}_{33}$), respectively, which are very well constrained experimentally. Therefore, the VLQs masses are approximated by: 
\begin{align}
m_T&\approx\frac{\left(\lambda^{\prime}_{q}\right)_{13}}{s_{_{\theta}R}^{u}}\frac{u_{_{R}}}{\sqrt{2}}&
m_B&\approx\frac{\left(\lambda^{\prime}_{q}\right)_{13}}{s_{_{\theta}R}^{d}}\frac{u_{_{R}}}{\sqrt{2}}
\end{align}

\noindent
The neutrinos mass matrix $\mathcal{M}_{\mathcal{N}}$ can be expressed in terms of the $3\times 3$ Dirac-type ($M_D$) block, the $3\times 3$ Majorana-type ($M_R$) block, the $1\times 3$ matrix $\sqrt{2}\,\lambda^{\prime}_l\, u_{_{R}}$ (describing the mixing of the chiral neutrinos with the vector-like neutrino) and the bare VLLs mass $M_L$. Hence, we write:
 \begin{align}
  \mathcal{M}_{\mathcal{N}}=
  \left(
  \begin{array}{cc!{\vrule width 0.4pt}c}
0 & M_D & 0\\
M_D^{T} & M_R & 0\\
\hline
0 & \sqrt{2}\,\lambda^{\prime}_l\,u_{_{R}}& 2\,M_L
\end{array}
\right)
\label{MN}
 \end{align}
The mass matrix $\mathcal{M}_{\mathcal{N}}$ can be diagonalized by the following bi-unitary transformations:
\begin{align}
\mathcal{M}^{\mathcal{N}}_{diag}=U^{^{\mathcal{N}}}_L \mathcal{M}^{\mathcal{N}} \left(U^{^{\mathcal{N}}}_R\right)^{\dagger}
\end{align}
As in the case of quarks and leptons, we assume that the VLN mixes only the neutrinos of the 3rd generation, i.e. $\lambda^{\prime}_l\equiv[0,\, \, 0, \,  \, (\lambda^{\prime}_l)_{13}]$ Therefore, these bi-unitary transformations $U^{^{\mathcal{N}}}_L$ and $U^{^{\mathcal{N}}}_R$ can be decomposed as follows:
\begin{align}
U^{\mathcal{N}}_{L}=\begin{pmatrix}
V^{\nu} && 0 \\
0 && 1
\end{pmatrix}
\begin{pmatrix}
{1\!\!1}_{5\times5} && 0 \\
0 && W_{L}^{\mathcal{N}}
\end{pmatrix}
&&
U^{\mathcal{N}}_{R}=
\begin{pmatrix}
V^{\nu} && 0 \\
0 && 1
\end{pmatrix}
\begin{pmatrix}
{1\!\!1}_{5\times5} && 0 \\
0 && W_{R}^{\mathcal{N}}
\end{pmatrix}
\label{mnvw}
 \end{align}
where $W_{R}^{\mathcal{N}}$ and $W_{L}^{\mathcal{N}}$ are $2\times2$ unitary matrices, while $V^{\nu}$ is a $6\times6$ unitary matrix that diagonalize the symmetric upper left block of $\mathcal{M}^{\mathcal{N}}$. We show in appendix~\ref{diagNmass} that:
\begin{align}
\mathcal{M}_{\mathcal{N}}^{(0)}&=
\begin{pmatrix}
(V^{\nu})^{\dagger} && 0 \\
0 && 1
\end{pmatrix} 
\mathcal{M}_{\mathcal{N}}
\begin{pmatrix}
V^{\nu} && 0 \\
0 && 1
\end{pmatrix}
&&\text{with}&
\mathcal{M}_{\mathcal{N}}^{(0)}&\approx 
\left(
  \begin{array}{cc!{\vrule width 0.4pt}c}
-\widetilde{M}_D^{2}\widetilde{M}_R^{-1}  &0 & 0\\
0 & \widetilde{M}_R & 0\\
\hline
0 & \sqrt{2}\, \,u_{_{R}} \lambda^{\prime}_l& 2\,M_L
\end{array}
\right)
\end{align}
At this stage, we obtain the light neutrino mass matrix $-\widetilde{M}_D^2 \widetilde{M}_R^{-1}$ and the heavy Majorana mass matrix $\widetilde{M}_R$ of the ordinary LRS~\cite{a}. We write:
\begin{align}
\widetilde{M}_D&= \text{diag}\left[m_{\nu_1}\, \, , m_{\nu_2}\, \, , m_{\nu_3}\right].\\
\widetilde{M}_R&= \text{diag}\left[m_{_{N_1}},\, \, m_{_{N_2}},\, \, M_{_{N_3}}\right].\\
-\widetilde{M}_D^2 \widetilde{M}_R^{-1}&= \text{diag}\left[m_{\nu_e},\, \,  m_{\nu_\mu},\, \,  m_{\nu_\tau}\right].
\end{align}
where $m_{\nu_e}, m_{\nu_\mu}$ and $m_{\nu_\tau}$ are the physical masses of the light neutrinos, $m_{_{N_1}}$ and $m_{_{N_2}}$ are the physical masses of the 1st and 2nd generation of the heavy Majorana neutrinos, while $M_{N_3}$ is a bare mass associated to 3rd generation Majorana neutrino. The physical mass of the latter particle receive a contribution from the mixing with the VLN as we will show in the following. Let's express the matrix $\mathcal{M}_{\mathcal{N}}^{(0)}$ as:
\begin{align}
\mathcal{M}_{\mathcal{N}}^{(0)}&=
\left(
  \begin{array}{ccc!{\vrule width 0.4pt}cc}
-\widetilde{M}_D^{2}\widetilde{M}_R^{-1}  &0 & 0&0&0\\
0 & m_{_{N_1}}&0 &0 & 0\\
0 & 0&m_{_{N_2}} &0 & 0\\
\hline
0 & 0 & 0& M_{_{N_3}}  & 0\\ 
0 & 0&0 & \sqrt{2}\, \,u_{_{R}} (\lambda^{\prime}_l)_{13}& 2\,M_L
\end{array}
\right)
\end{align}
The lower right $2\times2$ block can be diagonalized by the $2\times2$ bi-unitary matrices $W_{L}^{\mathcal{N}}$ and $W_{R}^{\mathcal{N}}$, cf.~ eq.~(\ref{mnvw}). We have:
 \begin{align}
 \mathcal{M}_{_{\mathcal{N}}}^{diag}=
 \begin{pmatrix}
{1\!\!1}_{5\times5} && 0 \\
0 && (W_{L}^{\mathcal{N}})^{\dagger}
\end{pmatrix}
 \mathcal{M}_{_{\mathcal{N}}}^{(0)}
 \begin{pmatrix}
{1\!\!1}_{5\times5} && 0 \\
0 && W_{R}^{\mathcal{N}}
\end{pmatrix} &&
\text{with}&&
 W_{L,R}^{\mathcal{N}}=\begin{pmatrix}
c_{_{\theta}L,R}^{_{\mathcal{N}}} & - s_{_{\theta}L,R}^{_{\mathcal{N}}} \\
s_{_{\theta}L,R}^{_{\mathcal{N}}} & c_{_{\theta}L,R}^{_{\mathcal{N}}}
\end{pmatrix}
\end{align}
Proceeding the same way as in the case of the quarks, we show that the mixing angles in terms of the Yukawa couplings, the vev $u_{_R}$ and the VLQ bare mass are given by:
\begin{align}
 \tan(2\theta_{R}^{_{\mathcal{N}}})=\frac{2\sqrt{2}\left(\tilde{\lambda}^{\prime}_{l}\right)_{13} u_{_{R}} M_{L}}{4 M_{L}^2-\left(\lambda^{\prime}_{l}\right)^2_{13} u^2_{_{R}}/2-m_{_{N_3}}^2}
&&
 \tan(\theta_{L}^{_{\mathcal{N}}})=\frac{m_{_{N_3}}}{4 m_{N}} \tan \theta_{R}^{_{\mathcal{N}}}
\end{align}

\noindent
We show that the 3rd generation heavy Majorana ($N_3$) and the VLL ($N$) masses are given by:
\begin{align}
m_{N_3}&\simeq 2\sqrt{2} M_L M_{_{N_3}}/\left[4M_L^2+M_{_{N_3}}^2+2\lambda^{\prime 2}_l u_{_{{R_3}}}^2-\sqrt{\left(4M_L^2-M_{_{N_3}}^2\right)^2+4 \lambda^{\prime 2}_l \left(4M_L^2+M_{_{N_3}}^2\right) u_{_{R}}^2+4\tilde{\lambda}^{\prime 4}_l u_{_{R}}^4}\right]^{1/2}.\label{mN3:eq}\\
m_N&\simeq \left[4M_L^2+M_{_{N_3}}^2+\tilde{\lambda}^{\prime 2}_l u_{_{R}}^2-\sqrt{\left(4M_L^2-M_{_{N_3}}^2\right)^2+4 \tilde{\lambda}^{\prime 2}_l \left(4M_L^2+M_{_{N_3}}^2\right)u_{_{R}}^2+\tilde{\lambda}^{\prime 4}_l u_{_{R}}^4}\right]^{1/2}/\sqrt{2}.\label{mN:eq}
\end{align}
 with $\tilde{\lambda}^{\prime 2}_l=(\tilde{\lambda}^{\prime }_l)_{13}^2$. For very small $(\tilde{\lambda}^{\prime }_l)_{13}$, we get: $m_{N_3}\approx M_{N_3}$ and $m_N\approx2 M_L$. Multiplying the masses of the third generation neutrino, heavy neutrino and VLL, allow to write the see-saw like relation:
\begin{align}
 m_{\nu_\tau} m_{N_3} m_N= -2\, M_L\, (\widetilde{M}_{D})_{33}^2\equiv -2\, M_L\, m_{\nu_3}^2.
 \label{ses}
\end{align}
where the first two generations are still governed by the LRSM see-saw relation:
\begin{align}
 m_{\nu_{e}} m_{N_{1}}&= -(\widetilde{M}_{D})_{11}^2\equiv -m_{\nu_1}^2&
 m_{\nu_{\mu}} m_{N_{2}}= -(\widetilde{M}_{D})_{22}^2\equiv -m_{\nu_2}^2.
\end{align}

\noindent
Also, in the case of charged leptons, we consider mixing with only 3rd generation (as assumed for quarks and neutrinos). Thus, we can write:
 \begin{align}
\begin{pmatrix}
\tau_{L,R}\\
E_{L,R}
\end{pmatrix}=
\begin{pmatrix}
c_{_{\theta}L,R}^l && - s_{_{\theta}L,R}^l \\
s_{_{\theta}L,R}^l && c_{_{\theta}L,R}^l
\end{pmatrix}
\begin{pmatrix}
\tau^{\prime}_{L,R}\\
E^{\prime}_{L,R}
\end{pmatrix}
\end{align}
The expressions of the mixing angles in terms of the mass matrix elements are,
\begin{align}
 \tan (2\theta_{R}^{l})=\frac{\sqrt{2}\left(\lambda^{\prime}_{l}\right)_{13}u_{_{R}} M_{L}}{M_{L}^2-\left(\lambda^{\prime}_{l}\right)^2_{13}u^2_{_{R}}/2-\left(y^{l}_{33} s_{\beta}+\tilde{y}^{l}_{33} c_{\beta}\right)^2\, v^2_{_{EW}}/2}
&&
 \tan(\theta_{L}^{l})=\frac{m_{\tau}}{m_{E}} \tan(\theta_{R}^{l})
 \label{eqEtau}
\end{align}
where $m_{\tau}$ and $m_{E}$ are the masses of the $\tau$ and the charged VLL $E$, respectively. They can be expressed as:
\begin{align}
 m_\tau= c_{_{\theta}L}^l c_{_{\theta}R}^l \left(y^{l}_{33} s_\beta+\tilde{y}^{l}_{33} c_\beta\right)\frac{v_{_\text{EW}}}{\sqrt{2}}-s_{_{\theta}L}^l c_{_{\theta}R}^l \left(\lambda^{\prime}_{l}\right)_{13}\frac{u_{_{R}}}{\sqrt{2}}+s_{_{\theta}L}^l s_{_{\theta}R}^l M_L\nonumber\\
 m_E= s_{_{\theta}L}^l s_{_{\theta}R}^l \left(y^{l}_{33} s_\beta+\tilde{y}^{l}_{33} c_\beta\right)\frac{v_{_\text{EW}}}{\sqrt{2}}+c_{_{\theta}L}^l s_{_{\theta}R}^l \left(\lambda^{\prime}_{l}\right)_{13}\frac{u_{_{R}}}{\sqrt{2}}+c_{_{\theta}L}^l c_{_{\theta}R}^l M_L
\end{align}

\section{Phenomenology of $W^{\prime}$ production and decay at the LHC}
In the dedicated model, the 2nd-generation extra-gauge bosons ($W^{\prime\prime}$ and $Z^{\prime\prime}$) are heavier than those of the 1st-generation extra-gauge bosons ($W^{\prime}$ and $Z^{\prime}$), as in the pattern of symmetry breaking we required that the vev $u_{_{R}}$ to be grater than $v_{_R}$ ($u_{_{R}}>v_{_R}$). In this section, we study the production and decay of the lightest charged extra gauge boson in this model into VLQs or HNs.

\label{W_pheno}
\subsection{Production and decay of $W^{\prime}$ into VLQs and HNs}
 We are interested in the production and subsequent decay of $W^{\prime}$ (lightest extra charged gauge boson) into VLQs ($T$ and $B$) in association with ordinary quarks; or into Majorana HNs in association with standard leptons. We recall that the exploration of the heavy charged vector bosons decaying through these channels was the focus of CMS collaboration in several recent publications, see for example refs.~\cite{CMS:2022tdo, CMS:2021dzb, CMS:2017xcw}. \\

\noindent
In this study, we set the masses of the VLQs and the HNs to half of $W^{\prime}$ mass as adopted in refs.~\cite{CMS:2022tdo,CMS:2021dzb, CMS:2017xcw}~\footnote{Actually, in CMS publications~\cite{CMS:2022tdo,CMS:2021dzb}, they adopted several mass hierarchies. We chose half of $W^{\prime}$ mass for VLQs and HNs as this lead to the highest branching ratios of $W^{\prime}$ decaying to these particles among the other scenarios.}. We consider the following processes:
\begin{align}
pp&\rightarrow \{W^{\prime}\}\rightarrow t \bar{B}+\bar{t} B &\text{with}&& B&\rightarrow b Z. \label{process1}\\
pp&\rightarrow \{W^{\prime}\}\rightarrow b \bar{T}+\bar{b} T &\text{with}&& T&\rightarrow t(Z/h/\gamma)t. \label{process2}\\
pp& \rightarrow \{W^{\prime}\} \rightarrow l_i N_i &\text{with}&& N_i&\rightarrow l_i  jj\, (i=1,2,3).
\label{process3}
\end{align}
We recall that the $B$ decay solely into $bZ$ as the partial widths of the other decay channels are proportional to  the mixing angle $s_{_{\theta}R}^d$, which is relatively suppressed as we will see. In the process (\ref{process3}), we assumed that the HNs decay into charged lepton of the same generation and two jets.\\

\noindent
Moreover, we assume that the scalars of the model are very heavy (several TeV's), therefore the scalar sector is decoupled. We work in the four-flavor scheme (4FS), thus all ordinary quarks are assumed to be massless except the top and the bottom.

\subsection{Analysis of heavy particles decays}
Let's analyze the different decay channels of the heavy states involved in the reactions (\ref{process1}), (\ref{process2}) and (\ref{process3}).
\subsubsection{Decay of $W^{\prime}$ gauge boson}
\noindent
 The
partial widths of all possible decay channels of $W^{\prime}$ are:

\begin{align}
 &\Gamma\left(W^{\prime}\rightarrow j j\right)=\frac{c_1^2\,g^2}{16\pi}\,\left(1+c^2_{13}c^2_{23}\right)\,m_{_{W^{\prime}}}
 &
  &\Gamma\left(W^{\prime}\rightarrow t j \right)=\frac{\,c_1^2\,g^2}{32\pi\,m^5_{_{W^{\prime}}}}\,\left(c^2_{23}s^2_{13}+s^2_{23}\right)\, \Gamma^{W^{\prime}}_{tj}
  \nonumber\\
  &\Gamma\left(W^{\prime}\rightarrow bj\right)=\frac{\,c_1^2\,g^2}{32\pi\,m^5_{_{W^{\prime}}}}\, \left(c^2_{23}s^2_{13}+s^2_{23}\right)\, \, \Gamma^{W^{\prime}}_{bj}
  &
  &\Gamma\left(W^{\prime}\rightarrow tb\right)=\frac{\,c_1^2\,g^2}{32\pi\,m^5_{_{W^{\prime}}}}\,\left(c^2_{13}c^2_{23}+s_{_{\theta}R}^u s_{_{\theta}R}^d\right)\, \Gamma_{bt}^{W^{\prime}}
  \nonumber\\
   &\Gamma\left(W^{\prime}\rightarrow NE\right)=\frac{c_1^2\,g^2}{48\pi\,m^5_{_{W^{\prime}}}}\Bigl[\Gamma_{EN}^{W^{\prime}}
   +6\, \tilde{\Gamma}_{EN}^{W^{\prime}}\Bigr]
   &
   &\Gamma\left(W^{\prime}\rightarrow  N_{1}e\right)=\frac{c_1^2\,g^2}{96\pi\,m^5_{_{W^{\prime}}}}\,\Gamma_{eN_1}^{W^{\prime}}
   \nonumber\\
   &\Gamma\left(W^{\prime}\rightarrow  N_{2} \mu\right)=\frac{c_1^2\,g^2}{96\pi\,m^5_{_{W^{\prime}}}}\, \Gamma_{\mu N_2}^{W^{\prime}}
   &
   &\Gamma\left(W^{\prime}\rightarrow  N_{3} \tau\right)=\frac{c_1^2\,g^2}{96\pi\,m^5_{_{W^{\prime}}}}\, \Gamma^{W^{\prime}}_{\tau N_3}\nonumber\\
    &\Gamma\left(W^{\prime}\rightarrow t B\right)=\frac{c_1^2\,g^2}{32\pi\,m^5_{_{W^{\prime}}}}\, \left(s_R^{2d}-1\right)s_R^{2u}\, \Gamma_{tB}^{W^{\prime}}
    &
    &\Gamma\left(W^{\prime}\rightarrow b T\right)=\frac{c_1^2\,g^2}{32\pi\,m^5_{_{W^{\prime}}}}\,\left(s_R^{2u}-1\right)s_R^{2d}\, \Gamma_{bT}^{W^{\prime}}\nonumber\\
    &\Gamma\left(W^{\prime}\rightarrow  N \tau\right)=\frac{c_1^2\,g^2}{96\pi\,m^5_{W^{\prime}}}\, s_R^{2l}\, \Gamma_{\tau N}^{W^{\prime}} &
    &\Gamma\left(W^{\prime+}\rightarrow  h W^{+}\right)=\frac{c_1^2\,g^4\,k^2_1\,k^2_2}{192\,v^2_{_{EW}}\pi\,m^5_{W^{\prime}}}\, \tilde{\Gamma}_{hW}^{W^{\prime}}\nonumber\\
    &\Gamma\left(W^{\prime}\rightarrow  N_{3} \bar E\right)=\frac{c_1^2\,g^2}{96\pi\,m^5_{W^{\prime}}}\, \tilde{\Gamma}^{W^{\prime}}_{EN_3}&
    &\Gamma\left(W^{\prime}\rightarrow  l_i \nu_i\right)=\frac{c_1^2\,g^2}{96\pi\,m^5_{W^{\prime}}}\left(\frac{m_{\nu_i}}{m_{N_i}}\right) \,\Gamma^{W^{\prime}}_{l_i\nu_i}
 \end{align}
 with
\begin{align}
        \tilde{\Gamma}_{hW}^{W^{\prime}}&=\sqrt{\lambda_{hW}^{W^{\prime}}} \Bigl[\left(m^2_{h}-\left(m^2_{W}+m^2_{W^{\prime}}\right)\right)^2+8\,m^2_{W}m^2_{W^{\prime}}\Bigr]
        \\
    \tilde{\Gamma}^{W^{\prime}}_{EN_3}&=\Gamma^{W^{\prime}}_{EN_3}\left(s_L^{2{\cal{N}}}+s_R^{2{\cal{N}}}\right)+12\,\sqrt{\lambda^{W^{\prime}}_{EN_3}}\,m_{E}\,m_{N_3}\,m^2_{W^{\prime}}s_L^{{\cal{N}}}\,s_R^{{\cal{N}}}
    \label{widthWp}
\end{align}
and
\begin{align}
 \lambda_{bc}^{a}&=  \left(m_{a}^4+m_{b}^4+m_{c}^4-2m_{a}^2m_{b}^2-2m_{a}^2m_{c}^2-2m_{b}^2m_{c}^2\right)\nonumber\\
 \Gamma_{bc}^{a}&=\sqrt{\lambda_{bc}^{a}}\left[2 m^4_{a}-\left(m_b^2-m_c^2\right)^2-\left(m_c^2+m_b^2\right)m^2_{a}\right]
\end{align}
The partial widths, in eq.~(\ref{widthWp}), were obtained under the assumption that the VLQs mix only with the third generation quarks, and the left-handed mixing angles of VLFs are neglected according to eqs.~(\ref{eqQq}) and (\ref{eqEtau}). Moreover, due to the sensitivity of $\tau$ to any exotic mixing, we give $s_{_{\theta}R}^{l}$ and $s_{_{\theta}L,R}^{{\cal{N}}}$ small values to remain consistent with the LRSM, leading to insignificant enhancement to the partial widths. Therefore, the relevant free parameters controlling these partial widths, are thus the right-sector mixing angles between the VLQs and the third generation quarks (i.e. $s_{_{\theta}R}^{u,d}$), the extra gauge coupling $g_{_V}$ (contained in $c_1$) and the masses of the extra particles. Using the global limits on the mixing parameter $\kappa$ obtained in ref.~\cite{Benbrik:2024fku} from ATLAS and CMS Run I and Run II, we are able to put upper limits on the mixing angles of the right-sector. This can be achieved by making an analogy with the vertices describing the VLQs mixing with the third generation quarks and the Higgs from the effective field approach, cf.  eq.~(A.3) in the appendix $A$ of \cite{Benbrik:2024fku}~\footnote{We note also that the $T-t-h$ vertex looks quite similar to that one from the Little Higgs model~\cite{Han:2003wu}. However, the leading part in their case is the left-handed part since the $T$ VLQ has been included as a singlet.}. The Lagrangian describing the mixing of the VLQs with 3rd quark generation via the Higgs boson is the following:
\begin{align}
{\cal{L}}_{h}=&-\frac{m_t}{\sqrt{2}\,v_{_{EW}}}\,\bar t \left(\frac{m_t}{m_T} \tan\left(\theta_R^u\right) P_L+\tan\left(\theta_R^u\right) P_R\right)\,T\,h\nonumber\\&
-\frac{m_b}{\sqrt{2}\,v_{_{EW}}}\,\bar b \left(\frac{m_b}{m_B} \tan\left(\theta_R^b\right) P_L+\tan\left(\theta_R^d\right) P_R\right)\,B\,h+h.c
\end{align}

\noindent
In the left panel of figure \ref{p1}, we show allowed region of $s_{_{\theta}R}^{u,d}$ for several benchmark values of $m_{T,B}$, where the mixing angle of down sector is slightly smaller the one of the up sector. Thus, we can confidently choose $s_{_{\theta}R}^{u}=0.54$ and $s_{_{\theta}R}^{d}=0.275$ for the rest of the analysis. We note that the total width of $W^{\prime}$ could be affected by the right sector mixing angles, as shown in the right panel of figure \ref{p1}.\\

\begin{figure}[!h]
  \centering
  {\includegraphics[width=8cm,height=6.5cm]{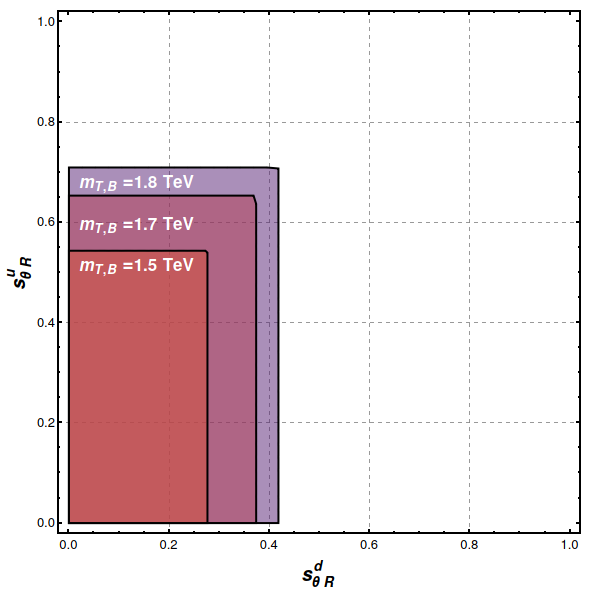}\label{p1}}
  {\includegraphics[width=8cm,height=6.5cm]{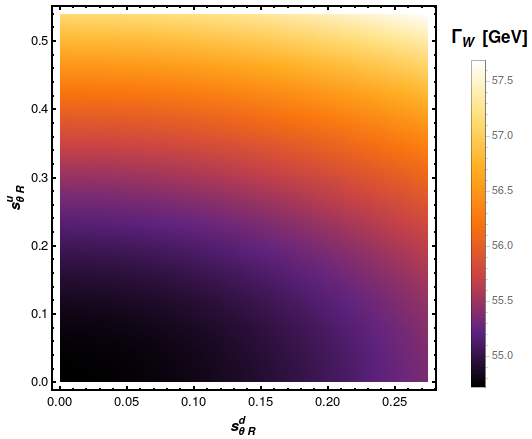}\label{p2}}
\caption{The $s_{_{\theta}R}^{u}-s_{_{\theta}R}^{d}$ allowed region derived from the experimental limits on the mixing parameter $\kappa$ by exploiting ATLAS and CMS data at the LHC during Run 1 and Run 2 for several benchmark values of $m_{T,B}=1.5$, $1.7$ and $1.8$ TeV~\cite{Benbrik:2024fku} (left panel). Scan of the total width of $W^{\prime}$ in the $s_{_{\theta}R}^{u}-s_{_{\theta}R}^{d}$ plan for $m_{W^{\prime}}=3$ TeV (right panel).}
  \label{p1}
\end{figure}

\noindent
Let's now discuss the variation of the width-to-mass ratio of $W^{\prime}$ ($\Gamma_{W^{\prime}}/m_{W^{\prime}}$). In figure~\ref{1}, we show the variation of $\Gamma_{W^{\prime}}/m_{W^{\prime}}$, in percentage, in term of $m_{W^{\prime}}$ for several benchmark values of $g_{_V}$.  We observe that $\Gamma_{W^{\prime}}/m_{W^{\prime}}$ increase with $g_{_V}$, where the largest value (about $3.8 \%$) correspond to $g_{_V}=3$. Therefore, the width-to-mass ratio of $W^{\prime}$ is narrow and one can use the narrow width approximation (NWA)~\cite{Deandrea:2021vje,Rekaik:2025qmt}. \\

\begin{figure}[!h]
  \centering
  {\includegraphics[width=9cm,height=7.5cm]{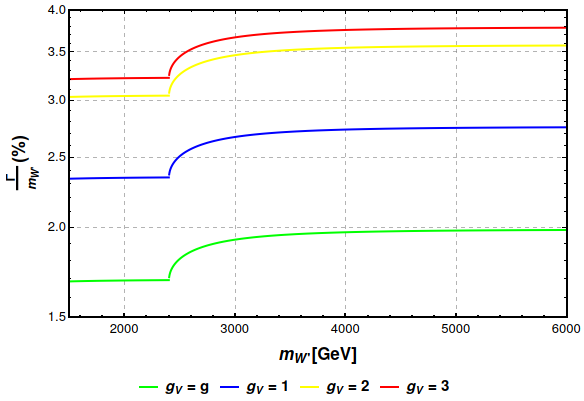}\label{1}}
  \caption{Percentage of variation of the total width of $W^{\prime}$ over their mass w.r.t $m_{W^{\prime}}$ for several benchmark values of $g_{_V}$.}
  \label{1}
\end{figure}

\begin{figure}[!h]
  \centering
  {\includegraphics[width=8.5cm,height=5.5cm]{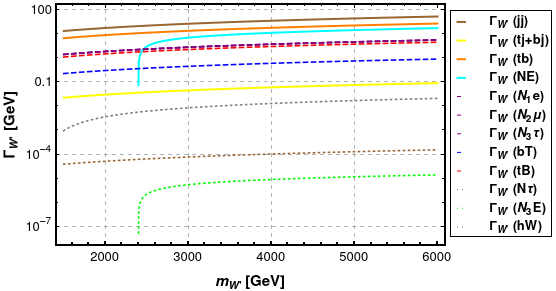}\label{3}}
  {\includegraphics[width=8.5cm,height=5.5cm]{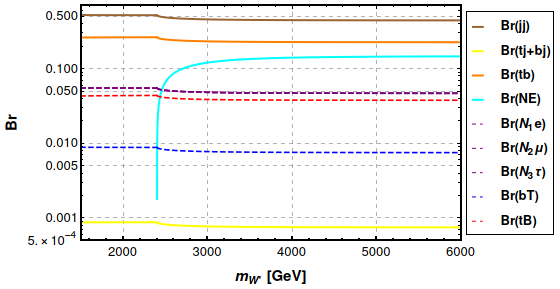}\label{4}}
  \caption{Variation w.r.t the mass $m_{W^{\prime}}$ of the partial widths (left) and branching ratios (right) of the all the $W^{\prime}$ allowed channels, for $g_{_V}=g$. }
  \label{2}
\end{figure}

\noindent
The variation of the partial widths and branching ratios of $W^{\prime}$ in term of $m_{W^{\prime}}$, for all the the kinematically allowed channels, are shown in figure~ \ref{2}. We observe that several channels have suppressed partial widths, see the dotted curves in the left panel, and therefore would be ignored. The dashed curves are associated to channels of interest in this work, which are $tB/bT/N_il$. In the right panel, we show the variation of the branching ratios of the unsuppressed channels w.r.t $m_{W^{\prime}}$. We observe that the $W^{\prime}$ decay is almost dominated by $jj$ and $tb$ channel, where both represent about $70\%$ of the decay rate. Regarding the branching ratios of the channels of interest are about $5\%$ for each generation of $N_il$, $4\%$ for $tB$ and $0.8\%$ for $bT$ final states. It is very important to note that the decay rate of the $W^{\prime}$ ito VLFs $EN$ is significant, on the order of $10\%$, making it a promising channel for future research.

\noindent
In our framework, the VLQs can decay to the 3rd quark generation and neutral SM bosons ($\gamma$, $Z$ and $h$), a property shared by many BSM model, see for example ref.~\cite{Vignaroli:2014bpa}. The partial widths of $T$ and $B$, which are independent of the parameter $g_{_V}$, are given in the following:

\begin{align}
 \Gamma\left(T\rightarrow \gamma t\right)&=\frac{g^2\,s^2_{\theta_{w}}}{36\pi\,m^3_T}\,c_{\theta R}^{2u}s_{\theta R}^{2u}\,\left(m^4_T-m^4_t\right)
&&  \Gamma\left(B\rightarrow \gamma b\right)=\frac{g^2\,s^2_{\theta_{w}}}{144\pi\,m^3_B}\,c_{\theta R}^{2d}s_{\theta R}^{2d}\,\left(m^4_B-m^4_b\right)
\nonumber\\
\Gamma\left(T\rightarrow h t\right)&=\frac{m_t^2\tan^2(\theta_R^{u})}{64\pi\,m^3_T\,v^2_{_{EW}}}\, \sqrt{\lambda^h_{tT}}\left(m^2_T+m^2_t-m^2_h\right)
&& \Gamma\left(B\rightarrow h b\right)=\frac{m_b^2\tan^2(\theta_R^{d})}{64\pi\,m^3_B\,v^2_{_{EW}}}\,\sqrt{\lambda^{b}_{Bh}}\left(m^2_B+m^2_b-m^2_h\right)\nonumber\\
\Gamma\left(T\rightarrow Z t\right)&=\frac{g^2\,s^4_{\theta_{w}}}{72\pi\,m^3_T\,m^2_Z\,c^2_{\theta_{w}}}\,c_{\theta R}^{2u}s_{\theta R}^{2u}\, \Gamma_{T}^{tZ}
&&\Gamma\left(B\rightarrow Z b\right)=\frac{g^2\,s^4_{\theta_{w}}}{288\pi\,m^3_B\,m^2_Z\,c^2_{\theta_{w}}}\,c_{\theta R}^{2d}s_{\theta R}^{2d}\, \Gamma_{B}^{bZ}
\label{Twidths}
\end{align}
with
\begin{align}
\Gamma_{_{VLQ}}^{qZ}&=\sqrt{\lambda^{qZ}_{_{VLQ}}}\left[\left(m^2_{_{VLQ}}-m^2_q\right)^2+\left(m^2_{_{VLQ}}+m^2_q\right)\,m^2_Z-2\,m^2_Z\right]
\end{align}

\begin{figure}[!h]
  \centering
  {\includegraphics[width=0.495\textwidth]{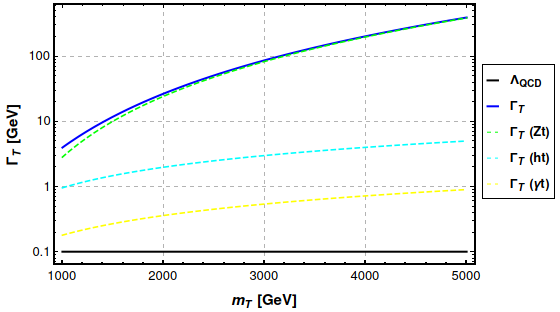}\label{9}}
  {\includegraphics[width=0.485\textwidth]{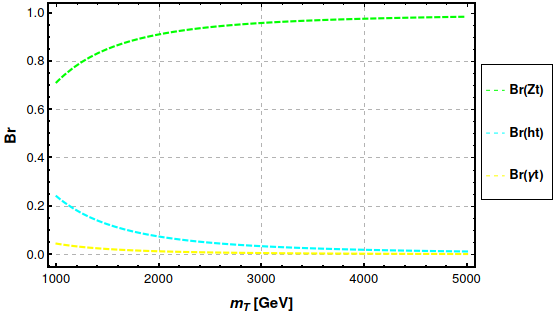}\label{10}}
  {\includegraphics[width=0.501\textwidth]{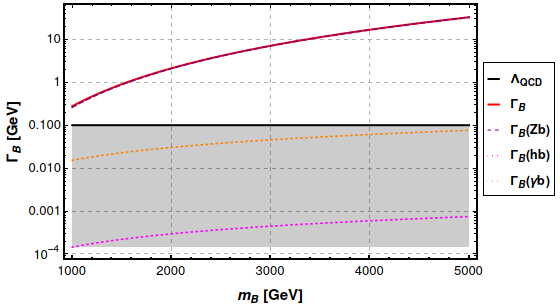}\label{11}}
  {\includegraphics[width=0.493\textwidth]{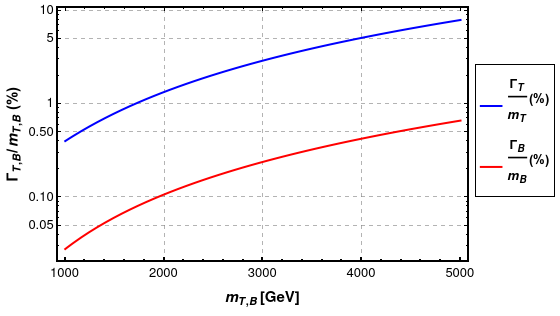}\label{12}}
  \caption{Total and partial widths of $T$ w.r.t $m_T$ (upper left), branching ratios of $T$ w.r.t $m_T$ (upper right), total and partial widths of $B$ w.r.t $m_B$ (lower left) and total width-to-mass ratios (in percentage) $T$ and $B$ quarks (lower right) for $s_{\theta R}^{u}=0.54$ and $s_{\theta R}^{d}=0.275$.}
  \label{p4}
\end{figure}

\noindent
From eq.~(\ref{Twidths}), we see that the partial widths of the VLQs are controlled only by the mixing angles $\theta_R^{q}$ and their masses, i.e. they are completely independent of $g_{_V}$. The variation of the total and partial widths of $T$ and $B$, in term of $m_{_{VLQ}}$, are shown in the upper and lower left panels of the figure \ref{p4}, respectively. We observe that the total width of $T$ is much larger than total width of $B$, since the latter one is proportional to $s_{_{\theta}R}^{d}$, which is much smaller than $s_{_{\theta}R}^{u}$, for which the former is proportional to. It's worth to notice that the partial widths of $B$ through $\gamma b$ and $hb$ are less than $\Lambda_{QCD}$ (under the black line)~\footnote{$\Lambda_{QCD}$ has been taken to be $100$ MeV \cite{Gari:1986vr}, as it was the case included by default in {\tt M{\scriptsize AD}G{\scriptsize RAPH}5}.}. Therefore, they have to be neglected which makes the $B$  decaying $100\%$ through $Zb$.\\

\noindent
In the upper left panel of figure~\ref{p4}, we show the variation of the branching ratios of $T$ as a function of its mass. We observe that, the channel $tZ$ dominates over the others, with branching ratio varying from $70\%$ (at $m_{_T}=1$ TeV) to about $97\%$ (at $m_{_T}=5$ TeV). The branching ratios of the channels $th$ and $t\gamma$ are approximately 25\% and 5\% for small masses, but they decree significantly for higher mass while remaining greater than $\Lambda_{QCD}$. The lower right panel of the same figure shows the variation of the total width-to-mass ratios (in percentage ) of both VLQs w.r.t their masses. We see that they are relatively small (not exceeding $10\%$ and $1\%$ for $T$ and $B$ quarks, respectively), which justifies the use of the NWA as we will show later.
\\

\begin{figure}[!h]
  \centering
  {\includegraphics[width=1.0\textwidth]{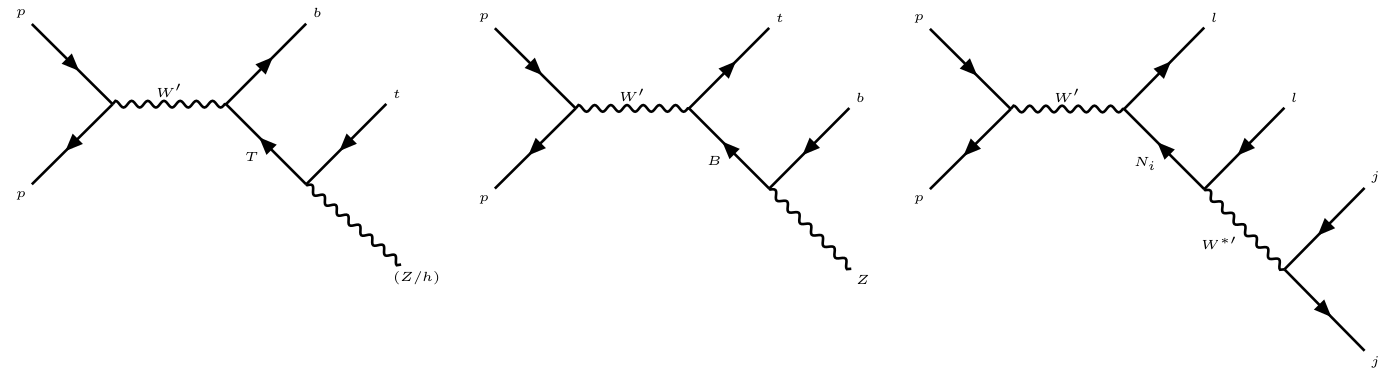}}
 \caption{Born Feynman diagrams of the resonant $W^{\prime}$ production and its decay into VLQs (left and middle) and  Majorana HNs (right).}
 \label{diag}
\end{figure}

\noindent
The production and decay to VLQs of the $W^{\prime}$ boson was examined by CMS in many publications, where the VLQ are assumed to decay onto a 3rd generation quark and a $Z$ and/or Higgs boson~\cite{CMS:2022tdo, CMS:2021dzb}, or to heavy neutrinos in association with ordinary leptons of the same generation~\cite{CMS:2017xcw}. We recall that the $T$ quark, in the context of our model, can decay to $tZ$ and $tH$ (and $t\gamma$ but can be neglected), while the $B$ decays $100\%$ to $bZ$, see the left and middle Feynman diagrams in figure~\ref{diag}. The right Feynman diagrame of the same figure, describe the Drell-Yan like production of a $W^{\prime}$ which subsequently decay into HNs and SM leptons of the same generation $N_i l$ (i.e. $N_1 e$, $N_2 \mu$ and $N_3 \tau$), where the branching ratios are assumed to be equal. According to the Majorana nature of the HNs, the final-state of theses channels could be two charged leptons has the same sign with tow jets, where the channels of interest that has been studied and will be examined in our model is when the HNs gives through an off-shell $W^{*\prime}$ an other lepton of the same generation with two jets. 

\noindent
Since the mediators of these processes are unstable, the common approach to regularise the propagators of the unstable particles is to use the Breit-Wigner (BW) approximation. Thus, the denominators of the squared propgators are replaced by:
\begin{align}
 BW(p^2)=\frac{1}{p^2-m^2+\Gamma^2\,m^2}
\end{align}
where $p$ is the momentum of the mediator, $m$ is their on-shell mass and $\Gamma$ is their total width.
Moreover, we have shown above that the total width-to-mass ratios of the $W^{\prime}$ boson and the VLQs are relatively samall ($\Gamma_{W^{\prime}}/m_{W^{\prime}}<4\%$, $\Gamma_{T}/m_{T}<10\%$ and $\Gamma_{B}/m_{B}<1\%$). Therefore, we can rely instead on the NWA by replacing $BW(p^2)$ by:
\begin{align}
 BW(p^2)\approx \frac{\pi}{\Gamma\,m}\delta\left(p^2-m^2\right)
\end{align}
This allows as the factorize the cross sections into a production of production and decay factors~\cite{Deandrea:2021vje}. Thus, we get:
\begin{align}
\sigma_{pp\rightarrow tb(Z/h)}\,&\simeq\,\sigma_{W^{\prime}}\times\,Br_{W^{\prime}\rightarrow Tb}\times\,Br_{T\rightarrow t(Z/h)}&
\sigma_{pp\rightarrow l l j j}\,&\simeq\,\sigma_{W^{\prime}}\times\,Br_{W^{\prime}\rightarrow l l j j}
\end{align}
where $\sigma_{W^{\prime}}$ is $W^{\prime}$ boson production cross section, which is sensitive to the coupling $g_{_V}$.\\

 \noindent
The numerical calculations in this paper has been done using {\tt M{\scriptsize AD}G{\scriptsize RAPH}5}  \cite{Alwall:2014hca}, where we used the {\tt NNPDF3.0}~\cite{NNPDF:2014otw} and fixed the factorization and renormalization scales to the sum of the transverse mass of the final state particles devided by 2.
\begin{figure}[h!]
  \centering
  {\includegraphics[width=0.495\textwidth]{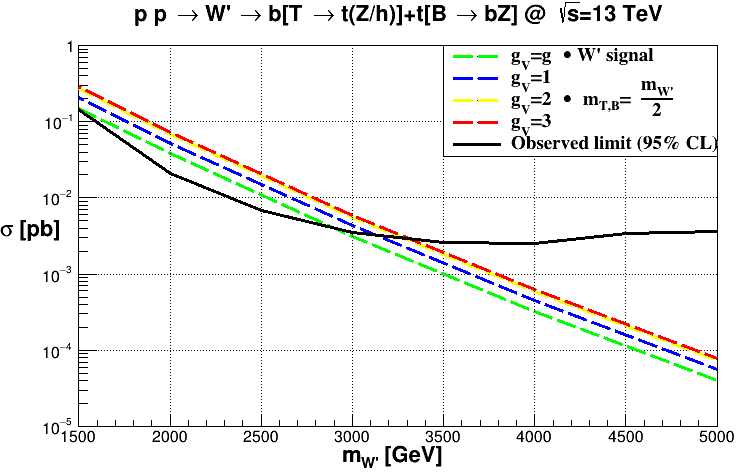}\label{cs1}}
  {\includegraphics[width=0.495\textwidth]{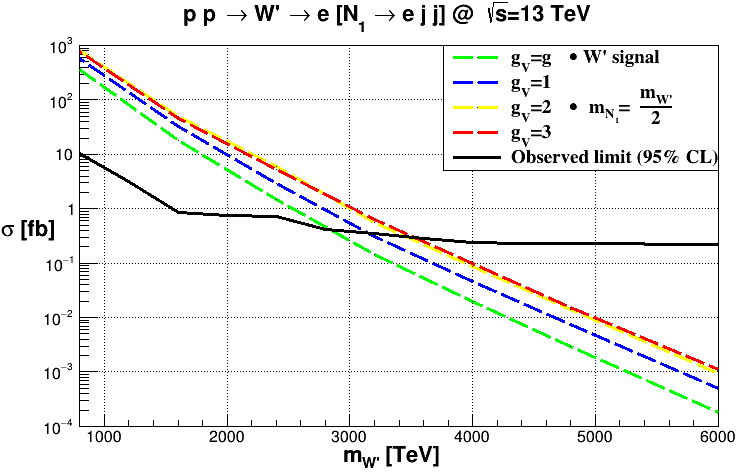}\label{cs2}}
  {\includegraphics[width=0.495\textwidth]{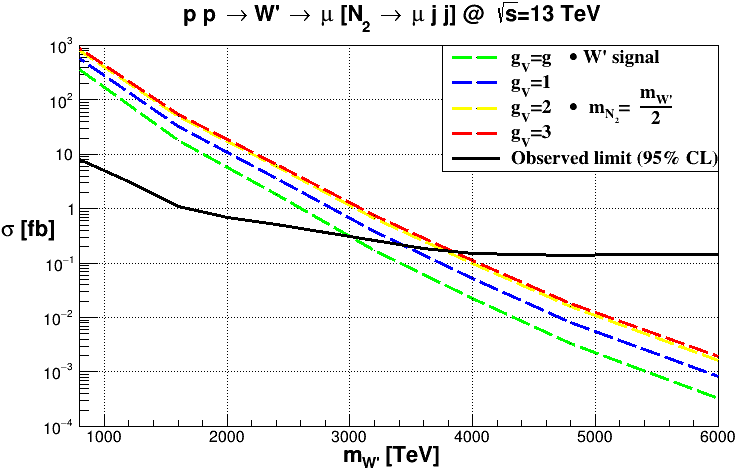}\label{cs3}}
  {\includegraphics[width=0.495\textwidth]{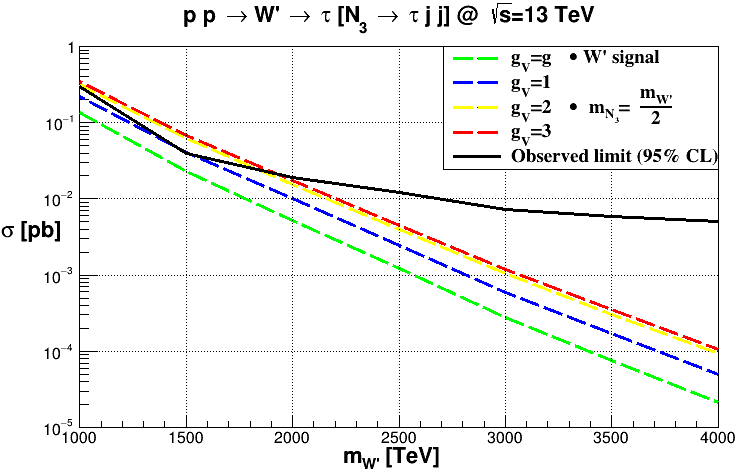}\label{cs4}}
  \caption{ The NWA of the LO cross sections for $W^{\prime}$  production and decay through the VLQs and the Drell-Yan like channels w.r.t $m_{W^{\prime}}$ for several benchmark values of $g_{_V}$ at $\sqrt{s}=13$ TeV. The black lines represent the experimental observed limits from CMS at $95\%$ CL \cite{CMS:2022tdo, CMS:2021dzb, CMS:2017xcw}.} 
  \label{p5}
\end{figure}
In figure~\ref{p5}, we show the variation of the total cross sections at LO for the production and decay of $W^{\prime}$ through VLQs (up left panel), and the 1st, 2nd and 3rd generation HNs in association with SM leptons of the same generation (up right panel, down left and right panel, respectively), for 4 values of the coupling $g_{_V}$ at $\sqrt{s}=13$ TeV. The masses of the VLQs and the HNs are assumed to half of $W^{\prime}$, a benchmark scenarios adopted in refs.~\cite{CMS:2022tdo, CMS:2021dzb, CMS:2017xcw}. To each panel, we add the upper limits on the total cross section at $95\%$ CL (black curves) from the CMS {\tt run II} data\footnote{\url{https://www.hepdata.net/record/ins2039384} (for VLQs) and \url{https://www.hepdata.net/record/ins1986733} (for HNs).}. We notice that the most powerful lower limit on the $W^{\prime}$ mass comes from its decay into HN and SM lepton of the 2nd generation (lower left panel). The lower limits  on $m_{W^{\prime}}$, for different values of $g_{_V}$, from the different examined channels are shown in table \ref{tab1}. We note that, these limits indirectly restrict the mass of $Z^{\prime}$ boson, cf.~(\ref{masseq}). We can see that the constraints on $m_{Z^{\prime}}$ are strong enough in way that the experimental bounds can be ignored.
\begin{table}[h!]
\begin{tabular}{|c|c|c|c|c|}
\hline
$\mathbf{g_{_V}}$  & \textbf{VLQs channel} & $\mathbf{N_1 e}$ \textbf{channel} & $\mathbf{N_2 \mu}$ \textbf{channel} & $\mathbf{N_3 \tau}$ \textbf{channel} \\
\hline
$g$ & $2.90$ $(6.27)$ TeV & $2.80$ $(6.06)$ TeV & $3.00$ $(6.49)$ TeV  & $\textbf{--}$ \\
\hline
$1.0$ & $3.10$ $(5.72)$ TeV & $3.20$ $(5.90)$ TeV & $3.40$ $(6.27)$ TeV & $\textbf{--}$ \\
\hline
$2.0$ & $3.25$ $(5.56)$ TeV & $3.45$ $(5.91)$ TeV & $3.75$ $(6.42)$ TeV & $1.85$ $(3.17)$ TeV \\
\hline
$3.0$ & $3.30$ $(5.58)$ TeV & $3.55$ $(5.98)$ TeV & $3.80$ $(6.41)$ TeV & $1.95$ $(3.29)$ TeV  \\
\hline
\end{tabular}
\caption{The $m_{W^{\prime}}$ limits ($m_{Z^{\prime}}$ indirect limits) for different benchmark values of $g_{_V}$ from the experimental data of ATLAS on $W^{\prime}$ production and decay through VLQs channels \cite{CMS:2022tdo}, and Drell-Yan like channels \cite{CMS:2021dzb, CMS:2017xcw}.}
\label{tab1}
\end{table}

\section{Top VLQ single production in association with top and bottom quarks}
\label{LHC_VLQ_singel}
It is well known that the single production of VLQs dominates over the pair production for higher VLQs mass. This is due the fact that the latter one is of pure QCD origin, thus the phase space become more restricted for higher mass, while the former one is of pure electroweak origin, where the couplings might be large for higher mass~\cite{Deandrea:2021vje,Rekaik:2025qmt}. For this reason, we focus on the single production 
of a $T$ quark in association with a top and bottom quarks. The $T$ quark can be produced singly through CC processes (via the exchange of $W^{\prime}$ and $W^{\prime\prime}$) or through NC processes (via the exchange of $Z^{\prime}$ and $Z^{\prime\prime}$ or a photon)~\footnote{Although, the NC processes could be mediated by the neutral scalars $H_1^0$ and $A_1^0$. However, their contribution is not relevant in our case, since we have assumed that the scalar sector is decoupled.}. The relevant processes are the following:
\begin{align}
 p p \rightarrow \left(W^{\prime}/W^{\prime\prime}\right)\rightarrow b T &&  p p \rightarrow \left(\gamma/Z/Z^{\prime}/Z^{\prime\prime}\right)\rightarrow t T
\end{align}

\begin{figure}[!h]
  \centering
  {\includegraphics[width=0.325\textwidth]{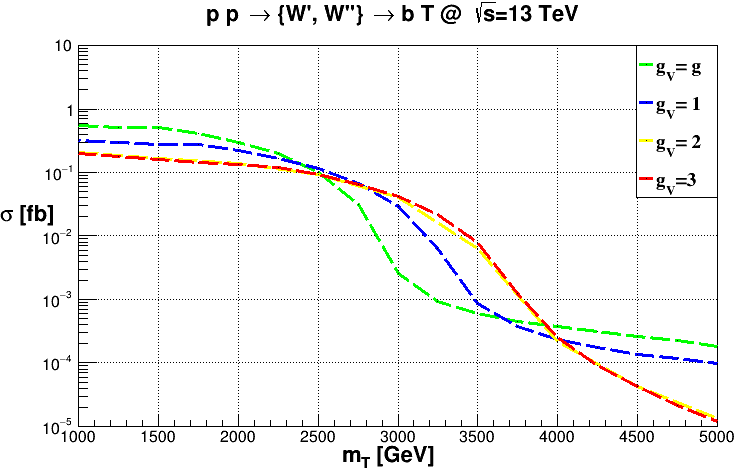}}
  \hfill
  {\includegraphics[width=0.325\textwidth]{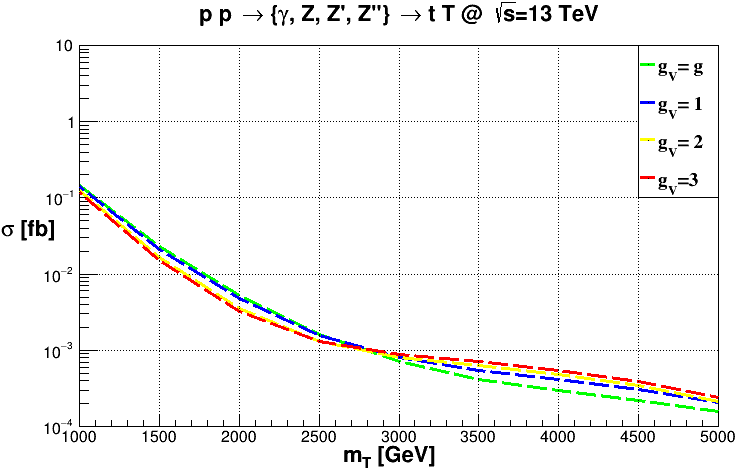}}
  \hfill
  {\includegraphics[width=0.325\textwidth]{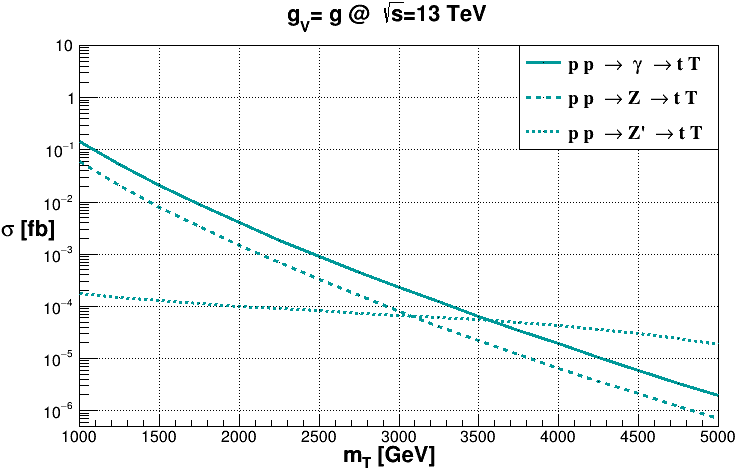}}
  \caption{(Left) LO cross sections for $T$ single production with $b$ quark. (Middle) LO cross sections for $T$ single production with $t$ quark. (Right) LO sub-processes cross sections for $T$ single production with $t$ quark. The calculations has been done at $\sqrt{s}=13$ TeV for several benchmark values of $g_{_V}$.}
  \label{fig2}
\end{figure}

\noindent
In figure.~\ref{fig2}, we provide the variation of the total cross section as a function of $m_{_T}$ for the CC process (right panel) and NC process (middle panel) at center-of-mass energy $\sqrt{s}=13$ TeV, for several benchmark values of the coupling $g_{_V}$, where we have fixed the mass of $W^{\prime}$ to their most restricted limits $m_{W^{\prime}}=3.00, 3.40, 3.75$ and $3.80$ TeV for $g_{_V}=g, 1, 2$ and $3$ (see table table~\ref{tab1}). We notice that the contributions of the second-generation gauge bosons ($W^{\prime\prime}$ and $Z^{\prime\prime}$) are negligible compared to the contribution of the other gauge bosons due to their large masses (for $g_{_V}=g$, $m_{_{W^{\prime\prime}}}$ and $m_{_{Z^{\prime\prime}}}$ are larger than 6.5 TeV, and their mass values are even larger for the other benchmarks of $g_{_V}$).\

\noindent
We observe that the CC cross sections (for different $g_{_V}$ values) decrease with increasing $m_{_T}$, where they exhibiting a sharp drop near the decay threshold of $W^{\prime}\rightarrow bT$, i.e. where $m_{W^{\prime}}\approx m_{_T}+m_{b}$. This behavior is explained as follows: {\it (i)} In the region where $m_{_T}<m_{_W^{\prime}}-m_{b}$ and given the small width-to-mass ratio of $W^{\prime}$ (less than $3.5\%$), it can effectively be treated as being produced on-shell and subsequently decays to $bT$. In this regime, the cross section factorizes to production and decay contributions, well approximated by $\sigma_{W^{\prime}}\times \text{Br}(W^{\prime}\rightarrow bT)$, where $\sigma_{W^{\prime}}$ is the cross section of $W^{\prime}$ production. {\it (ii)} In the region $m_{_T}>m_{_W^{\prime}}-m_{b}$, the decay $W^{\prime}\rightarrow bT$ becomes kinematically forbidden, making  the total width of $W^{\prime}$ drop sharply. Therefore, the $W^{\prime}$ must be treated as an off-shell state. We recall that the mass of $W^{\prime}$ itself depends on $g_{_V}$. Consequently, the position of the decay thresholds varies slightly for the different $g_{_V}$ values, as see in the green, blue, yellow and red curves of the left panel. From the right panel, we observe that the NC cross section is steadily decreasing w.r.t increasing $m_T$ as expected, due to the increasingly restriction of the phase space for high mass of the final state particles. We note that the cross section is dominated by the contributions of the sub-processes mediated by $\gamma$ and $Z$. This is demonstrated by the curves solid-cyan (photon only), dashed-cyan ($Z$ boson only) and dotted-cyan ($Z^{\prime}$ boson only), see the right panel of figure~\ref{fig2}.\\

\noindent
We observe that the CC cross sections are larger than the NC cross sections, particularly in the region where $m_{_W^{\prime}}>m_{_T}$, where the latter one decreases shapely while the former exhibits a less pronounced drop. For example, in the regime $m_{_T}\in [1,2.5]$ TeV with $g_{V}=g$, the NC cross section varies between $10^{-1}$ and $10^{-2}$ fb, whereas the CC cross section varies between  $5\times 10^{-1}$ and $10^{-1}$ fb. This observation makes the single production CC channel particularly promising, as it dominates not only over the pair-production processes as mentioned earlier but also over the NC reactions. The distinctive behavior of the CC cross section around the decay threshold is indicative of the presence of such a new mediator.\\

\begin{figure}[!h]
  \centering
  {\includegraphics[width=0.49\textwidth]{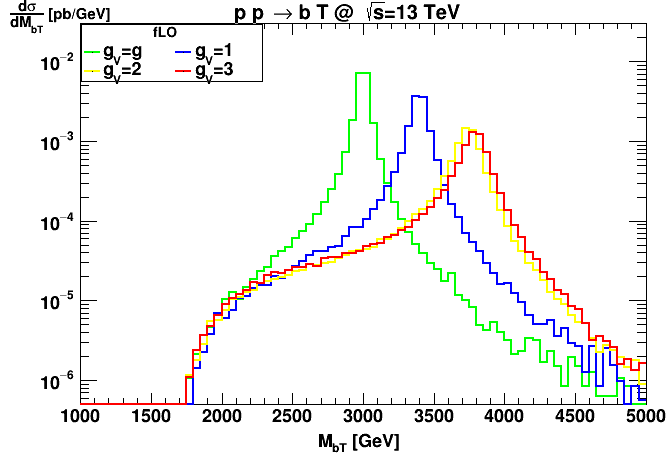}}
  \hfill
  {\includegraphics[width=0.49\textwidth]{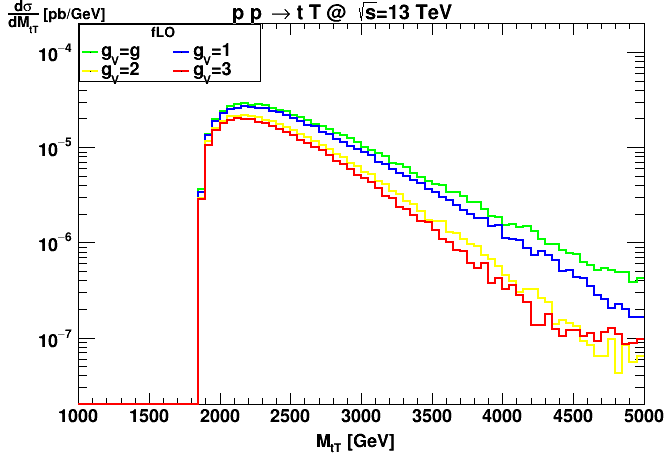}}
  \caption{LO invariant mass differential distribution of the the final-state for the CC process (upper left) and for the NC process (lower left). LO invariant mass differential distribution of the final-state leading jets finding for the CC process (upper right) and for the NC process (lower right).}
  \label{fig3}
\end{figure}

\noindent
One distinguishing features of the CC cross section, in comparison to the NC one, is the presence  of the $W^{\prime}$ decay threshold effects. To examine this closely, we generated the fixed leading order (fLO) invariant mass distributions of the final state particle at the parton level, assuming $m_{_T}=1.7$ TeV, see figure~\ref{fig3}. We observe that the NC differential distributions show no peak around $m_{_{Z^{\prime}}}$, as the contributions of the sub-processes mediated by the $Z^{\prime}$ are negligible compared to those mediated by $Z$ and $\gamma$, see the right panel of figure~\ref{fig2}. In contrast, the CC differential distributions show a clear peak near the decay threshold, i.e. around $m_{_{W^{\prime}}}= M_{tT}$. This demonstrates that the resonantly produced $W^{\prime}$ contribution is very significant.   

\begin{figure}[!h]
  \centering
  {\includegraphics[width=8.5cm,height=6.5cm]{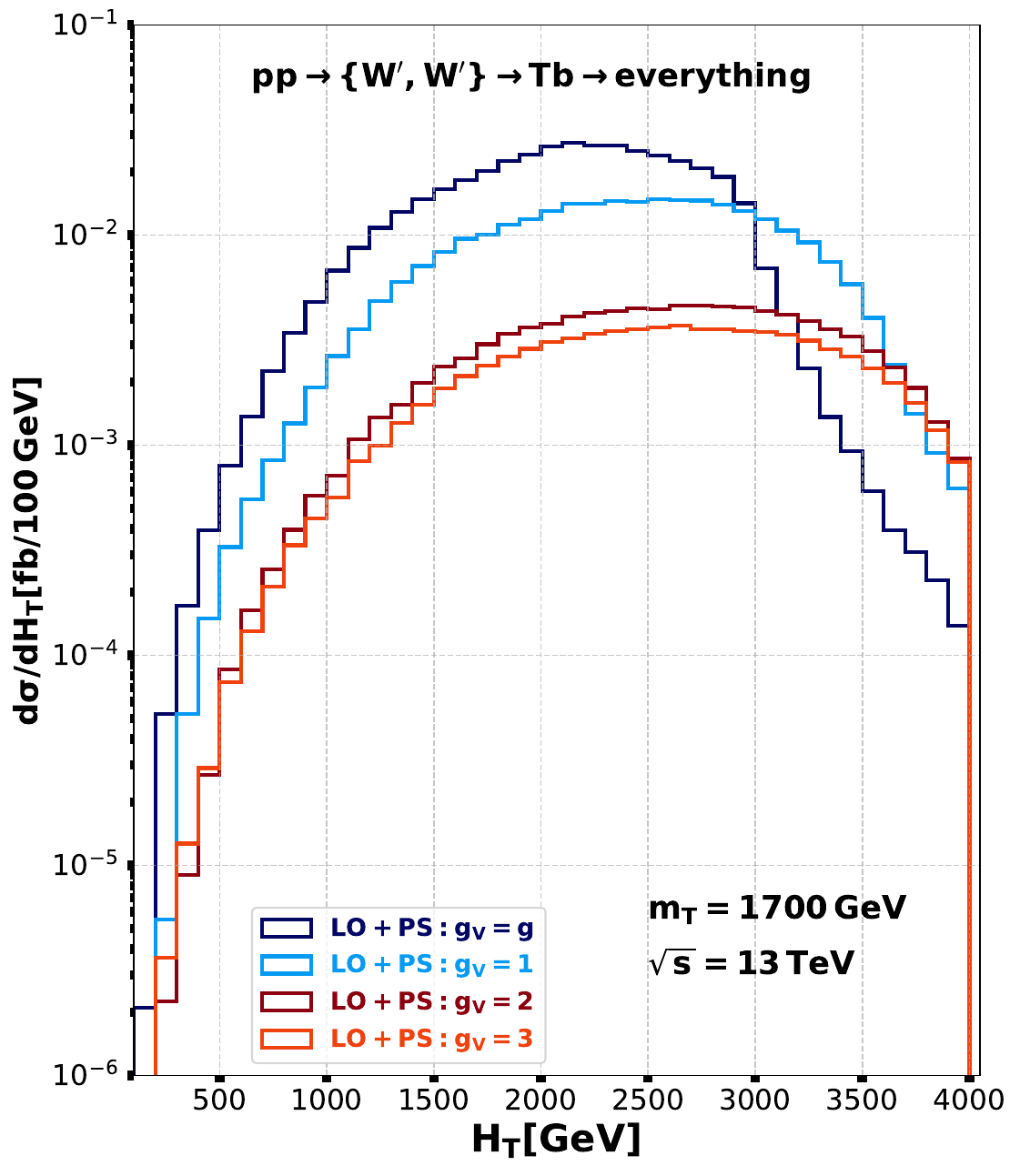}}
  \hfill
  {\includegraphics[width=8.5cm,height=6.5cm]{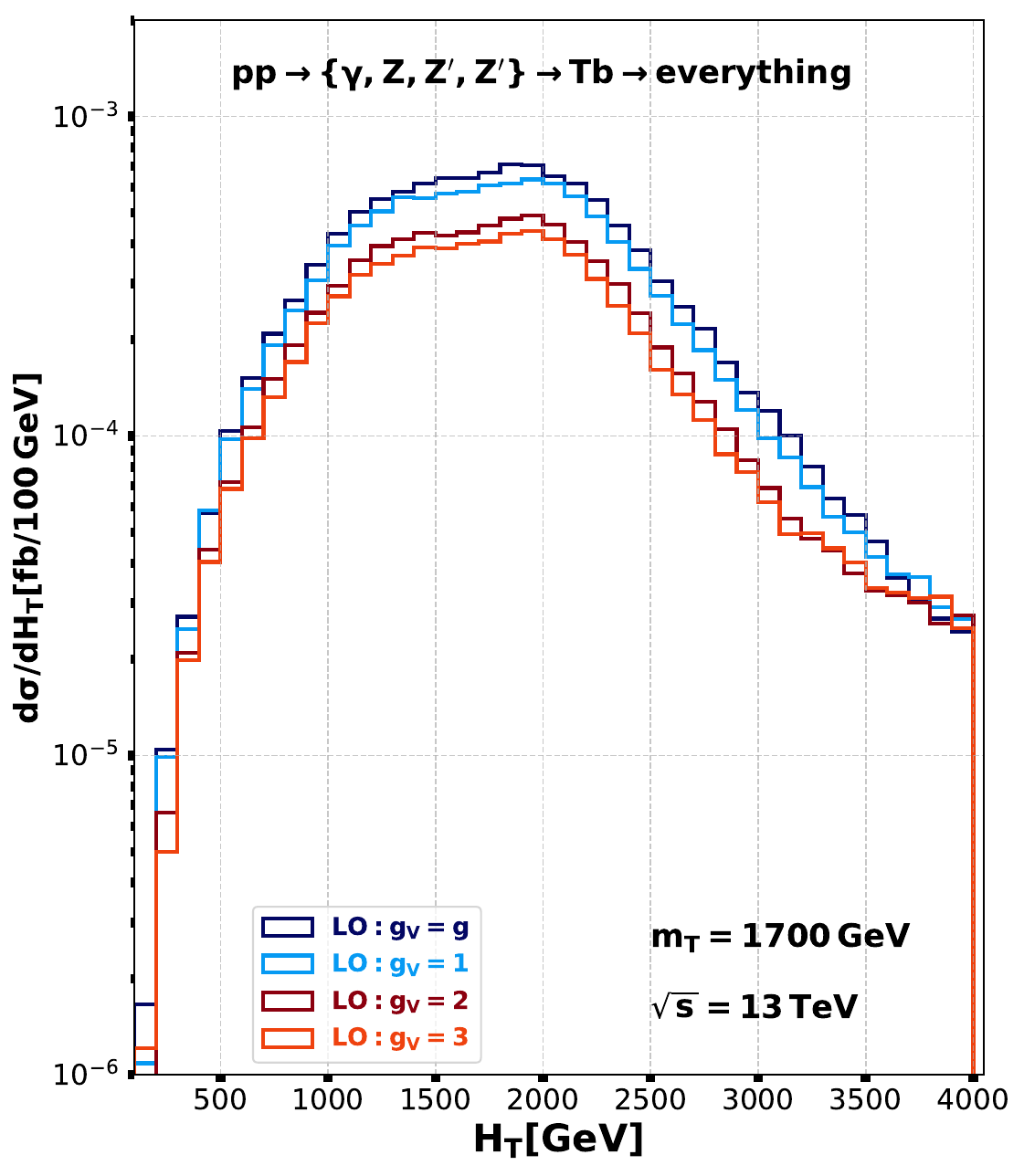}}
  \caption{\small LO+PS $H_T$ differential distribution.}
  \label{distHT}
\end{figure}

\begin{figure}[!h]
  \centering
  {\includegraphics[width=8.5cm,height=6.5cm]{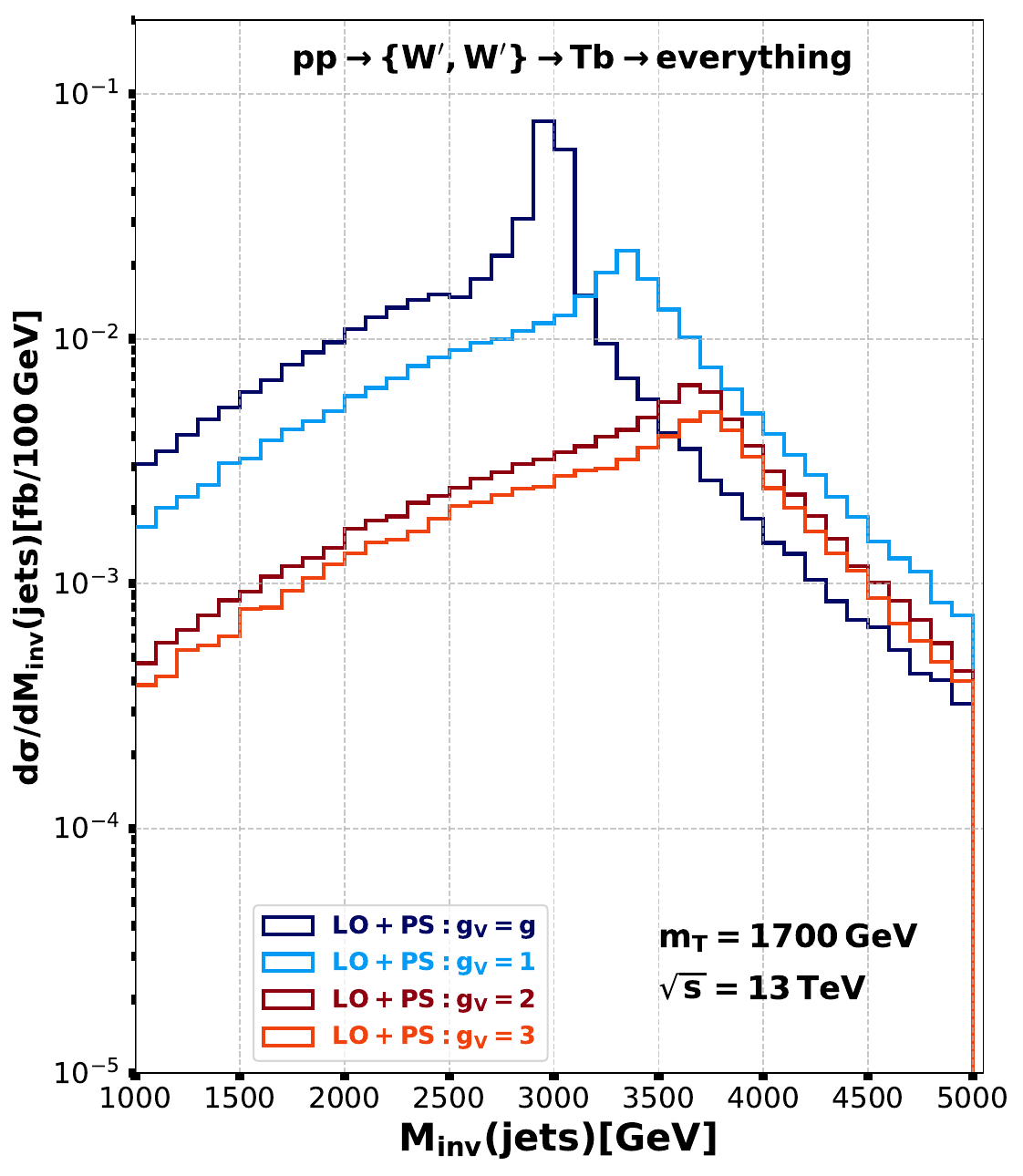}}
  \hfill
  {\includegraphics[width=8.5cm,height=6.5cm]{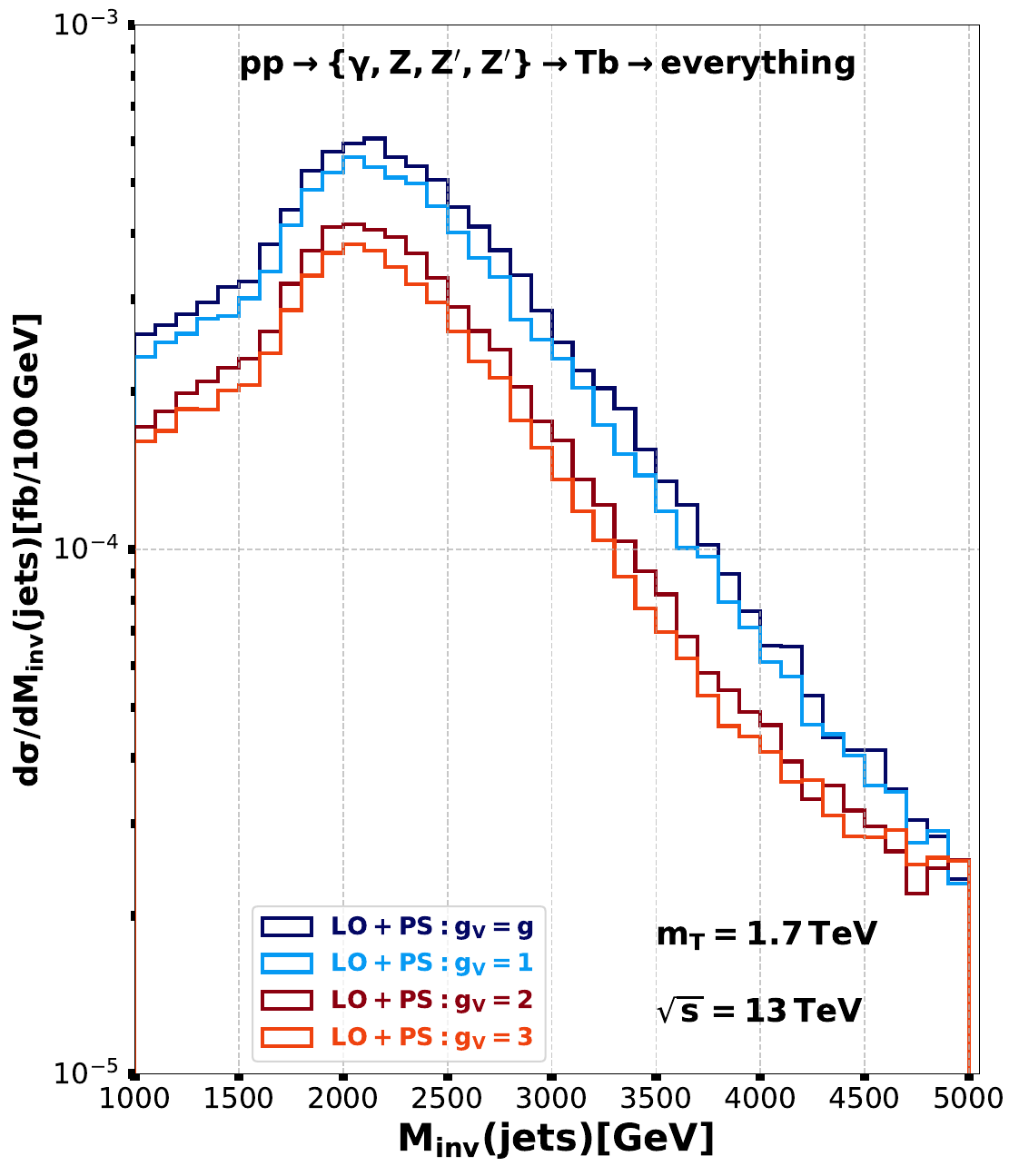}}
   \caption{\small LO+PS $M_{\text{inv}}$ of all jets differential distribution.}
  \label{distMinv}
\end{figure}

\begin{figure}[!h]
  \centering
  {\includegraphics[width=8.5cm,height=6.5cm]{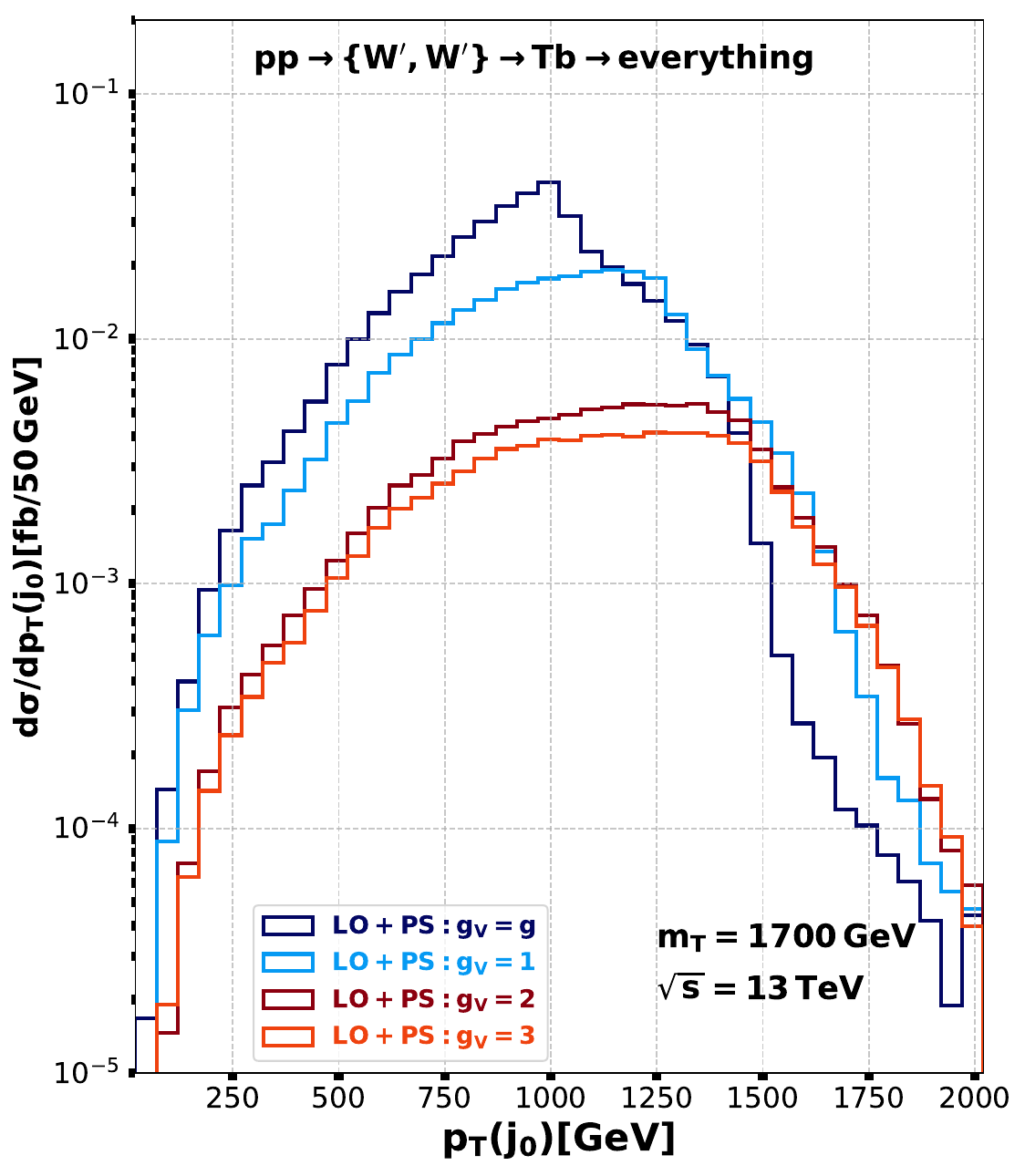}}
  \hfill
  {\includegraphics[width=8.5cm,height=6.5cm]{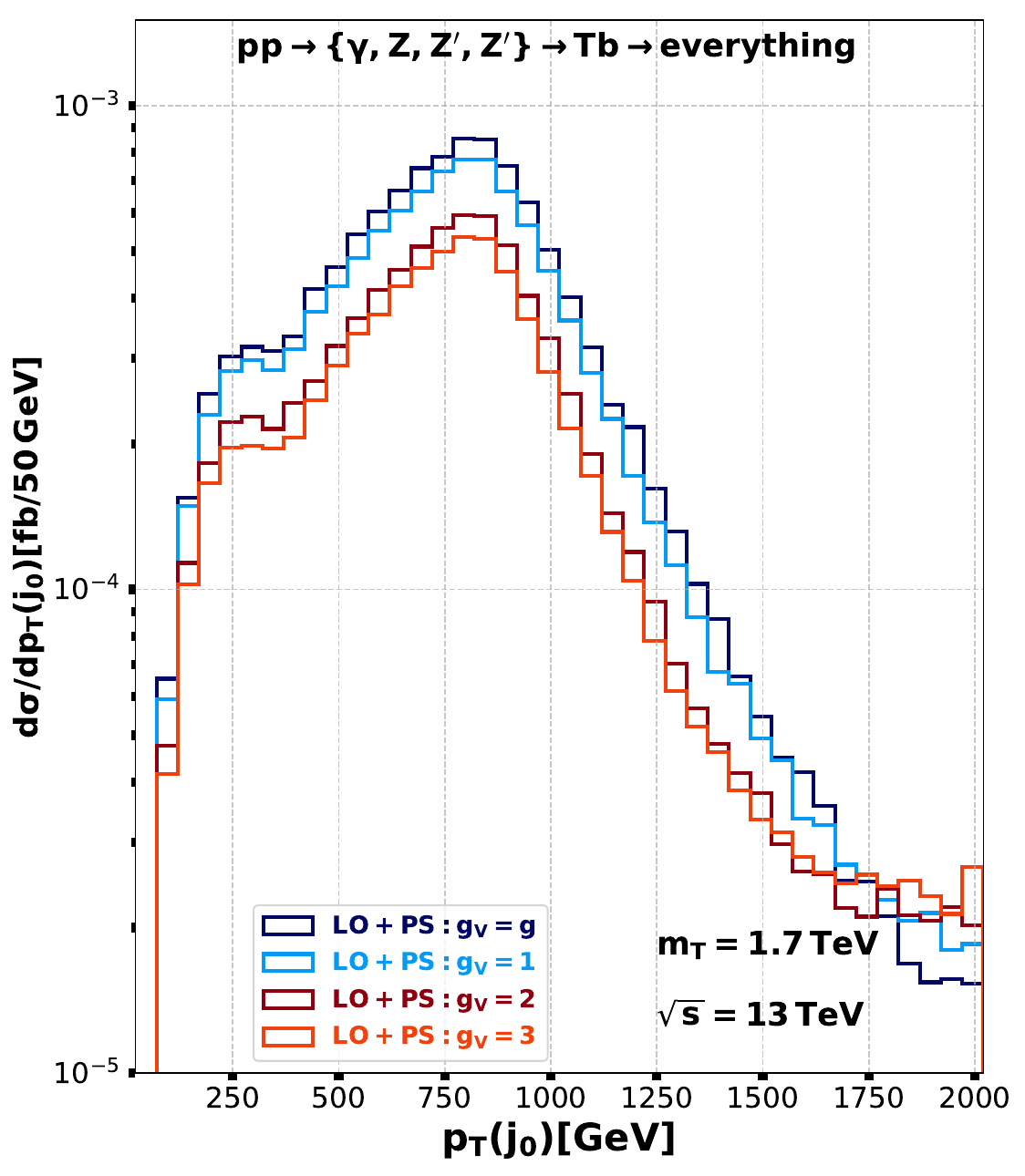}}
   \caption{\small LO+PS $p_{T}$ of the leading jet differential distribution.}
  \label{distpTj0}
\end{figure}

\begin{figure}[!h]
  \centering
  {\includegraphics[width=8.5cm,height=6.5cm]{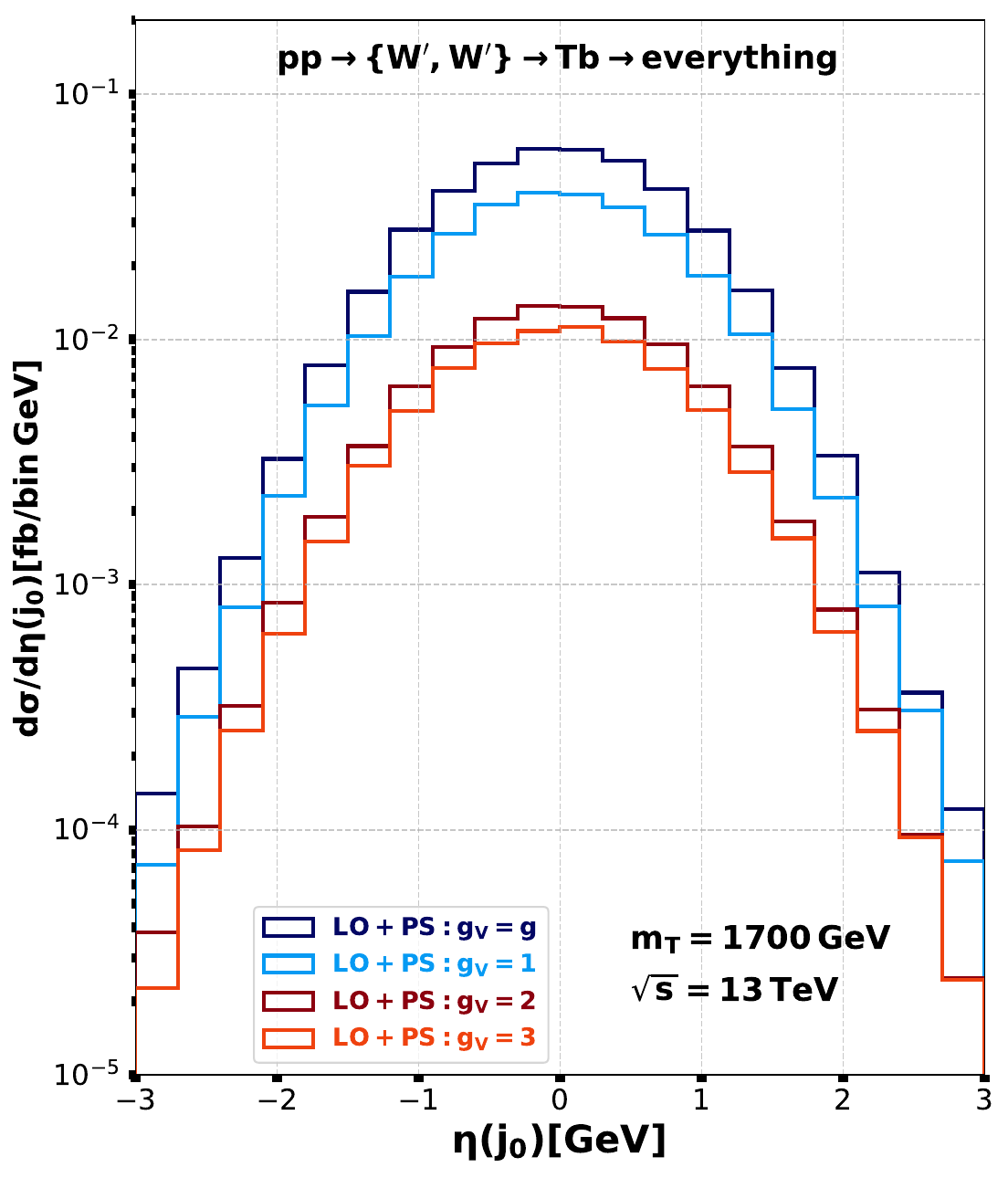}}
  \hfill
  {\includegraphics[width=8.5cm,height=6.5cm]{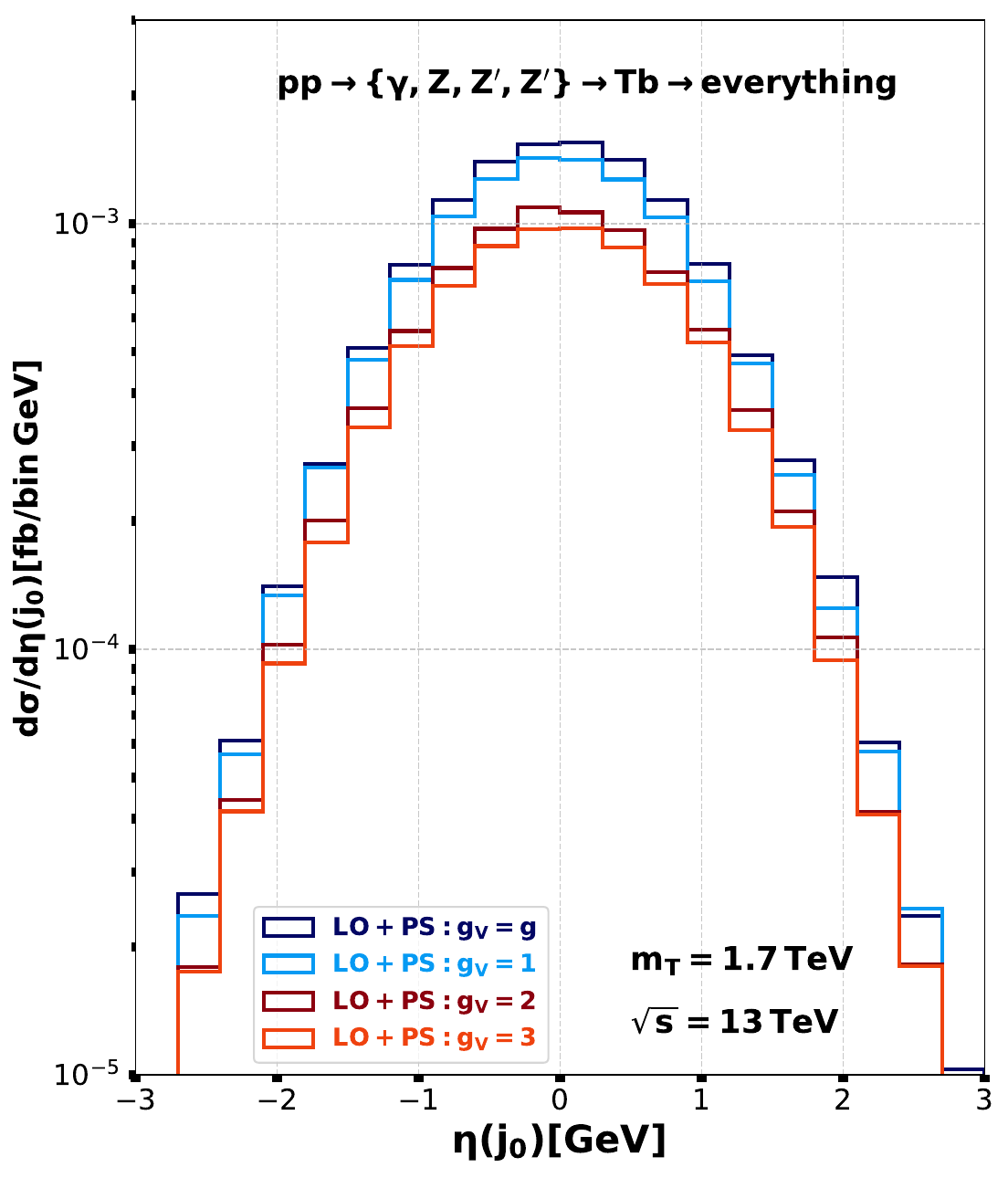}}
   \caption{\small LO+PS $\eta$ of the leading jet differential distribution.}
  \label{distetaj0}
\end{figure}

\noindent
Now, let's consider some LO matched to parton shower (LO+PS) differential cross sections at the reconstruction level for $m_{_T}=1.7$ TeV and $g_{_V}=g, 1, 2$ and $3$, see figure~\ref{distHT}, \ref{distMinv}, \ref{distpTj0} and \ref{distetaj0}.
We used {\tt M{\scriptsize AD}S{\scriptsize PIN}} \cite{Artoisenet:2012st} to automatically handle the decays of the final state particles, where both off-shell and spin correlation effects are retained, {\tt M{\scriptsize AD}A{\scriptsize NALYSIS}5} \cite{Conte:2014zja} to produce the differential distributions and {\tt F{\scriptsize AST}J{\scriptsize ET}} \cite{Cacciari:2006sm} for jet clustering.  To reduce the unwanted events, we applied the following cuts on the final-state jets: $P_{Tj}>20$, $|\eta_j|<5$ and $\Delta R_{jj}>0.4$\footnote{Here $j$ denotes both light-jets and b-jets.}.\\

\noindent
$\bullet$ In figure~\ref{distHT}, we show the scalar sum of the transverse momenta of all reconstructed jets distribution ($H_T$), in units of {\tt fb/100\, GeV}, for the CC (left panel) and NC (right panel) processes. We observe that the CC (NC) differential distributions have similar shapes but their magnitudes depend on the value of $g_{_V}$. In both cases, they are mostly populated at high $H_T$, where for the NC case they peak at $H_T\approx 1.8$ TeV for the 4 values of $g_{_V}$, i.e. the peak is independent of $g_{_V}$. In contrast, the positions of the peaks in the CC case are depend on $g_{_V}$, where they reach their maximum at $H_{_T}=2.2, 2.5, 2.7$ and $2.8$ TeV for $g_{_V}=g, 1, 2$ and $3$, respectively. This latter observation makes the $H_T$ distributions very helpful for guiding experimental research on the $W^{\prime}$ decaying into vector-like-quark.

\vspace{0.25cm}
\noindent
$\bullet$ In figure~\ref{distMinv}, we provide the all reconstructed final state jets invariant mass distributions ($M_{\text{inv}}(\text{jets})$) in units of {\tt fb/100\, GeV}. For the CC case (left panel), we observe that all the distributions peak at $m_{_{W^{\prime}}}=M_{\text{inv}}(\text{jets})$. The position of these peaks depends on the value of $g_{_V}$, as expected. This characteristic makes these distributions a very promising avenue for discovering the $W^{\prime}$ at colliders. Regarding the NC distributions (right panel), due to parton-shower effects, they are very small but not vanishing in the region $M_{\text{inv}}(\text{jets})<m_{_T}+m_t$ reach. They reach their maximum slightly above  $m_{_T}+m_t$. Here as well, the magnitude of the NC distribution is smaller than that of the CC case just as for the total cross section, which makes again the CC channel very promising for collider experiments. 

\vspace{0.25cm}
\noindent
$\bullet$ In figure~\ref{distpTj0}, we show the leading-jet transverse momentum distribution ($p_T(j_0)$), where $j_{0}$ is the highest reconstructed light-jet in $p_T$, in units of {\tt fb/50\, GeV}. The distributions for both processes are hard, as the jet $j_0$ must originate from the decay of a heavy state. The NC distributions (left panel), show peaks in the same position for different values of $g_{_V}$ (around $p_T(j_0)=0.9$ TeV). However, the CC distributions peak at different positions depending on the value of $g_{_V}$. For example,  peaks are observed at $p_T(j_0)=1.0, 1.2, 1.3$ and $1.4$ TeV for $g_{_V}=g, 1, 2$ and $3$, respectively. This latter observation suggest that these differential distributions could be used to constrain the coupling $g_{_V}$.

\vspace{0.25cm}
\noindent
$\bullet$ In figure~\ref{distetaj0}, we show the leading jet pseudo-rapidity distribution ($\eta(j_0)$), in units of {\tt fb/bin\_size}. In both cases (CC left panel, NC right panel), the magnitude of the distributions depends on $g_{_V}$, with this dependence being more pronounced for the CC processes.

\noindent
From the previous discussion, we conclude that the CC process provides a promising channel for discovering a $W^{\prime}$ boson that decays to a $T$-quark at collider experimental. Furthermore, the coupling $g_{_V}$ could be constrained using the studied distributions.


\section{Conclusion}
\label{conc}
In this paper, we introduced an extension of the LRSM with an extra $SU(2)$ gauge factor and one generation of the VLFs. To simplify the study, we assumed that the extra scalars are decoupled from the scale of interest and the VLFs mix with 3rd generation SM fermions. The model was studied in detail, starting from its gauge group, moving to the pattern of symmetry breaking, and ending with the diagonalization of mass matrices of the gauge bosons, scalars and fermions.\\

\noindent
One of the great achievements of this model is that it explains the smallness of the neutrinos masses, showing that the 1rst and 2nd generation neutrinos masses are governed by the seesaw relation of the LRSM, while the 3rd generation neutrino mass is controlled by a new seesaw relation which involves the VLN.\\ 

\noindent
We studied the production of a resonant $W^{\prime}$ gauge boson which decays to a $T$ quark and heavy Majorana neutrinos at the LHC, where the $T$ quark decays through $Zt$, $ht$ and $\gamma t$ channels. We exploited the {\tt run II} LHC data to set lower limits on the mass of $W^{\prime}$, and consequentially on the $Z^{\prime}$. We find that the most restrictive constraints come from the decay of the $W^{\prime}$ to the 2nd generation heavy Majorana neutrinos in association with $\mu$ channel.\\  

\noindent
We have investigated the $T$ single production in association with $t$ or $b$ quarks, where the mass of the $W^{\prime}$ is fixed to its lower experimental limit. We find that contributions of the $W^{\prime\prime}$, $Z^{\prime}$ and $Z^{\prime\prime}$ are effectively negligible and that the CC current contribution is significantly more important than the NC ones at the level of the total cross section and the differential distribution, where parton-shower is taken into account. \\

\noindent
It turns out that the neutral VLL could play the role of a viable dark matter candidate under certain assumptions on the Yukawa couplings, where it has been shown to have an excellent compatibility with the direct detection and indirect detection data. This work is addressed in detail a separate article~\cite{BBZ}.\\

\noindent
In this paper, all the masses of the extra scalars (neutral, charged, and doubly charged) were fixed at a high scale to ensure that they decouple from the scale of interest. We leave for future work the case where the scalars can be produced at the LHC, appearing either in the final states or as or virtual particles. 

\appendix

 \section{Minimization conditions}
\label{appA}
\noindent

In this appendix, we provide the explicit formulae of the eight minimization conditions, i.e. the eight first derivatives on the vevs and the phases $\theta_4$ and $\theta_L$ given in eq.~(\ref{minc1}). This conditions lead to the vev's seesaw relation discussed in section~\ref{sec3}. We have:
\begin{align}
 \frac{\partial V}{\partial u_{_R}}&=4\lambda_5 u_{_R}^3+\frac{u_{_R}}{2}\left[\left(k_1^2 + k_2^2\right) \left(\alpha_{10} + 2 \alpha_5\right)+ 8 k_1 k_2 \alpha_6 \, \cos\left(\theta_4\right)+ 2u_{_L}^2 \left(2 \alpha_4 + \alpha_9\right)+2\left(v_{_L}^2 + v_{_R}^2\right) \alpha_7 + v_{_R}^2 \alpha_8-
   4 \mu_4^2\right]\equiv0
   \label{minimcond1}
\\
 \frac{\partial V}{\partial u_{_L}}&=4\lambda_5 u_{_L}^3+\frac{u_{_L}}{2}\left[\left(k_1^2 + k_2^2\right) \left(\alpha_{10} + 2 \alpha_5\right)+ 8 k_1 k_2 \alpha_6 \, \cos\left(\theta_4\right)+ 2u_{_R}^2 \left(2 \alpha_4 + \alpha_9\right)+2\left(v_{_L}^2 + v_{_R}^2\right) \alpha_7 + v_{_L}^2 \alpha_8-
   4\mu_4^2\right]\equiv0
   \label{minimcond2}
\nonumber\\
\\
\frac{\partial V}{\partial k_1}&=k_1^3\lambda_1+k_1k_2^2 (\lambda_1+2\lambda_3)+k_2\left[k_2^2+3k_1^2\right]\lambda_4\cos(\theta_4)\nonumber\\
&+\frac{k_2}{2}\left[\left\{4\alpha_6(u_{_L}^2+u_{_R}^2)+2\alpha_2(v_{_L}^2+v_{_R}^2)-4\mu_2^2\right\}\cos(\theta_4)+v_{_L}v_{_R}\beta_1\cos(\theta_4-\theta_L)\right]\nonumber\\
&+\frac{k_1}{2}\left[(\alpha_{10}+2\alpha_{5}) (u_{_L}^2+u_{_R}^2)+\alpha_1(v_{_L}^2+v_{_R}^2)-2\mu_1^2+8\lambda_2 k_2^2\cos(2\theta_4)+2\beta_2 v_{_L}v_{_R}\cos(\theta_L)\right]
\label{minimcond3}\\
\nonumber\\
\frac{\partial V}{\partial k_2}&=k_2^3\lambda_1+k_2k_1^2 (\lambda_1+2\lambda_3)+k_1\left[k_1^2+3k_2^2\right]\lambda_4\cos(\theta_4)\nonumber\\
&+\frac{k_1}{2}\left[\left\{4\alpha_6(u_{_L}^2+u_{_R}^2)+2\alpha_2(v_{_L}^2+v_{_R}^2)-4\mu_2^2\right\}\cos(\theta_4)+v_{_L}v_{_R}\beta_1\cos(\theta_4-\theta_L)\right]\nonumber\\
&+\frac{k_2}{2}\left[(\alpha_{10}+2\alpha_{5}) (u_{_L}^2+u_{_R}^2)+(\alpha_1+\alpha_3)(v_{_L}^2+v_{_R}^2)-2\mu_1^2+8 \lambda_2 k_1^2\cos(2\theta_4)+2\beta_3 v_{_L}v_{_R}\cos(2\theta_4-\theta_L)\right]\equiv 0\nonumber\\
\label{minimcond4}
\\
\frac{\partial V}{\partial v_{_R}}&=\rho_1 v_{_R}^3+\frac{v_{_R}}{2}\left[\alpha_1 (k_1^2+k_2^2)+\alpha_3 k_2^2-2\mu_3^2+2\alpha_7(u_{_L}^2+u_{_R}^2)+\alpha_8 u_{_R}^2+\rho_3 v_{_L}^2+4\alpha_2\, k_1k_2\cos(\theta_4)\right]\nonumber\\
&+\frac{v_{_L}}{2}\left[\beta_1 k_1k_2\cos(\theta_4-\theta_L)+\beta_2 k_1^2\cos(\theta_L)+\beta_3 k_2^2\cos(2\theta_4-\theta_L)\right]\equiv 0\nonumber\\
\label{minimcond5}
\\
\frac{\partial V}{\partial v_{_L}}&=\rho_1 v_{_L}^3+\frac{v_{_L}}{2}\left[\alpha_1 (k_1^2+k_2^2)+\alpha_3 k_2^2-2\mu_3^2+2\alpha_7(u_{_L}^2+u_{_R}^2)+\alpha_8 u_{_L}^2+\rho_3 v_{_R}^2+4\alpha_2\, k_1k_2\cos(\theta_4)\right]\nonumber\\
&+\frac{v_{_R}}{2}\left[\beta_1 k_1k_2\cos(\theta_4-\theta_L)+\beta_2 k_1^2 \cos(\theta_L)+\beta_3 k_2^2 \cos(2\theta_4-\theta_L)\right]\equiv 0
\label{minimcond6}
\end{align}

\begin{align}
\frac{\partial V}{\partial \theta_4}&=-k_1k_2\left[\lambda_4(k_1^2+k_2^2)-2\mu_2^2+2\alpha_6(u_{_L}^2+u_{_R}^2)+\alpha_2 (v_{_L}^2+v_{_R}^2)\right] \sin(\theta_4)-4k_1^2k_2^2\lambda_2\sin(2\theta_4)\nonumber\\
&+\frac{v_{_L}v_{_R}}{2}k_2\left[\beta_1 k_1 \sin(\theta_4-\theta_L)+2\beta_3 k_2\sin(2\theta_4-\theta_L)\right]\equiv 0
\label{minimcond7}\\
\nonumber\\
\frac{\partial V}{\partial \theta_L}&=\frac{v_{_L}v_{_R}}{2}\left[\beta_1 k_1 k_2 \sin(\theta_4-\theta_L)-\beta_2 k_1^2 \sin(\theta_L)+\beta_3 k_2^2 \sin(2\theta_4-\theta_L)\right]\equiv 0
\label{minimcond8}
\end{align}

\noindent
 The four conditions $\partial V/\partial k_1=0$, $\partial V/\partial k_2=0$, $\partial V/\partial v_{_R}=0$ and $\partial V/\partial u_{_R}=0$ can be seen as linear equations of the 4 mass parameters $\mu_1^2$, $\mu_2^2$, $\mu_3^2$ and $\mu_4^2$. Thus, they can be exploited express them in terms of the potential couplings and vevs. We find:

\begin{align}
\mu^2_1=&\frac{1}{2 (k_1^2-k_2^2)}\, \left\{2 v_Lv_R\bigl[\beta_2k^2_1 \cos\left(\theta_L\right)-\beta_3 k^2_2 \cos\left(2\theta_4-\theta_L\right)-\alpha_3k^2_2 \left(v^2_L+v^2_R\right)\bigr]\right\}\nonumber\\
&+2k_1 k_2 \lambda_4 \cos\left(\theta_4\right)+\lambda_1 \left(k^2_1+k^2_2\right)+\frac{1}{2}\left(u^2_{_{L}}+u^2_{_{R}}\right)\left(\alpha_{10}+2\alpha_{5}\right)+\frac{1}{2}\left(v^2_{_{L}}+v^2_{_{R}}\right) \alpha_{1}\label{mu1param}
\\
\nonumber\\
\mu_2^2=&k_1k_2 \sec\left(\theta_4\right)\left[2 \lambda_2 \cos\left(2\theta_4\right)-\lambda_3\right]+\left(u^2_{_{L}}+u^2_{_{R}}\right)\,\alpha_6+\frac{1}{2}\left(k^2_1+k^2_2\right)\lambda_4+\frac{1}{2}\left(v^2_L+v^2_R\right)\left[\frac{\alpha_3 k_1 k_2 \sec\left(\theta_4\right)}{2\left(k^2_1-k^2_2\right)}+\alpha_2\right]\nonumber\\
 &+\frac{k_1k_2v_Lv_R}{2\left(k^2_1-k^2_2\right)} \bigl[\beta_3 \cos\left(2\theta_4-\theta_L\right)-\beta_2 \cos\left(\theta_L\right)\bigr] \sec\left(\theta_4\right)+\frac{1}{4}\beta_1v_Lv_R \sec\left(\theta_4\right) \cos\left(\theta_4-\theta_L\right)
 \end{align}
 
\begin{align}
\mu_3^2=&2\alpha_2k_1k_2 \cos\left(\theta_4\right)+\frac{1}{2}\alpha_3k^2_2+\frac{1}{2}\alpha_1\left(k^2_1+k^2_2\right)+\frac{1}{2}\rho_3v^2_L+\rho_1 v^2_R+\alpha_7\left(u^2_{_{L}}+u^2_{_{R}}\right)+\frac{1}{2}\alpha_8 u^2_{_{R}} \nonumber\\
 &+\frac{v_{_L}}{2v_{_R}} \bigl[ \beta_2k^2_1 cos\left(\theta_L\right)+\beta_1k_1k_2 \cos\left(\theta_4-\theta_L\right)+\beta_3k^2_2 \cos\left(2\theta_4-\theta_L\right)\bigr]
 \label{mu3param}
  \end{align}
 
\begin{align}
\mu_4^2=&\frac{1}{4} \bigl[\left(k1^2 + k2^2\right) \left(\alpha_{10} + 2 \alpha_5\right) +
2 \left(v_{L}^2 + v_{R}^2\right) \alpha_7+v_{R}^2 \alpha_8 +2 u_{_L}^2 \left(2 \alpha_4 + \alpha_9\right) + 8 u_{_R}^2 \lambda_5 + 8 k1 k2 \alpha_6 \cos\left(\theta_4\right)\bigr]
 \label{mu4param}
 \end{align}
 \section{Diagonalization of neutrinos mass matrix}
 \label{diagNmass}
 \noindent
The Dirac and Majorana mass matrices ($M_D$ and $M_R$) in the upper left block of $\mathcal{M}_{\mathcal{N}}$ (which we denote $\widetilde{\mathcal{M}}_{\mathcal{N}}$), cf.~ eq.~(\ref{MN}), can be diagonalized as the following:
\begin{align}
\widetilde{\mathcal{M}}_{\mathcal{N}}^{(0)}&=V_{_{\mathcal{N}}}^{\dagger} \widetilde{\mathcal{M}}_{\mathcal{N}} V_{_{\mathcal{N}}}
&&\text{with}&
V_{_\mathcal{N}}&=
\begin{pmatrix}
 V_{_\mathcal{N}}^{L}&0\\
 0&V_{_\mathcal{N}}^{R}
\end{pmatrix}
&&\text{and}&
\widetilde{\mathcal{M}}_{\mathcal{N}}^{(0)}&=
\left(
  \begin{array}{cc}
0 & \widetilde{M}_D \\
\widetilde{M}_D & \widetilde{M}_R 
\end{array}
\right)
\end{align}
where $V_{_\mathcal{N}}$ and $V_{_\mathcal{N}}^{L/R}$ are, respectively, $6\times6$ and $3\times3$ unitary matrices (i.e. $\left(V_{_\mathcal{N}}\right)^{\dagger}\cdot V_{_\mathcal{N}}=1\!\!1$ and $\left(V_{_\mathcal{N}}^{L/R}\right)^{\dagger}\cdot V_{_\mathcal{N}}^{L/R}=1\!\!1$), and $M_R$ is assumed to be symmetric. Thus, we have:
\begin{align}
\widetilde{M}_D&=\left(V_{_\mathcal{N}}^{L}\right)^{\dagger}\cdot M_D\cdot V_{_\mathcal{N}}^{R}\equiv \text{diag}\left[m_{\nu_1}\, \, , m_{\nu_2}\, \, , m_{\nu_3}\right].\\
\widetilde{M}_R&=\left(V_{_\mathcal{N}}^{R}\right)^{\dagger}\cdot M_R\cdot V_{_\mathcal{N}}^{R}\equiv \text{diag}\left[M_{_{N_1}}\, \, , M_{_{N_2}}\, \, , M_{_{N_3}}\right].
\end{align}
\noindent
The matrix $\widetilde{\mathcal{M}}_{\mathcal{N}}^{(0)}$ is $6\times6$ symmetric matrix, therefore it can be approximately diagonalized by an approximate unitary transformation~\cite{a}. Thus, we write:
 \begin{align}
 \widetilde{\mathcal{M}}_{\mathcal{N}}^{(1)}=U_{^{\mathcal{N}}}^{\dagger}\widetilde{\mathcal{M}}_{\mathcal{N}}^{(0)} U_{_{\mathcal{N}}}
 &&
  U_{_{\mathcal{N}}}&\approx
  \begin{pmatrix}
1 & \widetilde{M}_D \widetilde{M}_R^{-1} \\
 -\widetilde{M}_D \widetilde{M}_R^{-1} & 1 \\
\end{pmatrix}
&
 \widetilde{\mathcal{M}}_{\mathcal{N}}^{(1)}\simeq
   \begin{pmatrix}
-\widetilde{M}_D^2 \widetilde{M}_R^{-1} & 0 \\
0 & \widetilde{M}_R \\ 
\end{pmatrix}
\end{align}
We note that, to obtain the approximate mass matrix $\mathcal{M}_{\mathcal{N}}^{(1)}$, we neglected terms of order $v_{_\text{EW}}^2/v_{_R}^2$ in $U_{_{\mathcal{N}}}$ and $\widetilde{\mathcal{M}}_{\mathcal{N}}^{(1)}$ are neglected. Thus, the matrix $V^{\nu}$ in eq.~(\ref{mnvw}) can be approximated by:
\begin{align}
V^{\nu}&\approx
\begin{pmatrix}
V_{_\mathcal{N}}^{L}\ & \widetilde{M}_D \widetilde{M}_R^{-1} \\
 -\widetilde{M}_D \widetilde{M}_R^{-1} & V_{_\mathcal{N}}^{R}\ \\
\end{pmatrix}
\end{align}

\section{Scalar sector rotations}
 \label{appD}
Under the mentioned assumptions in section \ref{sec5} of which $\left(v_{_{EW}},k_{-}\right)/\left(v_R,\mathfrak{u}_{_{R}}\right)\ll1$ and using the mass parameters relations with the quartic couplings in appendix \ref{appA}, it becomes clear that  $\delta_{R}^{0Im}$, $\phi_{2}^{0Im}$ and $\phi_{2}^{\pm}$ states are just the longitudinal components for the $Z^{\prime}$, $Z^{\prime\prime}$ and $W^{\prime\prime}$ gauge bosons, respectively. The terms of the blocks that are already diagonalized in the neutral mass matrix, which associated to $\phi_{1}^{0Re},\delta_{L}^{0Re}$ and $\phi_{1}^{0Im},\delta_{L}^{0Im}$ states. In fact those are nothing but the masses of the scalars $m^2_{H^0_3},m^2_{H^0_5}$ and the pseudo-scalars $m^2_{A^0_2},m^2_{A^0_3}$, respectively. The remnant non diagonal blocks are $M^2_{H^0_{4\times4}}$ and $M^2_{A^0_{2\times2}}$ matrices which associated to the real and imaginary states of the neutral fields. $M^2_{H^0_{4\times4}}$ is associated to the real basis $\{ \phi_{-}^{0}, \phi_{+}^{0},\delta_R^{0Re}, \phi_2^{0Re} \}$, it takes the form,
\subsection{Neutral scalars}
The neutral sector $12\times12$ mass matrix is reduced to $10\times10$ mass matrix, as the fields $\delta_{R}^{0\text{im}}$, $\phi_{2}^{0\text{im}}$ have vanishing eigenvalues (cf. eq.~(\ref{neutralBase})). These two field are interpreted as the Goldstone fields which provide the longitudinal components for the $Z^{\prime}$ and $Z^{\prime\prime}$ neutral gauge bosons. We note that the reduced mass matrix can be decomposed in 3 independent blocks associated, respectively, to the following basis: 
\begin{align}
\{\phi_{a}^{0}\}&\equiv \{\phi_{3}^{0\, \text{re}},\phi_{4}^{0\, \text{re}},\delta_{R}^{0\, \text{re}},\phi_{2}^{0\, \text{re}}\}\label{pho0a}\\
\{\phi_{b}^{0}\}&\equiv \{\delta_{L}^{0\, \text{re}},\phi_{1}^{0\, \text{re}},\delta_{L}^{0\, \text{im}},\phi_{1}^{0\, \text{im}}\}\label{pho0b}\\
\{\phi_{c}^{0}\}&\equiv \{\phi_{3}^{0\, \text{im}},\phi_{4}^{0\, \text{im}}\}\label{pho0c}
\end{align}

\noindent 
The fields $\phi_{3}^{0\, \text{re}}$ and $\phi_{4}^{0\, \text{re}}$ mix to form the new neutral scalars $\tilde{\phi}_{3}^{0\, \text{re}}$ and $\tilde{\phi}_{4}^{0\, \text{re}}$, defined by:
\begin{align}
\begin{pmatrix}
 \tilde{\phi}_{3}^{0\, \text{re}}\\
 \tilde{\phi}_{4}^{0\, \text{re}} 
\end{pmatrix}
&=
\begin{pmatrix}
 c_{\alpha}& s_{\alpha}\\
 -s_{\alpha}&  c_{\alpha}
\end{pmatrix}
\begin{pmatrix}
 \phi_{3}^{0\, \text{re}}\\
 \phi_{4}^{0\, \text{re}} 
\end{pmatrix}
\end{align}
where $c_{\alpha}=\cos(\theta)=k_1/v_{_{EW}}$ and $s_{\alpha}=\sin(\theta)=k_2/v_{_{EW}}$. Thus, the basis (\ref{pho0a}) becomes:
\begin{align}
 \{
 \phi_{-}^{0},
 \phi_{+}^{0},
 \delta_R^{0\, \text{re}},
 \phi_2^{0\, \text{re}}
\}=
 \{c_{\alpha}\, \phi_3^{0\, \text{re}}+s_{\alpha}\,\phi_4^{0\, \text{re}},
 c_{\alpha}\, \phi_4^{0\, \text{re}}-s_{\alpha}\,\phi_3^{0\, \text{re}},
 \delta_R^{0\, \text{re}},
 \phi_2^{0\, \text{re}}
\}\label{pho0aa}
\end{align}

\noindent
The elements of the mass matrix $M^{_{\{0,a\}}}$, in the new basis (\ref{pho0aa}), are given by:
\begin{align}
M^{_{\{0,a\}}}_{11}&=   2v_{_{EW}}^2 [\lambda_1+4 s_{\alpha}^2 c_{\alpha}^2 (2\lambda_2+\lambda_3)+4 s_{\alpha} c_{\alpha}\lambda_4],&
M^{_{\{0,a\}}}_{12}&= 2v_{_{EW}}^2(c_{\alpha}^2-s_{\alpha}^2) [2 s_{\alpha} c_{\alpha} \left(2\lambda_2+\lambda_3\right)+ \lambda_4],\nonumber\\
M^{_{\{0,a\}}}_{13}&=v_R v_{_{EW}} [\alpha_1+4 s_{\alpha} c_{\alpha} \alpha_2+s_{\alpha}^2 \alpha_2],&
M^{_{\{0,a\}}}_{14}&= u_{_{R}}v_{_{EW}}\bigl[\left(\alpha_{10}+2\alpha_5\right)+8 s_{\alpha} c_{\alpha} \alpha_6\bigr],\nonumber\\
M^{_{\{0,a\}}}_{22}&=\frac{v_{_R}^2 \alpha_3}{2(c_{\alpha}^2-s_{\alpha}^2)}+2\left(2\lambda_2+\lambda_3\right) (c_{\alpha}^2-s_{\alpha}^2),& 
M^{_{\{0,a\}}}_{23}&=\frac{1}{2}v_{_R} v_{_{EW}}\bigl[4\alpha_2 (c_{\alpha}^2-s_{\alpha}^2)+ 2 s_{\alpha} c_{\alpha} \alpha_3\bigr]\nonumber\\
M^{_{\{0,a\}}}_{24}&=4u_{_{R}} v_{_{EW}} \alpha_6 (c_{\alpha}^2-s_{\alpha}^2), &
M^{_{\{0,a\}}}_{33}&=2\rho_1v_{_R}^2,\nonumber\\
M^{_{\{0,a\}}}_{34}&=u_{_{R}} v_{_R}(2\alpha_7+\alpha_8),&
M^{_{\{0,a\}}}_{44}&=8 u_{_{R}}^2 \lambda_5.
\label{MassN0a}
\end{align}

\noindent
The mass matrix $M^{_{\{0,a\}}}$ is symmetric; thus, it can in principle be diagonalized by a unitary transformation, which transforms the scalar fields of the basis (\ref{pho0aa}) as:
\begin{align}
\begin{pmatrix}
h\\
H_1^{0}\\
H_2^{0} \\
H_4^{0} \\
\end{pmatrix}=
U_{H^{0}_{a}}
\begin{pmatrix}
 \phi_{-}^{0}\\
 \phi_{+}^{0}\\
 \delta_R^{0Re} \\
 \phi_2^{0Re}\\
\end{pmatrix}
\end{align}
where $h$ is the SM Higgs boson and $H_1^{0}$, $H_2^{0}$ and $H_3^{0}$ are new neutral scalars heavier than the ordinary Higgs boson. 

\noindent
However, since the four eigenvalues of this matrix are non-vanishing, such diagonalization becomes very complicated in the general case. Yet, under certain assumptions on the potential couplings and the scalar vevs, we can take the diagonal elements of $M^{_{\{0,a\}}}$ as a good approximation of the mass squared of the associated neutral scalar fields, and the unitary transformation can be approximated by the $4\times4$ identity matrix. We will prove this in the following. \\

\noindent
We introduce the new parameters $\xi$, $\varepsilon$ and $\varepsilon^{\prime}$, such that:
\begin{align}
\xi&=\frac{k_2}{k_1}\ll 1, & \varepsilon&=\frac{v_{_\text{EW}}}{v_{_R}}\ll 1, & \varepsilon^{\prime}&=\frac{v_{_R}}{u_{_R}}\lessapprox1.
\end{align}
We note that $c_{\alpha}$ and $s_{\alpha}$ are now expressed as: $c_{\alpha}=1/\sqrt{1+\xi^2}$ and $c_{\alpha}=\xi/\sqrt{1+\xi^2}$. We assume that all the dimensionless coefficients of the potential scale as $\mathcal{O}(1)$ except the following ones:
\begin{align}
\lambda_4&=\tilde{\lambda}_4 \xi \sim \mathcal{O}(\xi), & \alpha_i&=\tilde{\alpha}_i \xi \sim \mathcal{O}(\xi)\,\, (\text{for}\,\, i=2,5,6,7,8,10), &
\alpha_1&=\tilde{\alpha}_1 \xi^2 \sim \mathcal{O}(\xi^2).
\label{paramchoice}
\end{align}
\noindent
We expand the mass matrix $M^{_{\{0,a\}}}$ around $\xi=0$ and $\varepsilon=0$, we can write:
\begin{align}
M^{_{\{0,a\}}}&=M^{_{\{0,a\}}}_{0}+M^{_{\{0,a\}}}_{1}+M^{_{\{0,a\}}}_{2}+\mathcal{O}(\xi^3,\varepsilon^3).
\end{align}
with
\begin{align}
M^{_{\{0,a\}}}_{0}&=u_{_R}^2
\begin{pmatrix}
0&0&0&0\\
0& \alpha_3 \varepsilon^{\prime\, 2}/4 &0 &0\\
0& 0 & \rho_1 \varepsilon^{\prime\, 2} & 0\\
0&0&0& 4\lambda_5
\end{pmatrix}, &
M^{_{\{0,a\}}}_{1}&=u_{_R}^2
\begin{pmatrix}
0&0&0&\varepsilon\varepsilon^{\prime}\xi\, [\cdots]\\
0& 0 &\varepsilon\varepsilon^{\prime\, 2}\xi\, [\cdots] &\varepsilon\varepsilon^{\prime}\xi\, [\cdots]\\
0& \varepsilon\varepsilon^{\prime\, 2}\xi\, [\cdots]& 0 & \varepsilon^{\prime}\xi\, [\cdots]\\
\varepsilon\varepsilon^{\prime}\xi\, [\cdots]&\varepsilon\varepsilon^{\prime}\xi\, [\cdots]&\varepsilon^{\prime}\xi\, [\cdots]&0
\end{pmatrix},
\label{Ma001}
\end{align}
and
\begin{align}
M^{_{\{0,a\}}}_{2}&=u_{_R}^2
\begin{pmatrix}
\varepsilon^2\varepsilon^{\prime\, 2}[\lambda_1+\xi^2(2\lambda_2+\lambda_3+\tilde{\lambda}_4)]&\varepsilon^2\varepsilon^{\prime\, 2}\xi\, [\cdots]&\varepsilon\varepsilon^{\prime\, 2}\xi^2\, [\cdots]&\varepsilon\varepsilon^{\prime}\xi^2\, [\cdots]\\
\varepsilon^2\varepsilon^{\prime\, 2}\xi&\varepsilon^2\varepsilon^{\prime\, 2}(2\lambda_2+\lambda_3) (1-4\xi^2)+\frac{1}{2}\varepsilon^{\prime\, 2}\xi^2 \alpha_3 &0 &0\\
\varepsilon\varepsilon^{\prime\, 2}\xi^2\, [\cdots]& 0 & 0 & 0\\
\varepsilon\varepsilon^{\prime}\xi^2\, [\cdots]&0&0& 0
\end{pmatrix}.
\label{Ma02}
\end{align}
\noindent
To simplify the discussion, we provide only the full expansion of the diagonal elements and not the off-diagonal onces of the matrices given in eqs.~(\ref{Ma001}) and (\ref{Ma02}), where for the latter we add $[\cdots]$ to indicate this omission.\\

\noindent
We notice that the parameter $\xi$ is much smaller than the parameter $\varepsilon$ (i.e. $\xi \ll \varepsilon$). For example, for $k_2=1$ GeV, $v_{_\text{EW}}=246$ GeV and $v_{_R}=3000$ GeV, we obtain $\xi=0.004$ and $\varepsilon=0.082$. Thus, we can neglect all terms proportional to $\varepsilon^2\xi^2$, $\varepsilon\xi^2$, $\varepsilon^2\xi$, $\varepsilon\xi$, or in other words, all off-diagonal elements can be neglected, and the diagonal ones retained. We recall that this strategy (i.e. neglecting the off-diagonal elements) is adopted in the ordinary LRS models, see for example~\cite{P,Dev,Hm}.\\

\noindent
The fact that the off-diagonal elements of the mass matrix are very small means that the unitary transformation that diagonalize such matrix will have very small off-diagonal components, because this unitary transformation is constructed from the normalized eigenvectors of the mass matrix. As noted earlier, obtaining analytical expression of the unitary transformation is very complicated, since  we are dealing with four non-vanishing eigenvalues. However, we can demonstrate this numerically for the specific parameter choice given in eq.~(\ref{paramchoice}). For example, taking all dimensionless tiled (and non-tiled) parameters of order unity, and setting $\xi=0.04$, $\varepsilon=0.082$ and $\varepsilon^{\prime}=0.95$, the unitary matrix is given by:
\begin{align}
 U_{H^{0}_{a}}&=
 \begin{pmatrix}
  1.000000& -0.000714& -0.000002&- 0.000118\\
0.000714& 1.000000& -0.000673& -0.000165\\
0.000003& 0.000673& 0.999998& -0.001840\\
0.000118& 0.000166& 0.001840& 0.999998
\end{pmatrix}
\approx 
\begin{pmatrix}
  1 & 0 & 0 & 0 \\
  0 & 1 & 0 & 0 \\
  0 & 0 & 1 & 0 \\
  0 & 0 & 0 & 1 \\
\end{pmatrix}
\end{align}
Thus, we can take approximately $U_{H^{0}_{a}}$ as just the $4\times4$ identity matrix. Hence, we can write: 
\begin{align}
\begin{pmatrix}
h\\
H_1^{0}\\
H_2^{0} \\
H_4^{0} \\
\end{pmatrix}\approx
\begin{pmatrix}
 \phi_{-}^{0}\\
 \phi_{+}^{0}\\
 \delta_R^{0\, \text{re}} \\
 \phi_2^{0\, \text{re}}\\
\end{pmatrix}
\end{align}
In this case, the masses of the physical Higgs $h$, $H_1^{0}$, $H_2^{0}$ and $H_4^{0}$ correspond, respectively, to $M^{_{\{0,a\}}}_{11}$, $M^{_{\{0,a\}}}_{22}$, $M^{_{\{0,a\}}}_{33}$ and $M^{_{\{0,a\}}}_{44}$, cf. eq.~(\ref{MassN0a}). Note that if the parameters $\alpha_7$ and $\alpha_8$ are of order $\mathcal(1)$ (not $\mathcal(\xi)$ as assumed in eq.~(\ref{paramchoice})), then the off-diagonal element $M^{_{\{0,a\}}}_{34}$ ($M^{_{\{0,a\}}}_{43}$) are not any more negligible. In this case, the unitary transformation must be expressed as:
\begin{align}
  U_{H^{0}_{4\times4}}=\begin{pmatrix}
                        1 & 0 & 0 & 0 \\
                        0 & 1 & 0 & 0 \\
                        0 & 0 & c_{x} & s_{x} \\
                        0 & 0 & -s_{x} & c_{x}
                       \end{pmatrix} &&\text{with}&&
                       \tan\left({2\,x}\right)=\frac{u_{_{R}} v_{_R}\left(2\alpha_7+\alpha_8\right)}{4 u_{_{R}}^2 \lambda_5-v_{_R}^2\rho_1}
\end{align}
Thus, the physical Higgs bosons fields are defined by: $h\equiv \phi_{-}^{0}$, $H_1^{0}\equiv\phi_{+}^{0}$, $H_2^{0}\equiv c_{x}\delta_R^{0\, \text{re}}+s_{x}\phi_2^{0\, \text{re}}$ and $H_4^{0}\equiv -s_{x}\delta_R^{0\, \text{re}}+c_{x}\phi_2^{0\, \text{re}}$, where their masses are given by:
\begin{align}
m_{_h}&=2v_{_{EW}}^2 [\lambda_1+4 s_{\alpha}^2 c_{\alpha}^2 (2\lambda_2+\lambda_3)+4 s_{\alpha} c_{\alpha}\lambda_4],& 
m_{_{H_1^{0}}}&=\frac{v_R^2 \alpha_3}{2(c_{\alpha}^2-s_{\alpha}^2)}+2\left(2\lambda_2+\lambda_3\right) (c_{\alpha}^2-s_{\alpha}^2),\nonumber\\
m_{_{H_2^{0}}}&=\frac{4 [\cos(2\xi)-1]\lambda_5 u_{_R}^2+[\cos(2\xi)+1] \rho_1 v_{_R}^2}{2\cos(2\xi)},&
m_{_{H_4^{0}}}&=\frac{4 [\cos(2\xi)+1]\lambda_5 u_{_R}^2+[\cos(2\xi)-1] \rho_1 v_{_R}^2}{2\cos(2\xi)}.
\end{align}
with 
\begin{align}
\cos(2\xi)&=1/\left(1+\frac{u_{_{R}}^2 v_{_R}^2\left(2\alpha_7+\alpha_8\right)^2}{(4 u_{_{R}}^2 \lambda_5-v_{_R}^2\rho_1)^2}\right)^{1/2}.
\end{align}

\noindent
Now, let's focus on the basis $\{\phi_3^{0\, \text{im}},\phi_4^{0\, \text{im}}\}$ (cf. eq.~(\ref{pho0c})). The mass matrix associated to this basis is expressed as the following:
\begin{align}
M^2_{A^0}&=
\frac{v_{_R}^2}{2} \left[\alpha_3/(c_{\alpha}^2-s_{\alpha}^2)-4\varepsilon^2(2 \lambda_2-\lambda_3)\right]
\begin{pmatrix}
s_{\alpha}^2&
-s_{\alpha}c_{\alpha} \\
-s_{\alpha}c_{\alpha} &
c_{\alpha}^2
\end{pmatrix}
\end{align}

\noindent
This matrix can be diagonalized by the following exact unitary transformation:
\begin{align}
\begin{pmatrix}
G^{0}  \\
A_1^{0} \\
\end{pmatrix}=  \begin{pmatrix}
  c_{\alpha} & -s_{\alpha}\\
  s_{\alpha} & c_{\alpha}
 \end{pmatrix}
\begin{pmatrix}
 \phi_3^{0\, \text{im}}\\
 \phi_4^{0\, \text{im}}\\
\end{pmatrix}
\end{align}
where $G^0$ is a massless scalar which is identified with the Goldstone boson required to give mass to the $Z$ gauge boson, and $A_1^{0}$ is a neutral pseudo-scalar with the mass:
\begin{align}
m^2_{A^{0}_1}&=\frac{\alpha_3}{2 (c_{\alpha}^2-s_{\alpha}^2)}\, v_{_R}^2-2\, (2 \lambda_2-\lambda_3)\, v_{_{EW}}^2.
\end{align}

\noindent
The mass matrix associated the basis (\ref{pho0b}) is diagonal. Therefore, the real components are identified to neutral scalars (denoted $H^{0}_3$ and $H^{0}_5$) and the imaginary components to pseud-scalars (denoted $A^{0}_2$ and $A^{0}_3$). The masses of these bosons are given by:
\begin{align}
m_{_{H^{0}_3}}&=\frac{1}{2}\left[(\alpha_3-2\alpha_1) v_{_R}^2-\alpha_8 u_{_R}^2\right]=m_{_{A^{0}_2}}, &
m_{_{H^{0}_5}}&= (2\alpha_4+\alpha_9-4\lambda_5) u_{_R}^2-\frac{1}{2}\alpha_8 v_{_R}^2=m_{_{A^{0}_3}}.
\end{align}
\subsection{Singly and doubly charged scalars}
The fields $\phi_{2}^{\pm}$ (cf. eq.~(\ref{SingChargBase})) are massless, thus they are identified to the Goldstone bosons that will be absorbed to made the longitudinal components of the charged $W^{\prime\prime}$ gauge bosons. Therefore, The mass matrix of the singly charged scalars, is reduced to $5\times5$  matrix according to $\{\phi_{3}^{\pm},\phi_{4}^{\pm},\delta_{R}^{\pm},\phi_{1}^{\pm},\delta_{L}^{\pm}\}$, which decomposes into two independent blocks according to :
 \begin{align}
  m_{H^{\pm}}^2=
  \left(
  \begin{array}{ccc!{\vrule width 0.4pt}cc}
\frac{\alpha_3 s_{\alpha}^2 v_{_R}^2}{2 (c_{\alpha}^2-s_{\alpha}^2)} & -\frac{\alpha_3 s_{\alpha}c_{\alpha} v_{_R}^2}{2 (c_{\alpha}^2-s_{\alpha}^2)} & \frac{\alpha_3 s_{\alpha} v_{_{EW}} v_{_R}}{2\sqrt{2}}&0&0\\
-\frac{\alpha_3 s_{\alpha}c_{\alpha} v_{_R}^2}{2 (c_{\alpha}^2-s_{\alpha}^2)} & \frac{\alpha_3 c_{\alpha}^2 v_{_R}^2}{2 (c_{\alpha}^2-s_{\alpha}^2)}  & -\frac{\alpha_3 c_{\alpha} v_{_{EW}} v_{_R}}{2\sqrt{2}}&0&0\\
\frac{\alpha_3 s_{\alpha} v_{_{EW}} v_{_R}}{2\sqrt{2}}&-\frac{\alpha_3 c_{\alpha} v_{_{EW}} v_{_R}}{2\sqrt{2}}&
\frac{\alpha_3 (c_{\alpha}^2-s_{\alpha}^2) v_{_{EW}}^2}{4}&0&0\\
\hline
0 & 0& 0& m^2_{H^{\pm}_2}& 0\\
0&0&0&0&m^2_{H^{\pm}_3}
\end{array}
\right)
 \end{align}

\noindent
The diagonal lower $2\times2$ block provide the masses of the scalars $\phi_{1}^{\pm}\equiv H^{\pm}_2$ and $\delta_{L}^{\pm}\equiv H^{\pm}_3$, which are given by:
\begin{align}
m^2_{H^{\pm}_2}&=(2\alpha_4+\alpha_9-4\lambda_5)-\frac{1}{2} \alpha_8 v_{_R}^2,&
m^2_{H^{\pm}_3}&=\frac{1}{2}\left[-\alpha_8 u_{_R}^2-(2\alpha_1-\alpha_3) v_{_R}^2+\alpha_3 (c_{\alpha}^2-s_{\alpha}^2) v_{_{EW}}^2\right].
\end{align}

\noindent
The fact that the symmetric $3\times3$ upper block have two vanishing eigenvalues, makes the derivation of the exact unitary transformation that diagonalize it not complicated. It is easy to show that such transformation is given by:
\begin{align}
 \begin{pmatrix}
 G_{R_1}^{\pm}\\
 G_{R_2}^{\pm} \\
 H_1^{\pm}\\
\end{pmatrix}&=
\begin{pmatrix}
(c_{\alpha}^2-s_{\alpha}^2) \frac{v_{_{EW}}}{\sqrt{n_1}}& c_{\alpha}&\sqrt{2} s_{\alpha} \frac{v_{_R}}{\sqrt{n_2}}\\
0&s_{\alpha}&-\sqrt{2} c_{\alpha} \frac{v_{_R}}{\sqrt{n_2}}\\
-\sqrt{2} s_{\alpha} \frac{v_{_R}}{\sqrt{n_1}}&0&(c_{\alpha}^2-s_{\alpha}^2) \frac{v_{_{EW}}}{\sqrt{n_2}}
\end{pmatrix}
\begin{pmatrix}
 \phi_3^{\pm}\\
 \phi_4^{\pm}\\
 \delta_R^{\pm} \\
\end{pmatrix}
\end{align}
with $n_1=(c_{\alpha}^2-s_{\alpha}^2)^2 v_{_{EW}}^2+2 s_{\alpha}^2 v_{_R}^2$ and $n_2=(c_{\alpha}^2-s_{\alpha}^2)^2 v_{_{EW}}^2+2 v_{_R}^2$. The massless fields $G_{R_1}^{\pm}$ and $G_{R_2}^{\pm}$ are the Goldstone bosons required to generate the mass of the  $W$ and $W^{\prime}$ gauge bosons, whereas $H_1^{\pm}$ is a singly-charged scalar having the mass:
\begin{align}
m_{H_1^{\pm}}&=\frac{\alpha_3}{4} \left[(c_{\alpha}^2-s_{\alpha}^2) v_{_{EW}}^2+\frac{2}{c_{\alpha}^2-s_{\alpha}^2} v_{_R}^2\right].
\end{align}

\noindent
The mass matrix of the doubly charged scalars is diagonal. In the basis (\ref{DoubChargBase}), it is expressed as:
\begin{align}
m_{H^{\pm\pm}}^2=\frac{1}{2}
\begin{pmatrix}
 -\alpha_8 u_{_R}^2-(2\alpha_1-\alpha_3) v_{_R}^2+\alpha_3 (c_{\alpha}^2-s_{\alpha}^2) v_{_{EW}}^2&0\\
 0& \alpha_3 (c_{\alpha}^2-s_{\alpha}^2) v_{_{EW}}^2+\alpha_2 \frac{2}{c_{\alpha}^2-s_{\alpha}^2} v_{_R}^2
\end{pmatrix}
\end{align}
where the diagonal components correspond to the masses of the doubly charged scalars $m^2_{H_1^{\pm\pm}}\equiv\delta_L^{\pm\pm}$ and $m^2_{H_2^{\pm\pm}}\equiv\delta_R^{\pm\pm}$.\\

\section*{Acknowledgments}
We are very thankful to M. R. Hadj-Sadok for useful discussions and precious remarks.

\bibliographystyle{unsrt}
\bibliography{biblio}
\end{document}